\documentclass[letterpaper,twocolumn,10pt]{article}

\usepackage{zhanggroup}
\usepackage{tikz}
\usepackage{amsfonts}
\usepackage{xspace} 
\usepackage{amsmath}
\usepackage{mathrsfs}
\usepackage{amssymb}
\usepackage{amsmath}
\usepackage{amsthm}
\usepackage{multirow}
\usepackage{makecell}
\usepackage[sort&compress,numbers]{natbib}
\usepackage{caption}
\usepackage{subcaption}
\usepackage{float}
\usepackage{booktabs}
\usepackage{balance}
\usepackage{xcolor}
\usepackage{tcolorbox}
\usepackage{hyperref}
\hypersetup{
  colorlinks,
  linkcolor={blue!70!green},
  citecolor={green!70!blue},
  urlcolor={orange!70!red}
}

\usepackage[hang,flushmargin]{footmisc}
\usepackage[absolute]{textpos}

\newcommand{\mypara}[1]{\smallskip\noindent{\bf {#1}.}\xspace}

\begin{document}

\date{}
\pagestyle{plain} 

\author{
Xinlei He \ \ \ 
Xinyue Shen \ \ \ 
Zeyuan Chen \ \ \ 
Michael Backes \ \ \
Yang Zhang \ \ \ 
\\
\\
\textit{CISPA Helmholtz Center for Information Security} \ \ \
}

\title{MGTBench: Benchmarking Machine-Generated Text Detection}

\maketitle

\begin{abstract}

Nowadays, powerful large language models (LLMs) such as ChatGPT have demonstrated revolutionary power in a variety of natural language processing (NLP) tasks such as text classification, sentiment analysis, language translation, and question-answering.
Consequently, the detection of machine-generated texts (MGTs) is becoming increasingly crucial as LLMs become more advanced and prevalent.
These models have the ability to generate human-like language, making it challenging to discern whether a text is authored by a human or a machine.
This raises concerns regarding authenticity, accountability, and potential bias.
However, existing methods for detecting MGTs are evaluated using different model architectures, datasets, and experimental settings, resulting in a lack of a comprehensive evaluation framework that encompasses various methodologies.
Furthermore, it remains unclear how existing detection methods would perform against powerful LLMs.

In this paper, we fill this gap by proposing the first benchmark framework for MGT detection against powerful LLMs, named MGTBench.
Extensive evaluations on public datasets with curated texts generated by various powerful LLMs such as ChatGPT-turbo and Claude demonstrate the effectiveness of different detection methods.
Our ablation study shows that a larger number of words in general leads to better performance and most detection methods can achieve similar performance with much fewer training samples.
Additionally, our findings reveal that metric-based/model-based detection methods exhibit better transferability across different LLMs/datasets.
Furthermore, we delve into a more challenging task: text attribution, where the goal is to identify the originating model of a given text, i.e., whether it is a specific LLM or authored by a human.
Our findings indicate that the model-based detection methods still perform well in the text attribution task.
To investigate the robustness of different detection methods, we consider three adversarial attacks, namely paraphrasing, random spacing, and adversarial perturbations.
We discover that these attacks can significantly diminish detection effectiveness, underscoring the critical need for the development of more robust detection methods.
We envision that MGTBench will serve as a benchmark tool to accelerate future investigations involving the evaluation of powerful MGT detection methods on their respective datasets and the development of more advanced MGT detection methods.\footnote{Source code and datasets are available at \url{https://github.com/xinleihe/MGTBench}.}

\end{abstract}

\section{Introduction}
\label{section:introduction}

Large language models (LLMs), such as T5~\cite{RSRLNMZLL20}, GPT3~\cite{BMRSKDNSSAAHKHCRZWWHCSLGCCBMRSA20}, PaLM~\cite{CNDBMRBCSGSSTMRBTSPRDHPBAIGYDLGDMGMRFZILLZSSDAODPPLMCPLZWSDFCWMEDPF22}, and more powerful LLMs such as ChatGPT~\cite{chatgpt} and Claude~\cite{Claude}, have been a significant breakthrough in the field of natural language processing (NLP).
With huge numbers of parameters and being trained with massive amounts of data, LLMs have shown remarkable performance in various real-world applications, e.g., education, customer service, and finance.

While powerful LLMs have shown impressive qualities in terms of achieving remarkable performance in various tasks and generating human-like texts, several limitations and ethical concerns have to be taken into account.
First, LLMs may generate text that sounds realistic but may not be entirely accurate or factual~\cite{LHE22}.
Second, the misuse of LLMs may raise concerns in education, making a fair judgment impossible~\cite{S22}.
Also, it is difficult to trace back the machine-generated text to its source, which raises concerns about accountability, especially when the content is used to spread misinformation or propaganda~\cite{WMRGUHCGBKKBHSBBHRHILIG21}.

To address these issues, researchers have considered automatic detection methods that can identify the machine-generated text (MGT) from the human-written text (HWT).
Concretely, those methods can be concluded into two categories, i.e., metric-based methods and model-based methods.
For the metric-based methods~\cite{SBCAHWRW19,GSR19,MLKMF23}, metrics such as log-likelihood, word rank, and predicted distribution entropy are used to determine whether a text belongs to MGT or HWT.
Regarding the model-based methods~\cite{SBCAHWRW19,GZWJNDYW23,ZHRBFRC19,gptzero}, classification models are trained by both MGTs and HWTs.

Overall, existing MGT detection methods have been studied under different LLMs and different datasets, albeit in isolation.
Also, powerful LLMs such as ChatGPT demonstrate remarkable performance across diverse tasks, providing high-quality, human-like answers.
This prompts the need for a holistic evaluation of these methods against powerful LLMs.
To do this, we develop a comprehensive benchmark of MGT detection, namely MGTBench.
MGTBench follows a modular design, consisting of the input module, the detection module, and the evaluation module.
Until now, we have implemented ten MGT detection methods.
We also take advantage of the powerful LLMs including ChatGLM, Dolly, ChatGPT-turbo, GPT4All, StableLM, and Claude, which are the most popular and powerful LLMs (thus far), to produce MGTs on existing HWT datasets.
MGTBench enables researchers to benchmark various methods on different datasets.
Its modular design facilitates the integration of additional detection methods, as well as the plugging in of datasets and models.

\mypara{Evaluation}
We perform an extensive measurement study over the 13 detection methods against 6 LLMs.
Our measurement is performed on 3 benchmark datasets including Essay, WP, and Reuters.
Note that for each dataset, besides the texts written by humans (labeled as HWTs), we also query each LLM with the constructed prompt (see \autoref{table:dataset_prompts}) to obtain the MGT from the LLM (labeled as MGTs).

Extensive evaluations show that the LM Detector outperforms other detection methods and reaches the best performance.
For instance, to differentiate whether the texts are generated by humans or ChatGPT-turbo on Essay, the detection F1-score is 0.993 with LM Detector while only 0.968 with Log-Likelihood.
Our ablation study shows that an increased word count can enhance the detection performance, and 200 words is a sufficient length to achieve satisfactory performance.
We also find that most detection methods can achieve comparable performance with significantly fewer training samples.
For instance, with only 10 training samples, Log-Likelihood achieves 0.967 F1-score to differentiate HWTs from MGT generated by ChatGLM on Esay, which is close to the performance with full training samples (0.970).
Another interesting finding is that metric-based methods can better transfer to new LLMs while model-based methods are more capable of adapting to new datasets.
For example, on Essay, Log-Likelihood trained on MGTs generated by GPT4All can attain 0.983 F1-score in differentiating MGTs generated by ChatGLM, and OpenAI Detector trained on Essay can achieve 0.931 F1-score on WP when dealing with MGTs generated by ChatGPT-turbo.

Later, we consider a more challenging task, i.e., text attribution, where the goal is to identify the exact model to generate the text, e.g., either human or one of the LLMs.
Our evaluation shows that the model-based detection methods perform much better than the metric-based detection method in the text attribution task.
For example, on Essay, LM Detector achieves 0.927 F1-score while the F1-score is only 0.208 and 0.387 for Rank and GLTR.

Then, to quantify the robustness of different detection methods, we consider three adversarial attacks, namely paraphrasing, random spacing, and adversarial perturbation.
Evaluation results demonstrate that current detection methods are extremely vulnerable to adversarial attacks.
Take LRR as an example, with WP as the dataset and ChatGPT-turbo as the MGT generator, the detection F1-score degrades 0.418, 0.413, and 0.469 in response to paraphrasing, random spacing, and adversarial perturbation attacks (see \autoref{table:adv_attack_performance} for more details).
This pronounced vulnerability underscores the necessity to develop more robust methods for MGT detection.
Our code and models will be made publicly available.

In summary, we make the following contributions:
\begin{itemize}
    \item We propose MGTBench, a benchmarking framework for MGT detection/attribution against powerful LLMs.
    \item Our empirical evaluation shows that the LM Detector outperforms other detection methods in both MGT detection and text attribution tasks. Additionally, our ablation studies shed light on the unique characteristics of different detection methods.
    \item We systematically evaluate the robustness of different detection methods by introducing three adversarial attacks. Our findings indicate that those methods are highly susceptible to such attacks, which highlights the need for future research to develop more resilient MGT detection methods.
\end{itemize}

\section{Preliminary and Related Work}

\subsection{Text Generation with Large Language Models}
\label{subsection:preliminary_LLMs}

The recent advancements in LLMs can be traced back to the transformer architecture proposed by Vaswani et al.~\cite{VSPUJGKP17}, which introduces the self-attention mechanisms to allow the model to focus on different regions of the input sequence.
Later Radford et al.~\cite{RNSS18} develop the generative pre-trained transformer (GPT) based on the transformer architectures.
Being trained with a large corpus of text data, GPT achieves superior performance on a wide range of language generation tasks.
GPT2~\cite{RWCLAS19}, a larger version of GPT that contains more parameters and is trained on a larger corpus, is developed and achieves better performance than GPT.
GPT3~\cite{BMRSKDNSSAAHKHCRZWWHCSLGCCBMRSA20} is the third generation of GPT with over 175B parameters, which has been shown to be capable of generating coherent and contextually appropriate text even in situations where it is provided with minimal input or guidance.
Since November 2022, OpenAI releases ChatGPT~\cite{chatgpt}, which is trained based on the GPT-3.5 architectures and leverages Reinforcement Learning from Human Feedback (RLHF)~\cite{CLBMLA17,SOWZLVRAC20} to improve its generation ability.
ChatGPT shows revolutionary capabilities in generating coherent and relevant texts, which can be integrated into various applications, such as chatbots, customer service, and education.

Another notable model series in the field is the masked language model.
Bidirectional Encoder Representations from Transformers (BERT) developed by Devlin et al.~\cite{DCLT19} is one of the most representative models that is pre-trained using the masked language modeling task.
Liu et al.~\cite{LOGDJCLLZS19} develop RoBERTa, which leverages a robustly optimized BERT pre-training approach and can reach an even better performance than BERT in various tasks.
With the widespread use of LLMs for generating texts, concerns about authenticity, accountability, and potential bias have also been raised.

In this paper, we consider six representative powerful LLMs, including  ChatGPT-turbo, ChatGLM, Dolly, GPT4All, StableLM, and Claude.

\mypara{ChatGPT-turbo~\cite{chatgpt}}
ChatGPT is an advanced large language model built upon the GPT-3.5 architecture and specifically designed to generate highly human-like responses.
To achieve this capability, ChatGPT is fine-tuned with Reinforcement Learning from Human Feedback (RLHF)~\cite{SOWZLVRAC20} where human trainers actively engage in conversations with ChatGPT.
Here we consider the latest iteration of ChatGPT, namely ChatGPT-turbo.

\mypara{ChatGLM~\cite{chatglm}}
ChatGLM is another large language model based on the GLM (General Language Model) frameworks~\cite{DQLDQYT22}.
The training process involved a combination of supervised fine-tuning, feedback bootstrap, and RLHF, thus enabling it to generate responses that are in accordance with human-like patterns and preferences.

\mypara{Dolly~\cite{dolly}}
Dolly is an instruction-following large language model released by Databricks.
It is trained on around 15K instruction/response records, generated by Databricks employees in capability domains such as brainstorming, classification, closed QA, generation, information extraction, open QA, and summarization.

\mypara{GPT4All~\cite{gpt4all}}
GPT4All is an open-source assistant-style large language model.
It is trained over a massive curated corpus including word problems, story descriptions, multi-turn dialogue, and code.

\mypara{StableLM~\cite{stablelm}}
StableLM is an auto-regressive language model based on the NeoX transformer architecture~\cite{neox}.
It is fine-tuned on various chat and instruction-following datasets.

\mypara{Claude~\cite{Claude}}
Claude, developed by Anthropic, is an AI language model similar to OpenAI's ChatGPT.
It is designed to engage in natural language conversations, provide helpful responses, and assist users with a variety of tasks.

\subsection{Machine-Generated Text Detection}

To address the above-mentioned issues, researchers have developed various MGT detection methods~\cite{SBCAHWRW19,GSR19,MLKMF23,GZWJNDYW23,PSARKBJV23,BAA08,UMLZL21,ULSL20,ZHRBFRC19,IDCE20}.
Current detection methods can be divided into two categories, i.e., metric-based methods and model-based methods.
Generally speaking, metric-based methods leverage pre-trained LLMs to process the text and extract distinguishable features from it, e.g., the rank or entropy of each word in a text conditioned on the previous context.
In this paper, we consider eight metric-based detection methods, including Log-Likelihood, Rank, Log-Rank, Entropy, GLTR, DetectGPT, LRR, and NPR.

\mypara{Log-Likelihood~\cite{SBCAHWRW19}}
This approach leverages a language model to measure the token-wise log probability.
Concretely, given a text, we average the token-wise log probability of each word to generate a score for this text.
Note that a larger score denotes the text is more likely to be machine-generated.

\mypara{Rank~\cite{GSR19}}
For each word in a text, given its previous context, we can calculate the absolute rank of this word.
Then, for a given text, we compute the score of the text by averaging the rank value of each word.
Note that a smaller score denotes the text is more likely to be machine-generated.

\mypara{Log-Rank~\cite{MLKMF23}}
Slightly different from the Rank metric that uses the absolute rank, the Log-Rank score is calculated by first applying the log function to the rank value of each word.

\mypara{Entropy~\cite{GSR19}}
Similar to the Rank score, the Entropy score of a text is calculated by averaging the entropy value of each word conditioned with its previous context.
As mentioned by previous work~\cite{GSR19,MLKMF23}, the machine-generated text is more likely to have a lower Entropy score.

\mypara{GLTR~\cite{GSR19}}
GLTR is developed as a support tool to facilitate the labeling process of whether a text is machine-generated.
In our evaluation, we follow the suggestion of Guo et al.~\cite{GZWJNDYW23} and consider the Test-2 features (i.e., the fraction of words that rank within 10, 100, 1,000, and others).
Note that one can easily implement other sets of features in MGTBench.

\mypara{DetectGPT~\cite{MLKMF23}}
Mitchell et al.~\cite{MLKMF23} propose DetectGPT that measures the change of the model's log probability function by adding minor perturbation to the original text.
The intuition is that the text derived from an LLM has a tendency to be in the local optimal of the model's log probability function.
Therefore, any minor perturbation of model-generated text tends to have a lower log probability under the model than the original text, while minor perturbation of human-written text may have a higher or lower log probability than the original text.

\mypara{LRR~\cite{SZWN23}}
Su et al.~\cite{SZWN23} propose Log-\textbf{L}ikelihood Log-\textbf{R}ank \textbf{R}atio (LRR), which combines Log-Likelihood and Log-Rank as they provide complementary information for the given text.

\mypara{NPR~\cite{SZWN23}}
Similar to DetectGPT, \textbf{N}ormalized \textbf{P}erturbed Log-\textbf{R}ank (NPR) also introduces perturbation to the original text.
The motivation for NPR is that both MGTs and HWTs show vulnerability to minor disturbances, as indicated by a rise in the Log-Rank score following such perturbations.
However, this effect is more pronounced in MGTs, as they exhibit a greater increase in the Log-Rank score after disturbances, implying a higher NPR (noise-to-perturbation ratio) score for MGTs compared to HWTs.

Regarding the model-based methods, a classification model is usually trained using a corpus that contains both HWTs and MGTs.
By doing this, the classification model is expected to have the capability in identifying MGTs from a given corpus.
In this paper, we consider five model-based methods.

\mypara{OpenAI Detector~\cite{SBCAHWRW19}}
The OpenAI Detector is used to detect GPT2-generated output, which was created by fine-tuning a RoBERTa model using outputs from the largest GPT2 model (i.e., with 1.5B parameters).
This model is capable of predicting whether a given text was machine-generated or not.

\mypara{ChatGPT Detector~\cite{GZWJNDYW23}}
ChatGPT Detector is developed by Guo et al.~\cite{GZWJNDYW23} to distinguish human-written texts from ChatGPT-generated texts.
The model was created by fine-tuning a RoBERTa model using the HC3~\cite{GZWJNDYW23} dataset.
The detector is based on the RoBERTa model.
The authors provide two ways to train the RoBERTa model.
The first one only leverages the pure answered text, and the second one leverages the question-answer text pair to jointly train the model.
In our evaluation, we consider the first one to be consistent with other detection methods.

\mypara{ConDA~\cite{BKML23}}
Bhattacharjee et al.~\cite{BKML23} develop Contrastive Domain Adaptation (ConDA), which leverages the representation power of contrastive learning to acquire domain-invariant representations

\mypara{GPTZero~\cite{gptzero}}
GPTZero is an MGT analyzer tool that uses two main measures, i.e., perplexity and burstiness, to determine whether a text is machine-generated or written by a human.\footnote{\url{https://www.makeuseof.com/gptzero-detect-ai-generated-text/}.}
GPTZero provides the publicly accessible API to produce a confidence score about how likely a text is generated by the machine.

\mypara{LM Detector}
Besides the previous methods, the detector can also be built by fine-tuning the pre-trained language model (LM) with an extra classification layer.
Here we take the BERT model as an example to evaluate its efficacy following Ippolito et al.~\cite{IDCE20}.

Compared to previous work~\cite{UMLZL21,PSARKBJV23}, our study offers notable advancements in the field of MGT detection by integrating more detection methods into our benchmarking framework.
Moreover, we extend the evaluation to more powerful LLMs such as ChatGPT-turbo and Claude.
This broader scope enables us to gain deeper insights into the performance and robustness of various detection methods when applied to a wider range of LLMs.
By incorporating these enhancements, our work significantly contributes to the existing literature and paves the way for further advancements in MGT detection research.

\section{MGTBench}

In this section, we introduce MGTBench, a modular framework designed to benchmark MGT detection methods.
Currently, we have provided reference implementations of the 8 metric-based detection methods and easy-to-use APIs for the 5 model-based methods we mentioned before.

\subsection{Modular Design}

MGTBench consists of three different modules, including the \textit{input module}, \textit{detection module}, and \textit{evaluation module}.

\mypara{Input Module}
In the input module, we provide specific dataset pre-processing functions for different datasets and our code base is easy to cope with datasets from HuggingFace, which facilitates the future developments of different users.

\mypara{Detection Module}
This module implements different metric-based and model-based detection methods with a standardized input/output format.
Currently, we support ten different detection methods.

\mypara{Evaluation Module}
This module is used to evaluate the performance of different detection methods.
Now, we provide five different evaluation metrics, including accuracy, precision, recall, F1-score, and AUC, which are the commonly used metrics to evaluate classification performance.
We also support sample-level logging for detailed analysis.

\subsection{Using MGTBench}

To be best of our knowledge, MGTBench is the most comprehensive benchmark tool for MGT detection against powerful LLMs.
Users can leverage MGTBench on their own dataset for a comprehensive risk assessment of potential MGTs in the dataset.
On the other hand, researchers can leverage MGTBench as a tool to evaluate new MGT detection/generation methods.
As MGTBench follows a modular design, its input and evaluation modules can be easily re-used by new detection methods.
Also, new detection methods can be easily implemented within the standardized API provided by MGTBench.
Moreover, MGTBench integrates well with HuggingFace, given the fact that many model-based detection methods have published or are willing to publish their models into HuggingFace, MGTBench can be seamlessly updated to fit the new model-based detection methods.
MGTBench is under continuous development and we will include more detection methods as well as analysis tools in the future.

\section{Experimental Settings}
\label{section:experimental_setting}

\subsection{Datasets}

\begin{table}[ht]
\centering
\caption{The prompts we used to obtain MGTs from LLMs. Following Verma et al.~\cite{VFTK23}, we set $K$ to the number of words in each HWT rounding to the nearest 100 and acquire the $<$prompt$>$/$<$headline$>$ by querying ChatGPT-turbo.}
\label{table:dataset_prompts}
\begin{tabular}{l | p{6cm} }
\toprule
\textbf{Dataset} & \textbf{Prompt}\\
\midrule
\textbf{Essay} &  Write an essay in $K$  words to the prompt $<$prompt$>$ \\
\midrule
\textbf{WP} & Write a story in $K$ words to the prompt $<$prompt$>$ \\
\midrule
\textbf{Reuters} & Write a news article in $K$ words with the following headline $<$headline$>$\\
\bottomrule
\end{tabular}
\end{table}

In this paper, we consider three datasets provided by Verma et al.~\cite{VFTK23}, namely Essay, WP, and Reuters.
Note that for each dataset, besides the HWTs, it also contains MGTs generated by ChatGPT-turbo and Clude.
And we additionally generate MGTs from other LLMs as well.

\mypara{Essay}
This dataset has 1,000 samples extracted from essays available on IvyPanda, encompassing both high school and university-level essays across various academic disciplines.
The authors first query ChatGPT-turbo to generate the $<$prompt$>$ corresponding to the essay.
Then, the constructed prompt would be leveraged to query different LLMs and obtain the generated essays.

\mypara{WP}
This dataset has 1,000 samples that are derived from the subreddit \texttt{r/WritingPrompts} where users share creative writing prompts and craft stories in response to these prompts.
Following the constructed prompts, we can then query different LLMs to generate stories.

\mypara{Reuters}
This dataset is derived from the Reuters 50-50 authorship identification dataset~\cite{HS06}, which consists of 1,000 (news) articles by 50 journalists (20 articles per journalist).
As the article does not have the headline, the authors first prompt ChatGPT-turbo to generate a $<$headline$>$ for each article.
We also leverage this headline to construct the prompt and query LLMs to obtain MGTs.

We show the prompt we used to query LLMs in \autoref{table:dataset_prompts}.
For each entry in a dataset, we have the human text and six LLM-generated texts (see \autoref{subsection:preliminary_LLMs} for more details about the six LLMs).
We only keep entries with more than 1 word for human and LLM-generated texts.
Then, we randomly split 80\% of the entries as the training set and the rest as the testing set.

\subsection{Tasks}

In our evaluation, our primary focus is on the MGT detection task, which involves detecting whether a given text is generated by a human or a machine (i.e., LLM).
To accomplish this, we establish a binary classification task for texts generated by humans and each LLM, such as Human vs. ChatGPT-turbo, Human vs. ChatGLM, etc.
Note that we also consider a more complex task, namely text attribution (see \autoref{section:text_attribution}).
This task seeks to pinpoint the precise model responsible for generating the text.
Essentially, when presented with a text, our goal is to determine whether it is generated by the human or one of the six LLMs.
This task can be viewed as a seven-category classification scenario.

\subsection{Detection Methods}

For metric-based methods, we use GPT2-medium as the base model in our experiments since it can already reach good performance with limited cost.
Given the metrics extracted with the GPT2-medium, we additionally build a logistic regression model on top of it to provide concrete predictions.
For model-based methods, we directly use the publicly available pre-trained models from HuggingFace (OpenAI Detector, ChatGPT Detector, and LM Detector) or GitHub (ConDA).
Concretely, for OpenAI Detector, we use the RoBERTa-base version of it as it usually gives better detection performance.
For ChatGPT Detector, we leverage the provided RoBERTa-base model of it.
For LM Detector, we leverage the distilled BERT-base model as it has superior performance and modest expenditure.
For ConDA, we leverage the RoBERTa-base model targeted on ChatGPT.\footnote{\url{https://github.com/AmritaBh/ConDA-gen-text-detection}.}
Note that for OpenAI Detector, ChatGPT Detector, and ConDA, the pre-trained models are already optimized for the MGT detection task (binary classification), therefore, we do not further fine-tune them unless otherwise mentioned.
It is also worth mentioning that, in the text attribution task, we fine-tune all detection methods as the number of classes increases from 2 to 7.

\begin{table*}[!htb]
\centering
\caption{The performance (F1-score) of different detection methods. Here OpenAI-D, ChatGPT-D, and LM-D denote the OpenAI Detector, ChatGPT Detector, and LM Detector. We follow this naming rule in the following tables/figures in this paper. * means we only sample part of the data (1/8) for testing.}
\label{table:performance}
\begin{tabular}{l l | c c c c c c}
\toprule
\textbf{Dataset} & \textbf{Method} & \textbf{ChatGLM} & \textbf{Dolly} & \textbf{ChatGPT-turbo} & \textbf{GPT4All} & \textbf{StableLM} & \textbf{Claude} \\
\midrule
\multirow{13}{*}{\textbf{Essay}}   
 & Log-Likelihood & 0.970 & 0.866 & 0.968 & 0.923 & 0.665 & 0.834 \\
 & Rank & 0.740 & 0.737 & 0.915 & 0.843 & 0.667 & 0.772 \\
 & Log-Rank & 0.983 & 0.865 & 0.966 & 0.923 & 0.692 & 0.814 \\
 & Entropy & 0.806 & 0.683 & 0.874 & 0.699 & 0.566 & 0.771 \\
 & GLTR & 0.988 & 0.848 & 0.954 & 0.925 & 0.756 & 0.806 \\
 & LRR & 0.982 & 0.810 & 0.925 & 0.904 & 0.748 & 0.746 \\
 & NPR & 0.956 & 0.865 & 0.218 & 0.927 & 0.740 & 0.238 \\
 & DetectGPT & 0.891 & 0.844 & 0.227 & 0.908 & 0.704 & 0.236 \\
 & GPTZero* & 0.923 & 0.880 & 0.980 & 0.943 & 0.486 & 0.870 \\
 & ConDA & 0.668 & 0.069 & 0.000 & 0.260 & 0.663 & 0.664 \\
 & OpenAI-D & 0.921 & 0.724 & 0.353 & 0.863 & 0.774 & 0.009 \\
 & ChatGPT-D & 0.923 & 0.630 & 0.742 & 0.815 & 0.491 & 0.057 \\
 & LM-D & 1.000 & 0.997 & 0.993 & 0.997 & 0.997 & 0.980 \\
\midrule
\multirow{13}{*}{\textbf{WP}}
 & Log-Likelihood & 0.980 & 0.794 & 0.841 & 0.934 & 0.786 & 0.773 \\
 & Rank & 0.840 & 0.760 & 0.797 & 0.891 & 0.781 & 0.709 \\
 & Log-Rank & 0.985 & 0.807 & 0.819 & 0.929 & 0.832 & 0.751 \\
 & Entropy & 0.800 & 0.662 & 0.770 & 0.766 & 0.644 & 0.731 \\
 & GLTR & 0.983 & 0.766 & 0.800 & 0.935 & 0.861 & 0.733 \\
 & LRR & 0.980 & 0.774 & 0.728 & 0.930 & 0.875 & 0.656 \\
 & NPR & 0.970 & 0.801 & 0.352 & 0.905 & 0.764 & 0.521 \\
 & DetectGPT & 0.812 & 0.719 & 0.608 & 0.808 & 0.695 & 0.517 \\
 & GPTZero* & 0.980 & 0.732 & 0.980 & 1.000 & 0.148 & 0.818 \\
 & ConDA & 0.585 & 0.039 & 0.075 & 0.674 & 0.667 & 0.000 \\
 & OpenAI-D & 0.980 & 0.776 & 0.093 & 0.948 & 0.937 & 0.029 \\
 & ChatGPT-D & 0.880 & 0.528 & 0.352 & 0.795 & 0.616 & 0.044 \\
 & LM-D & 0.998 & 0.950 & 0.990 & 0.983 & 0.966 & 0.970 \\
\midrule
\multirow{13}{*}{\textbf{Reuters}}
 & Log-Likelihood & 0.972 & 0.381 & 0.926 & 0.697 & 0.659 & 0.798 \\
 & Rank & 0.650 & 0.413 & 0.847 & 0.665 & 0.635 & 0.648 \\
 & Log-Rank & 0.990 & 0.373 & 0.944 & 0.735 & 0.701 & 0.785 \\
 & Entropy & 0.477 & 0.553 & 0.703 & 0.668 & 0.620 & 0.694 \\
 & GLTR & 0.987 & 0.556 & 0.946 & 0.742 & 0.750 & 0.772 \\
 & LRR & 0.992 & 0.590 & 0.948 & 0.796 & 0.766 & 0.715 \\
 & NPR & 0.950 & 0.790 & 0.284 & 0.843 & 0.751 & 0.560 \\
 & DetectGPT & 0.866 & 0.782 & 0.270 & 0.821 & 0.756 & 0.558 \\
 & GPTZero* & 0.980 & 0.485 & 0.936 & 0.980 & 0.611 & 0.750 \\
 & ConDA & 0.664 & 0.137 & 0.000 & 0.667 & 0.000 & 0.667 \\
 & OpenAI-D & 0.985 & 0.713 & 0.954 & 0.900 & 0.903 & 0.000 \\
 & ChatGPT-D & 0.968 & 0.650 & 0.931 & 0.898 & 0.617 & 0.019 \\
 & LM-D & 1.000 & 0.995 & 0.995 & 1.000 & 0.995 & 0.993 \\
\bottomrule
\end{tabular}
\end{table*}

\subsection{Evaluation Metrics}

MGTBench supports various metrics to evaluate performance, including accuracy, precision, recall, F1-score, and AUC (area under the ROC curve).
In our evaluation, unless otherwise mentioned, we use F1-score as the main evaluation metric.

\section{Evaluation}
\label{section:evaluation}

We first present the experiment results on the MGT detection task.
As shown in \autoref{table:performance}, we observe that metric-based methods such as Log-Likelihood, Log-Rank, GLTR, and LRR have relatively good performances on different datasets as well.
For instance, Log-Likelihood reaches 0.970, 0.980, and 0.972 F1-score on Essay, WP, and Reuters when distinguishing ChatGLM-generated texts from HWTs.
On the other hand, NPR and DetectGPT reach less satisfying performance.
For instance, on WP, to distinguish MGTs generated by ChatGPT-turbo from HWTs, the F1-score is only 0.352 and 0.608, respectively.
We suspect the reason is that the metric changes less when the perturbation is conducted on longer texts, which leads to lower performance.

\begin{figure*}[!t]
\centering
\begin{subfigure}{0.64\columnwidth}
\includegraphics[width=\columnwidth]{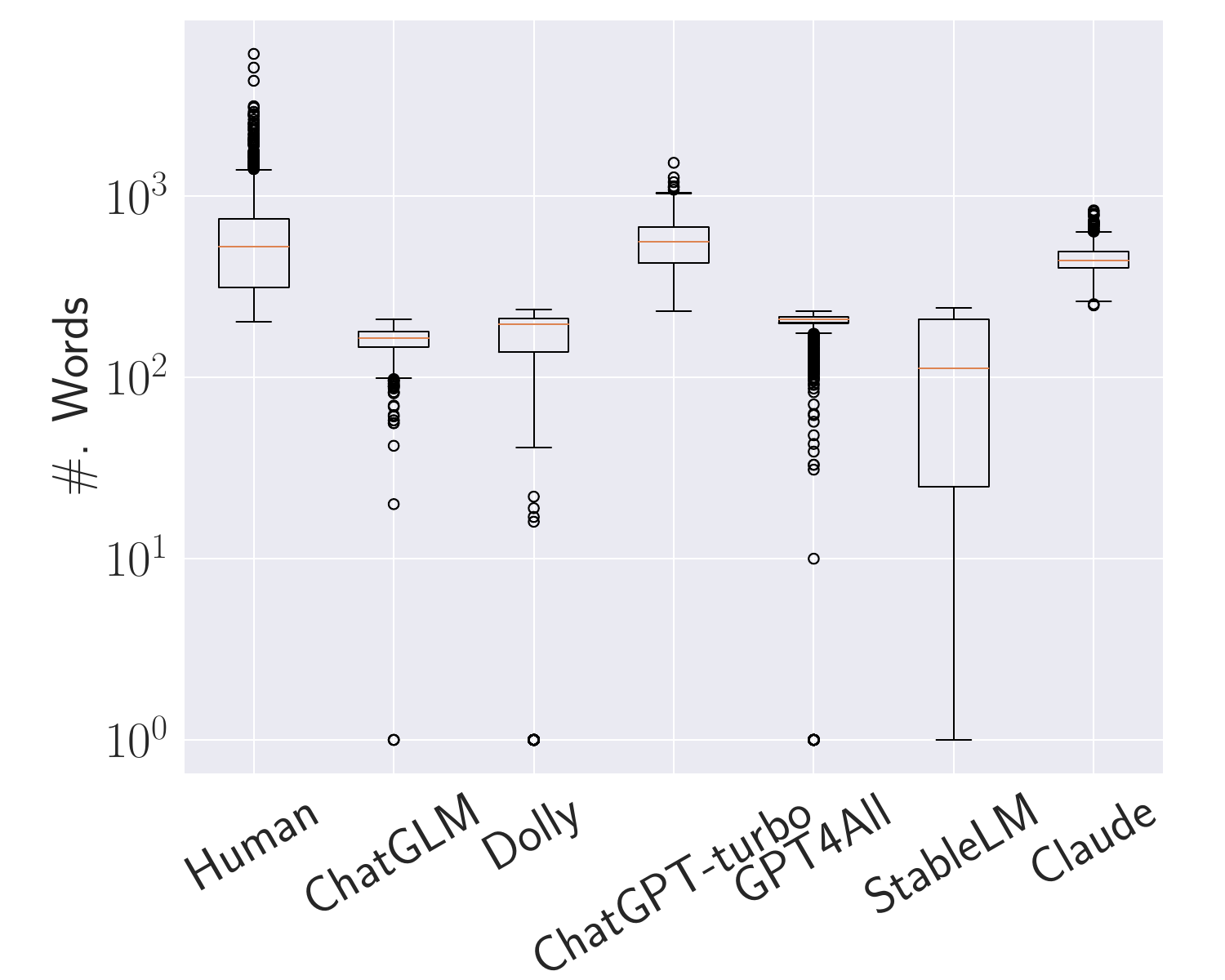}
\caption{Essay}
\label{figure:Essay_word_cnt}
\end{subfigure}
\begin{subfigure}{0.64\columnwidth}
\includegraphics[width=\columnwidth]{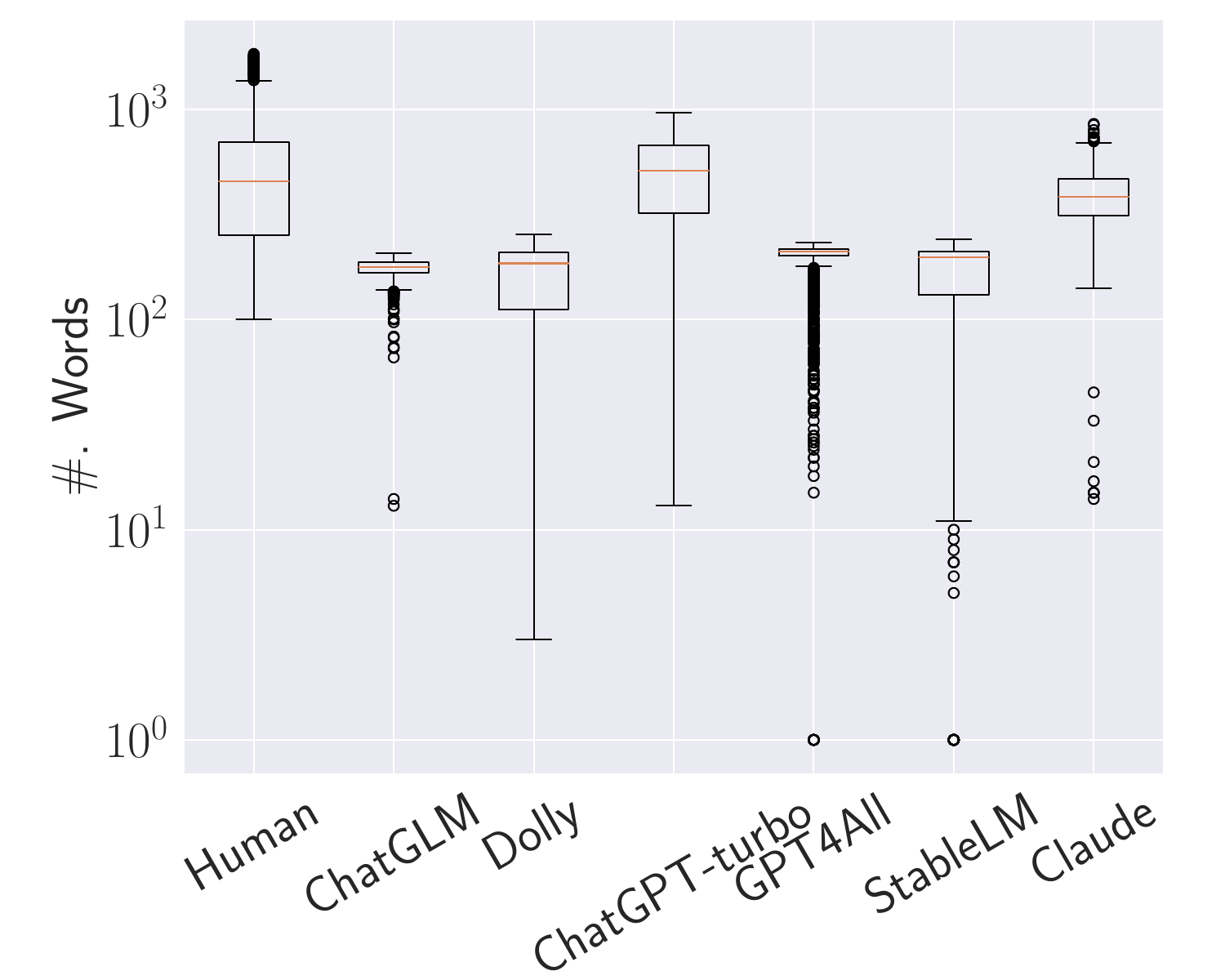}
\caption{WP}
\label{figure:WP_word_cnt}
\end{subfigure}
\begin{subfigure}{0.64\columnwidth}
\includegraphics[width=\columnwidth]{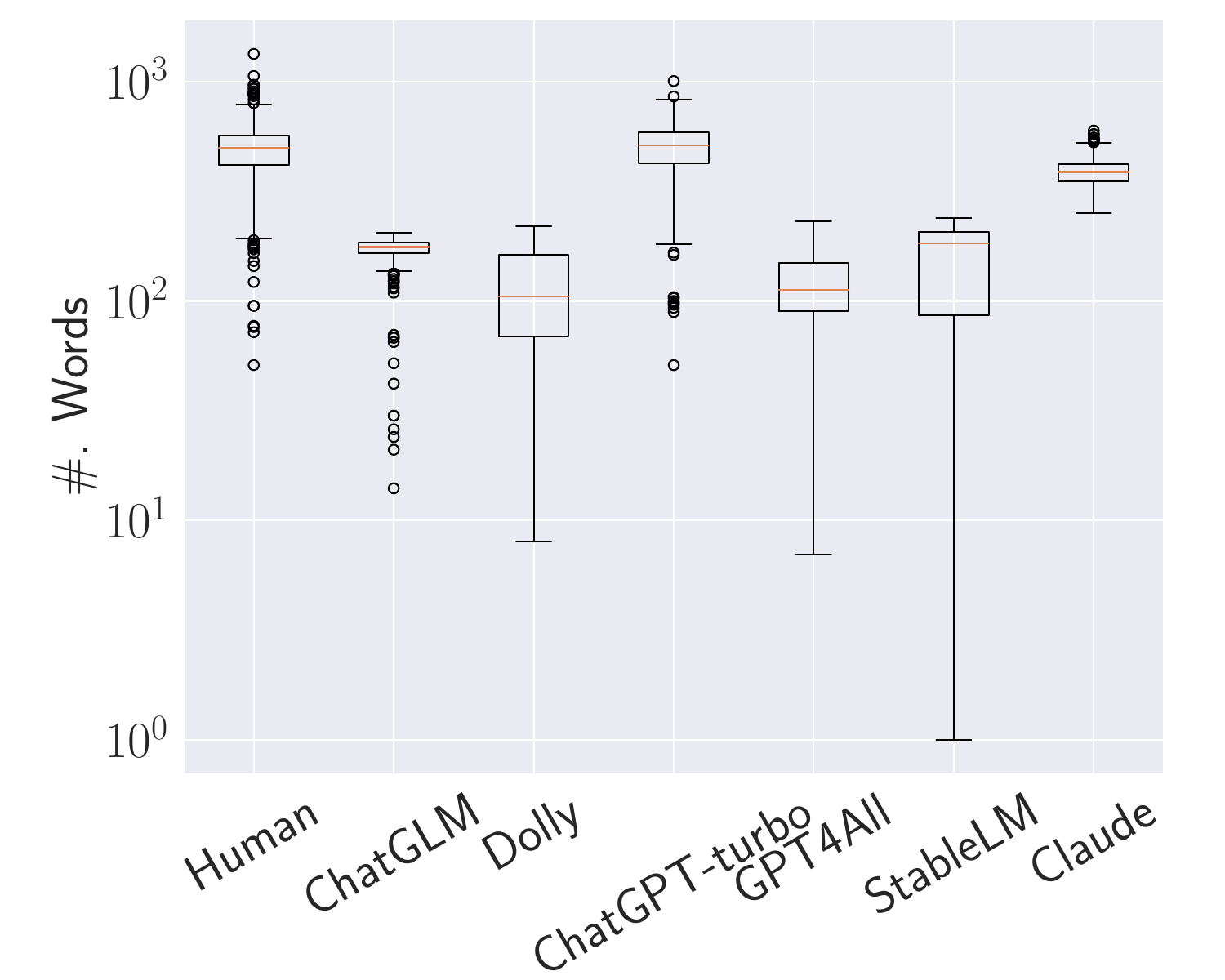}
\caption{Reuters}
\label{figure:Reuters_word_cnt}
\end{subfigure}
\caption{The distribution of \#. words for HWTs and MGTs on different datasets.}
\label{figure:ablation_word_cnt}
\end{figure*}

\begin{figure*}[!t]
\centering
\begin{subfigure}{0.64\columnwidth}
\includegraphics[width=\columnwidth]{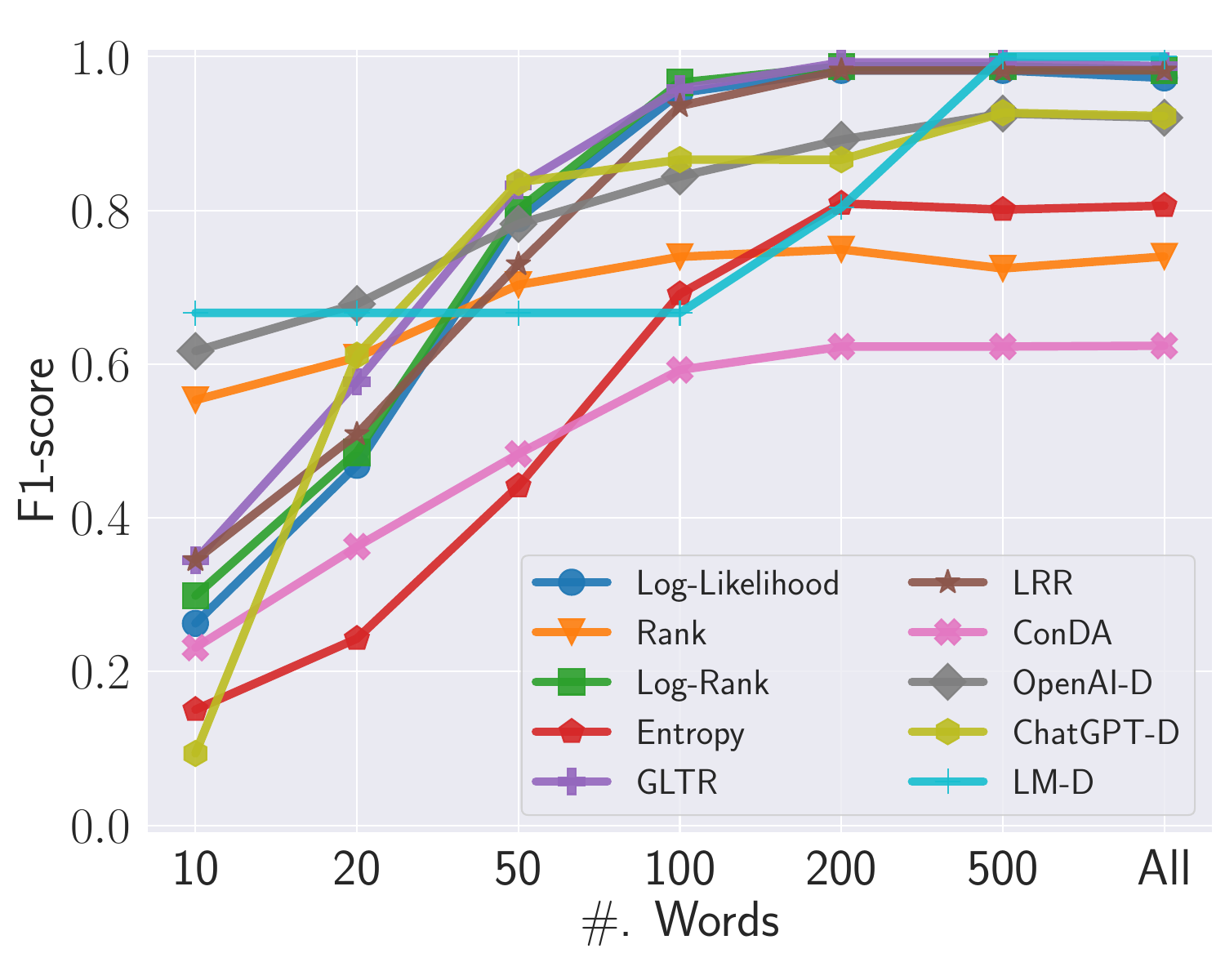}
\caption{ChatGLM}
\label{figure:Essay_ChatGLM_fewer_words}
\end{subfigure}
\begin{subfigure}{0.64\columnwidth}
\includegraphics[width=\columnwidth]{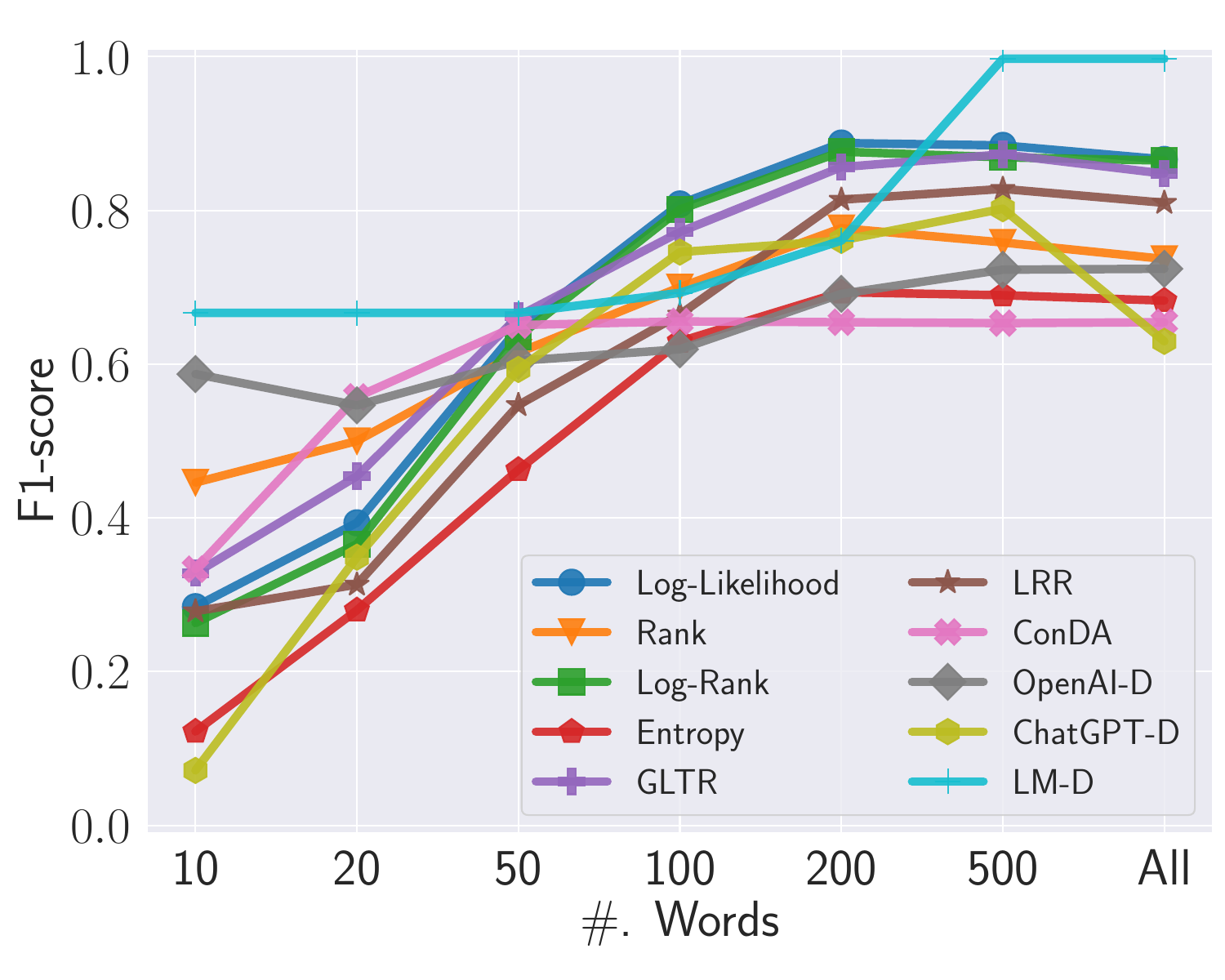}
\caption{Dolly}
\label{figure:Essay_Dolly_fewer_words}
\end{subfigure}
\begin{subfigure}{0.64\columnwidth}
\includegraphics[width=\columnwidth]{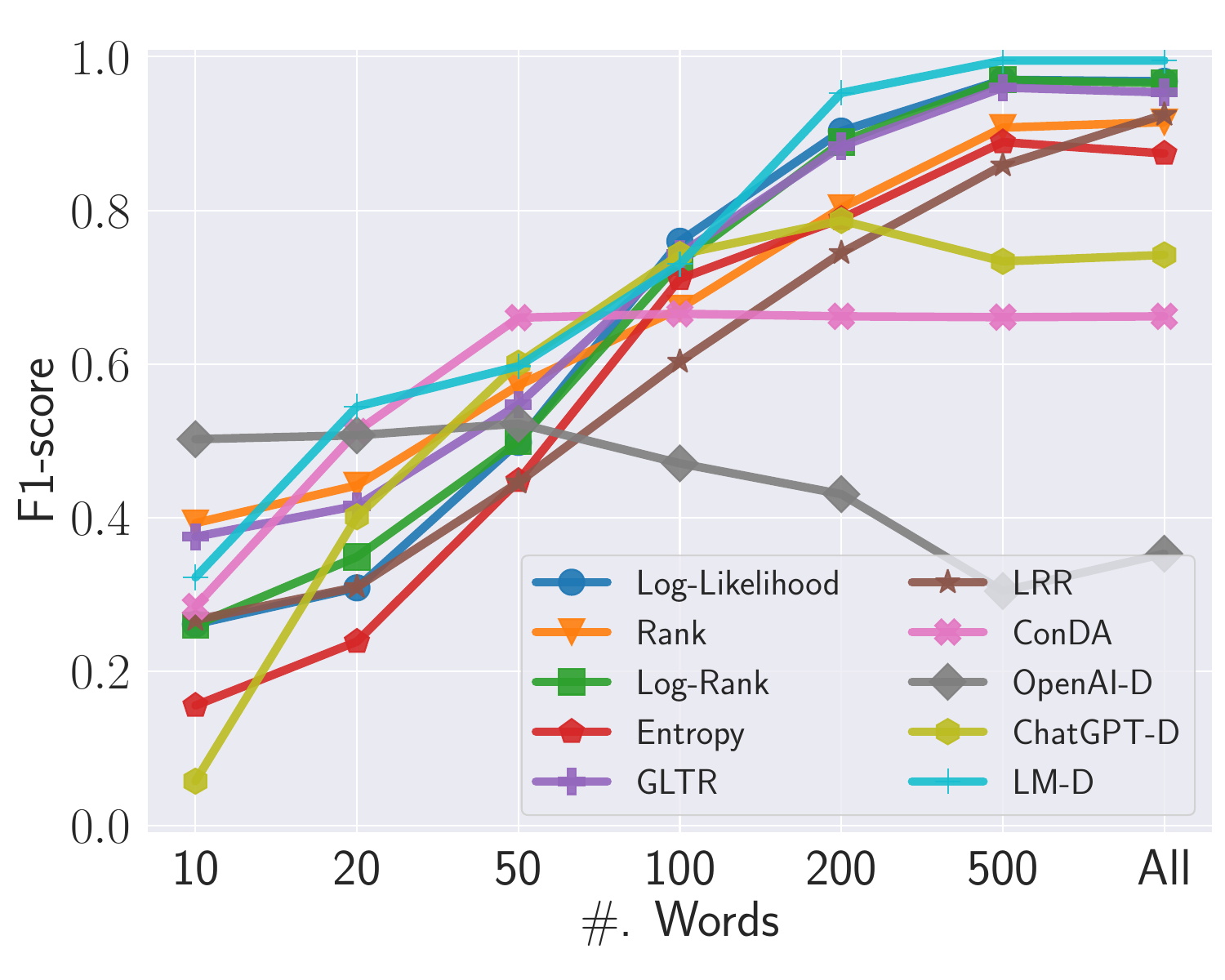}
\caption{ChatGPT-turbo}
\label{figure:Essay_ChatGPT-turbo_fewer_words}
\end{subfigure}
\begin{subfigure}{0.64\columnwidth}
\includegraphics[width=\columnwidth]{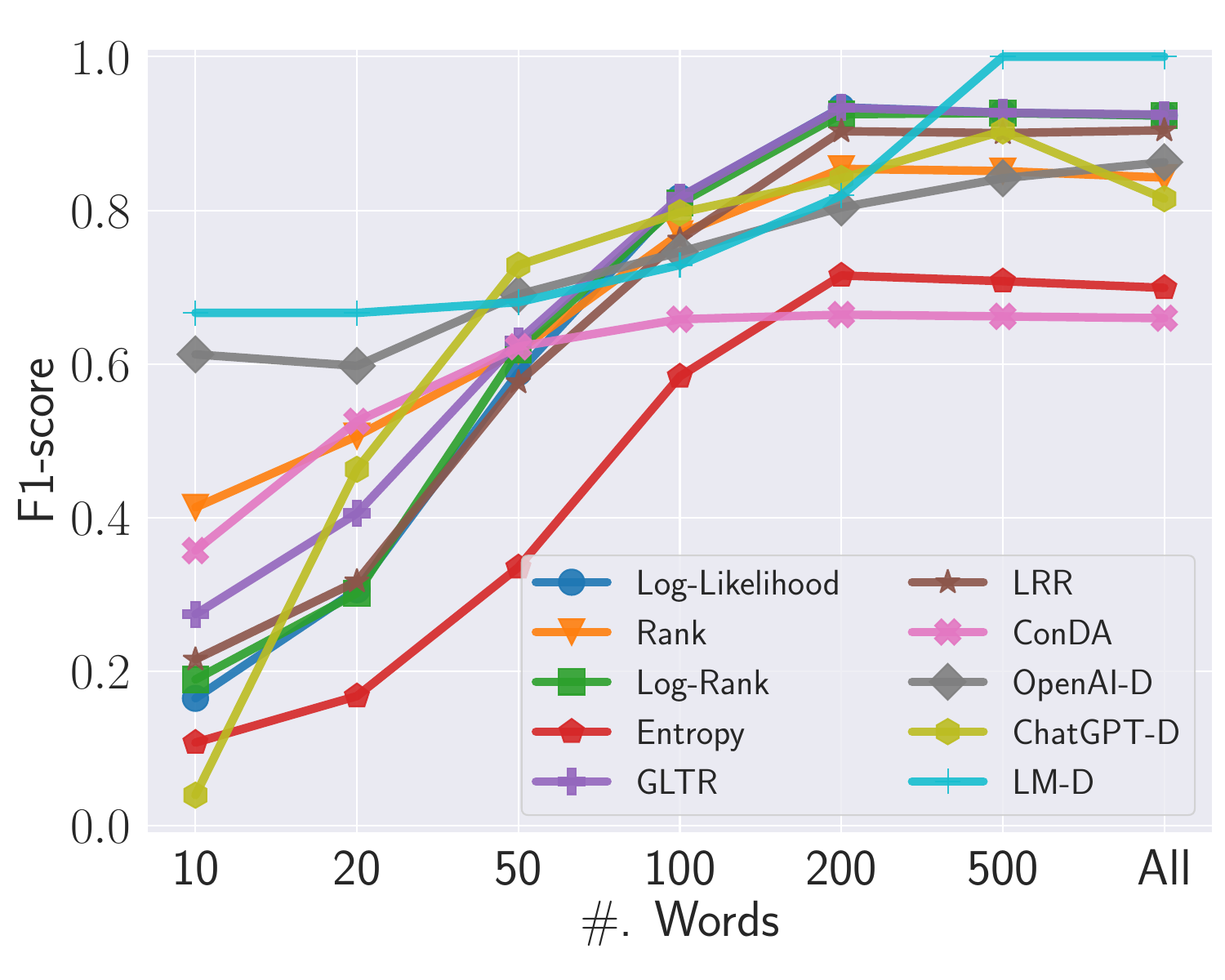}
\caption{GPT4All}
\label{figure:Essay_GPT4All_fewer_words}
\end{subfigure}
\begin{subfigure}{0.64\columnwidth}
\includegraphics[width=\columnwidth]{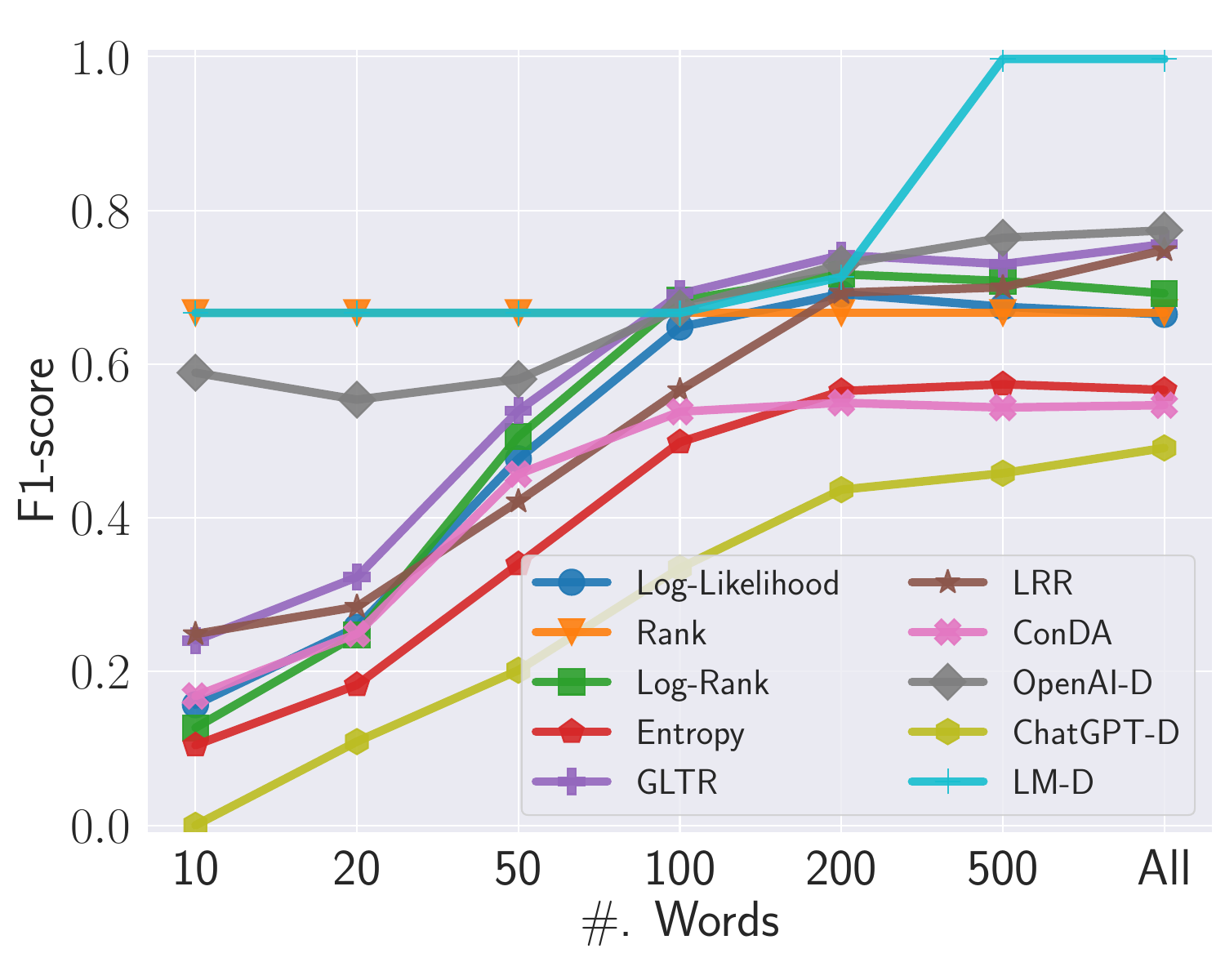}
\caption{StableLM}
\label{figure:Essay_StableLM_fewer_words}
\end{subfigure}
\begin{subfigure}{0.64\columnwidth}
\includegraphics[width=\columnwidth]{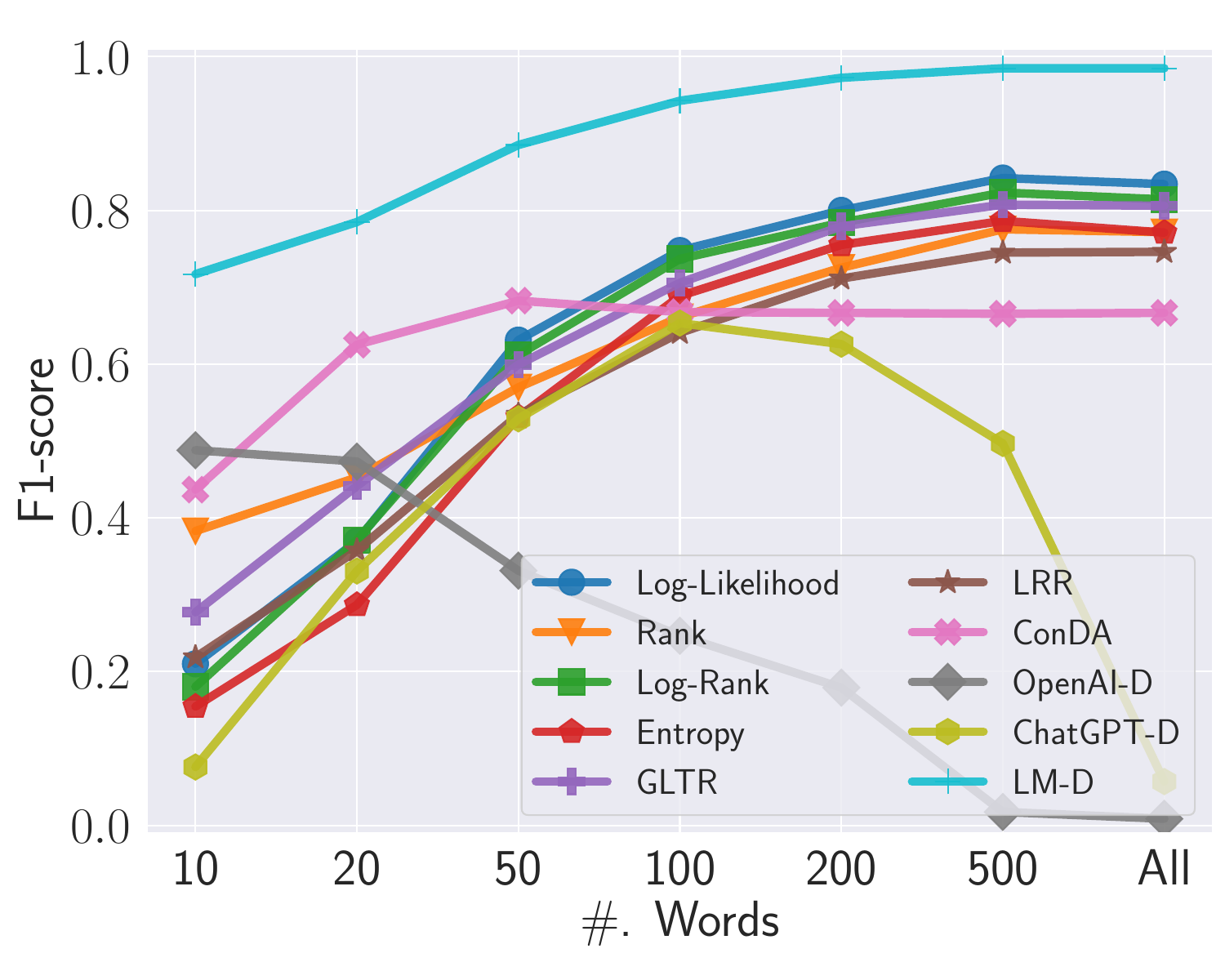}
\caption{Claude}
\label{figure:Essay_Claude_fewer_words}
\end{subfigure}
\caption{The F1-score of different detection methods under texts with different maximum number of words on Essay.}
\label{figure:ablation_fewer_words_Essay}
\end{figure*}

\begin{figure*}[!htbp]
\centering
\begin{subfigure}{0.64\columnwidth}
\includegraphics[width=\columnwidth]{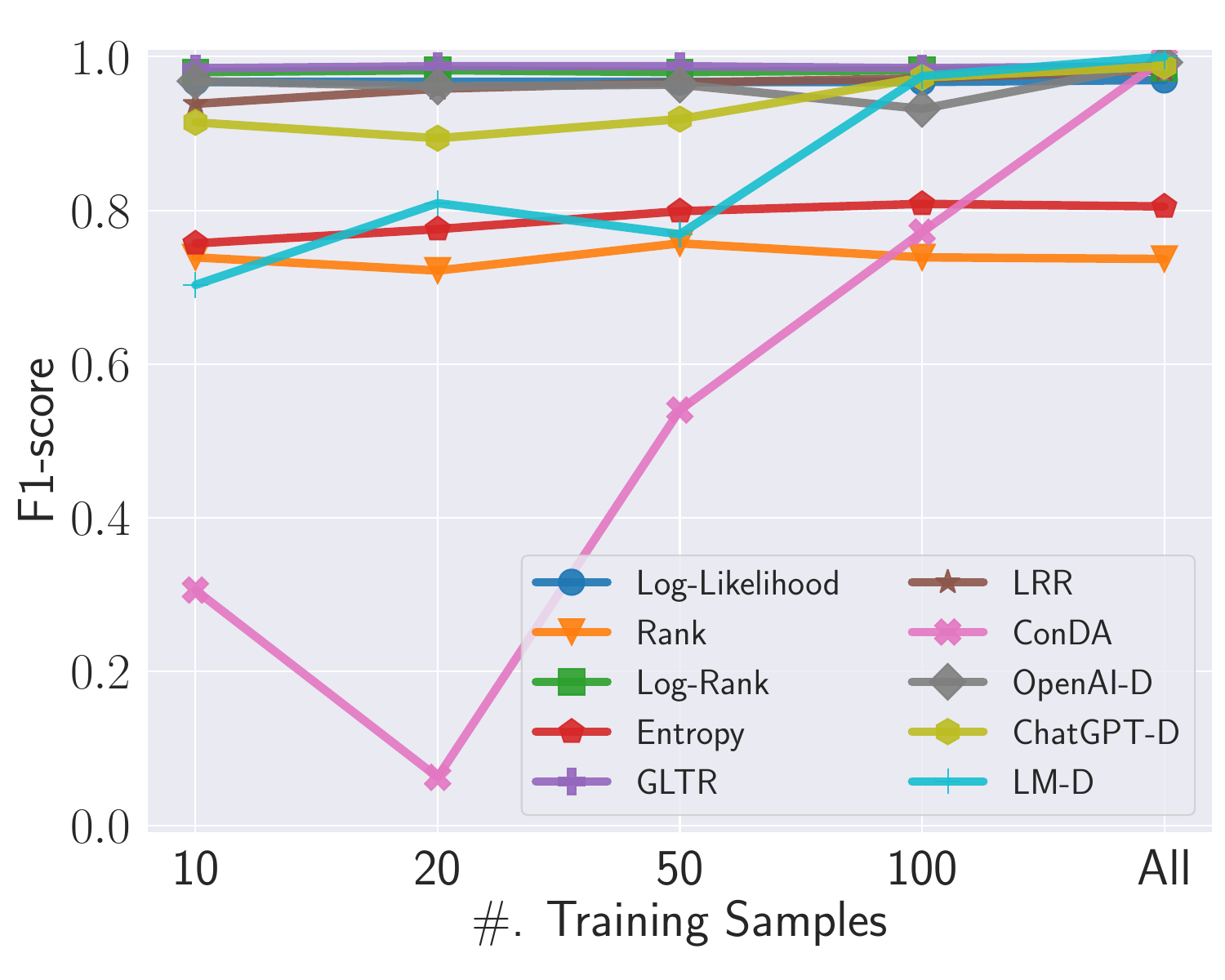}
\caption{ChatGLM}
\label{figure:ablation_fewer_sample_Essay_ChatGLM}
\end{subfigure}
\begin{subfigure}{0.64\columnwidth}
\includegraphics[width=\columnwidth]{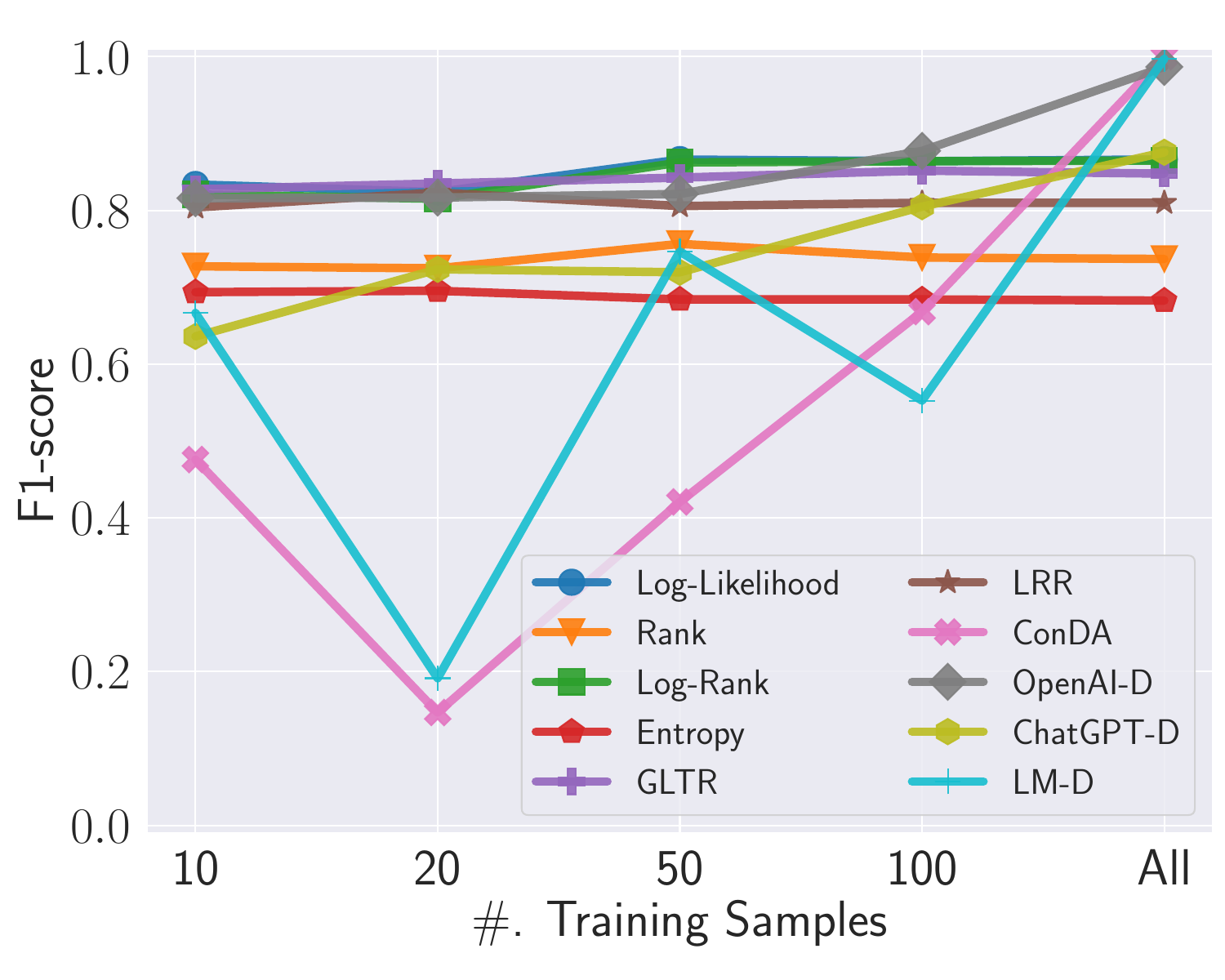}
\caption{Dolly}
\label{figure:ablation_fewer_sample_Essay_Dolly}
\end{subfigure}
\begin{subfigure}{0.64\columnwidth}
\includegraphics[width=\columnwidth]{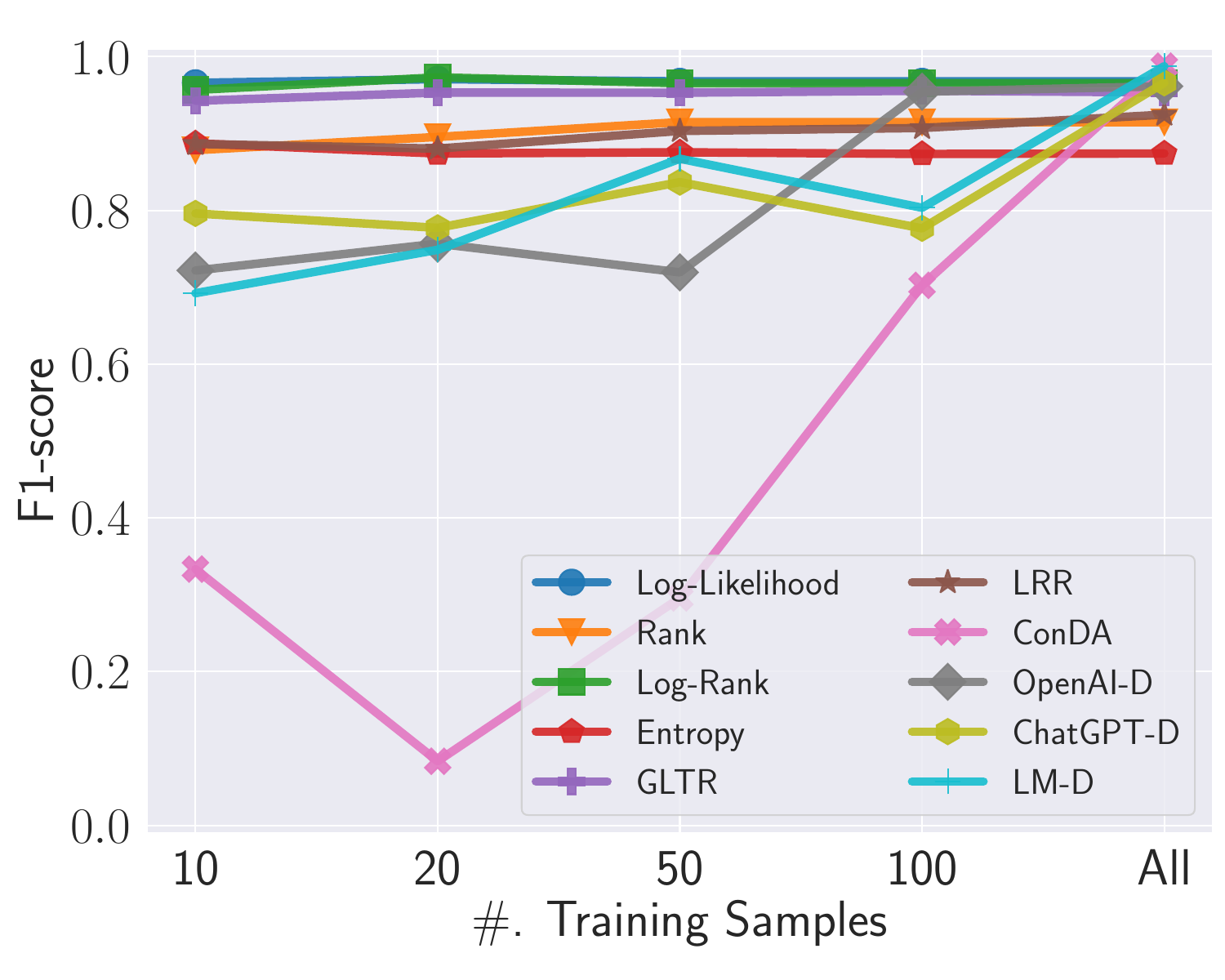}
\caption{ChatGPT-turbo}
\label{figure:ablation_fewer_sample_Essay_ChatGPT-turbo}
\end{subfigure}
\begin{subfigure}{0.64\columnwidth}
\includegraphics[width=\columnwidth]{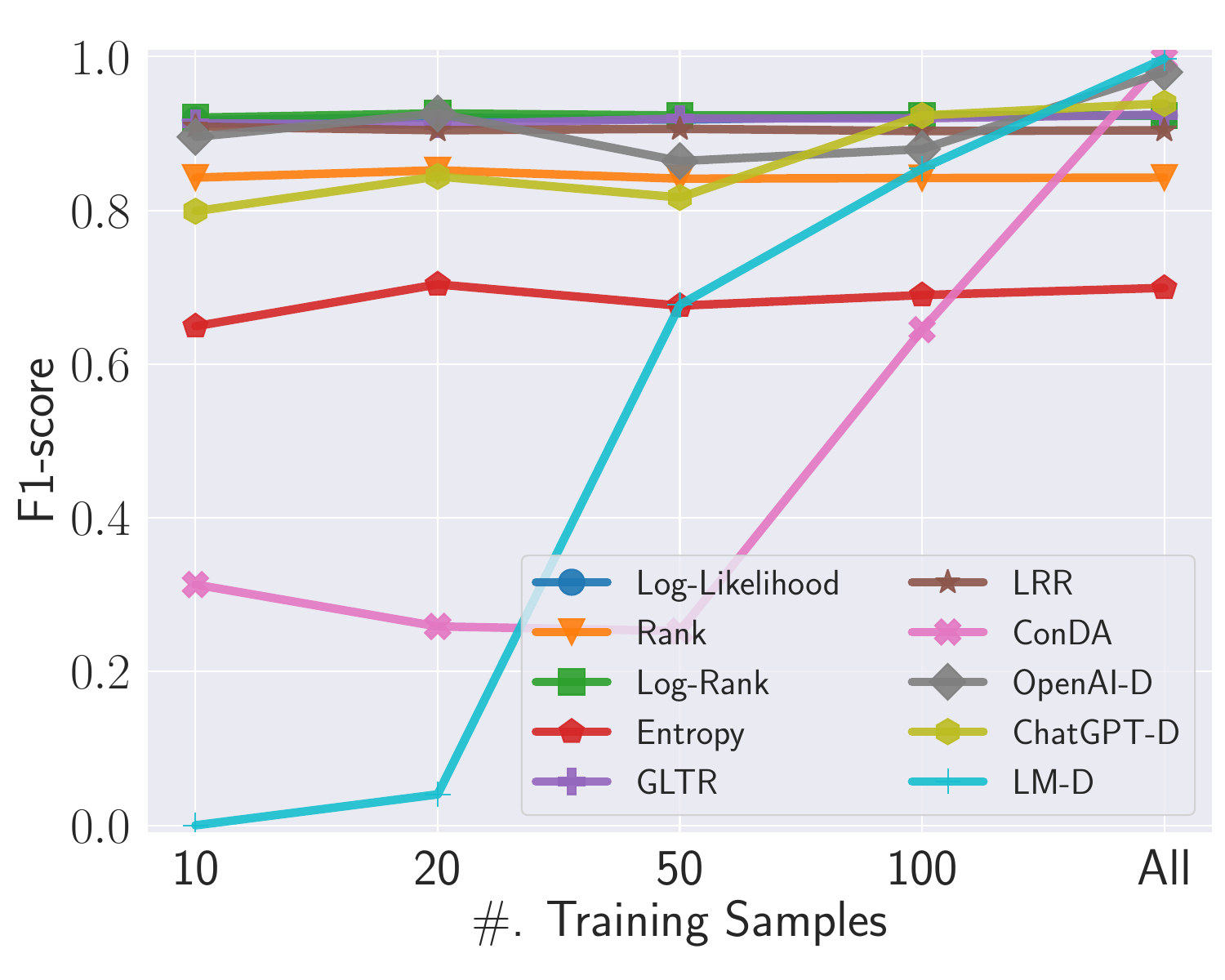}
\caption{GPT4All}
\label{figure:ablation_fewer_sample_Essay_GPT4All}
\end{subfigure}
\begin{subfigure}{0.64\columnwidth}
\includegraphics[width=\columnwidth]{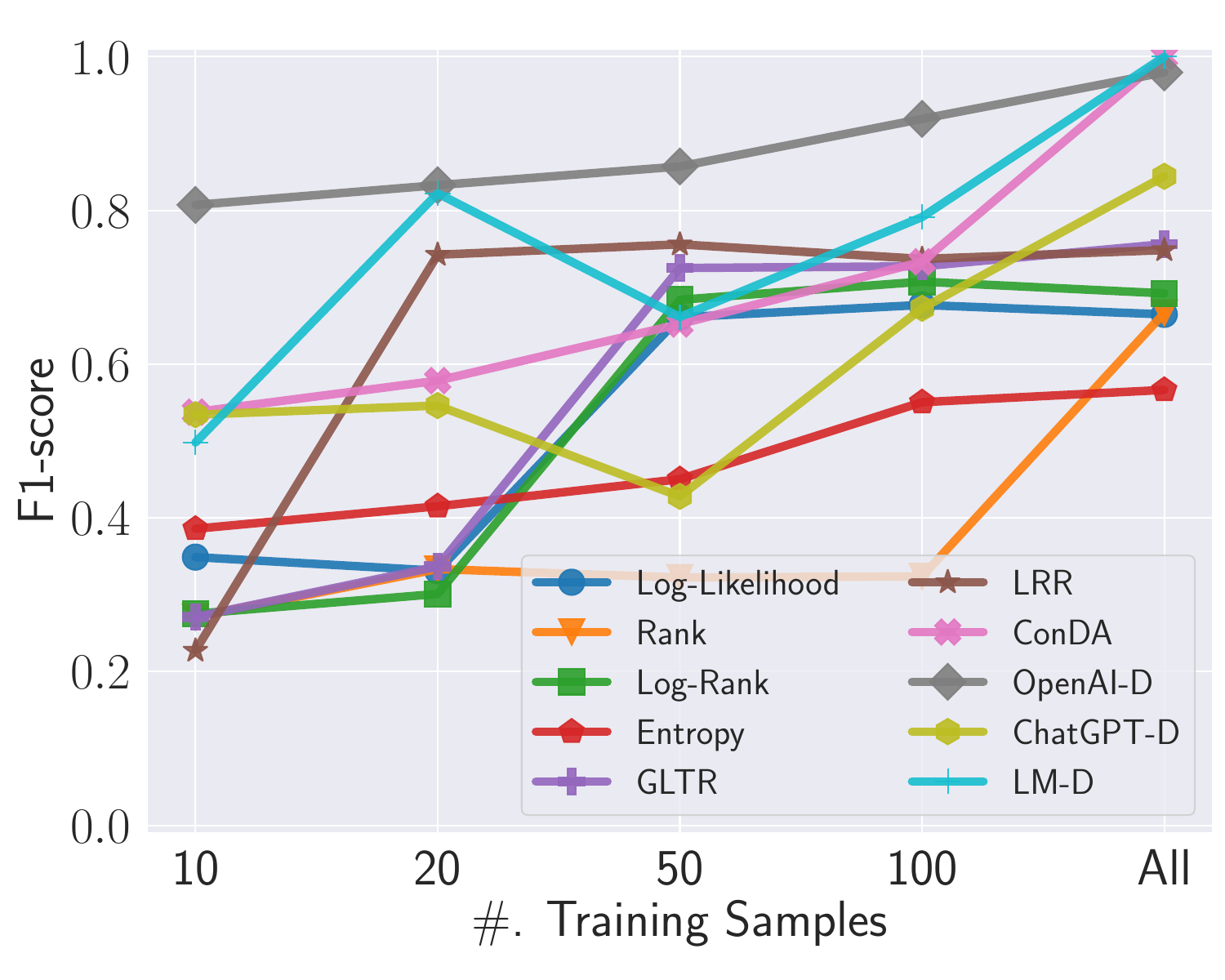}
\caption{StableLM}
\label{figure:ablation_fewer_sample_Essay_StableLM}
\end{subfigure}
\begin{subfigure}{0.64\columnwidth}
\includegraphics[width=\columnwidth]{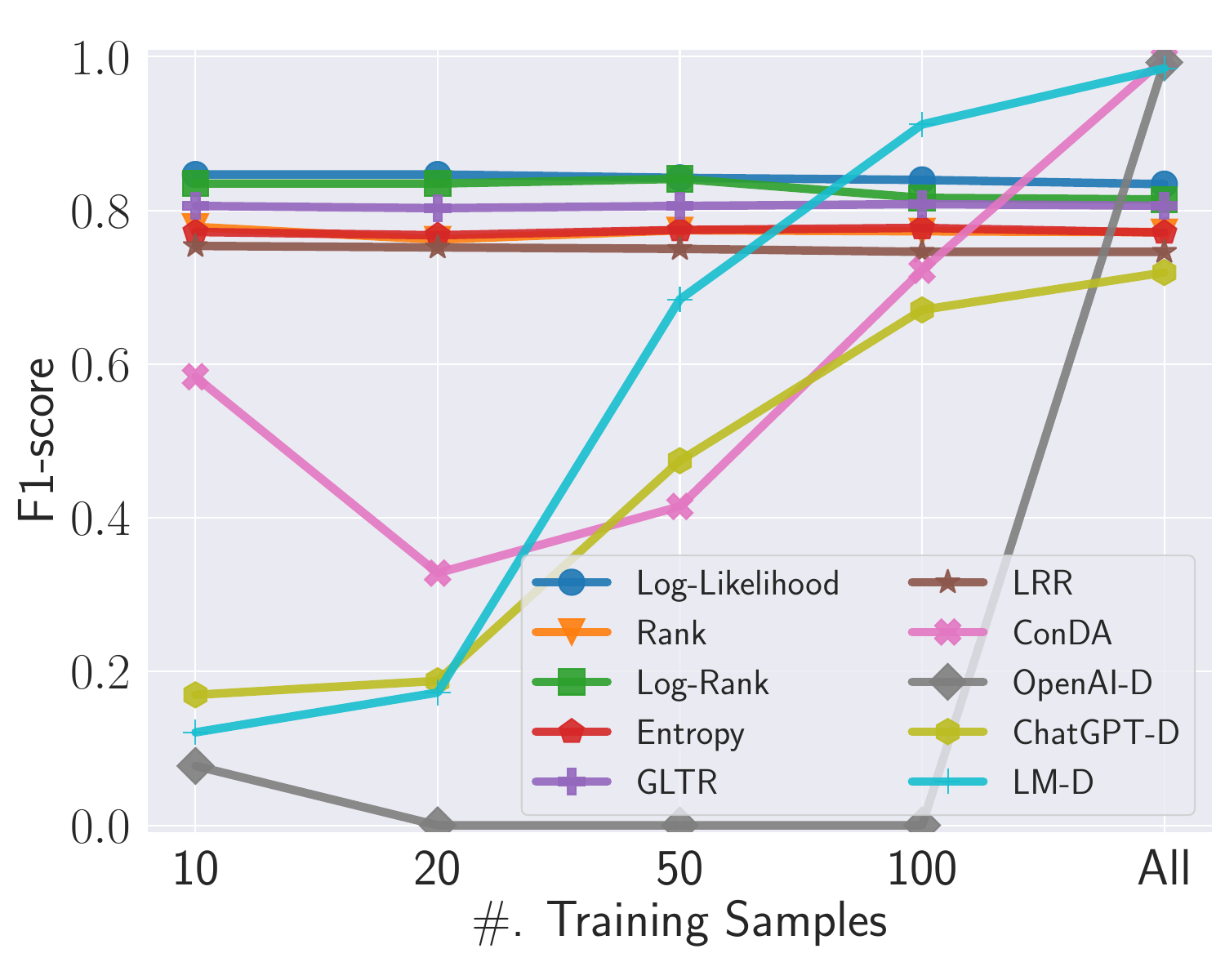}
\caption{Claude}
\label{figure:ablation_fewer_sample_Essay_Claude}
\end{subfigure}
\caption{The F1-score of different detection methods with different numbers of training samples on Essay.}
\label{figure:ablation_fewer_samples_Essay}
\end{figure*}

Regarding other model-based methods, we observe that LM Detector achieves the best performance across different datasets/LLMs in general.
For instance, on Essay, the LM detector achieves 0.993 F1-score against ChatGPT-turbo.
This is expected as the LM detector is trained on the corpus with both HWTs and MGTs, and the LM's ability to capture the context and coherence information greatly aids in differentiating the pattern variations between these two types of texts.
We also observe that the ChatGPT detector shows better performance in differentiating HWTs from MGTs when the models are from the GPT family, i.e., ChatGLM, ChatGPT-turbo, and GPT4All.
For instance, on Essay, the F1-score is 0.923, 0.742, and 0.815 for ChatGLM, ChatGPT-turbo, and GPT4All, respectively.
We suspect the reason is that the ChatGPT detector is trained on the corpus with both human answers and ChatGPT-generated answers, which can better capture the ``machine'' pattern on the GPT model family.
On the other hand, we find that the OpenAI Detector's performance is less satisfying in the detection task (e.g., 0.751 F1-score on Essay with Dolly-generated texts as MGTs).
This might be credited to the fact that this detector is trained on MGTs produced by GPT2 while the MGTs produced by larger LLMs such as ChatGPT-turbo or Claude have higher quality, which makes it harder to be detected.
Regarding ConDA, it is the weakest detection method.
One possible reason is that ConDA is trained with corpus with shorter length~\cite{BKML23}, which makes it harder to transfer to longer texts.
We later show that after fine-tuning, ConDA, OpenAI Detector, and ChatGPT Detector can achieve comparable or even better performance than LM Detector (see \autoref{figure:ablation_fewer_samples_Essay}).
Lastly, we find that GPTZero, a commercial MGT detection API, reaches good performance except for Dolly on WP and Reuters as well as StableLM on all datasets.
By checking the distribution of \#. words (shown in \autoref{figure:ablation_word_cnt}), we find that the MGTs has fewer \#. words in these cases, which leads to lower performance.
It is also mentioned on GPTZero's website.
Note that for GPTZero, we only sample 1/8 testing data for evaluation, which roughly matches the monthly word limit for the API subscription starting plan (\$49.99).\footnote{\url{https://app.gptzero.me/app/api-subscription}.}

\mypara{Detection Efficiency}
We then quantify the time cost of each detection method on different datasets.
Here we consider ChatGPT-turbo as the LLM for a case study since other LLMs show similar time costs.
As shown in \autoref{table:time_cost}, we can observe that most of the detection methods have similar time costs except NPR, DetectGPT, and GPTZero, which cost significantly higher time than the others.
This is because NPR and DetectGPT require multiple rounds of perturbation to the text to get a good estimation of the metric's change.
Also, GPTZero needs to query the public API, where the internet latency has to be taken into account.

In general, we consider LM Detector and Log-Rank as better detection methods as they achieve the best detection performance.
Moreover, their associated time cost is relatively low, making them comparable to other metric-based detection methods.
Note that later we omit NPR, DetectGPT, and GPTZero as they usually have less satisfying performance and take much longer time/monetary costs than the others.

\begin{table}[!htbp]
\centering
\caption{Time cost (seconds) to differentiate texts generated by ChatGPT or humans.}
\label{table:time_cost}
\scalebox{0.95}{
\begin{tabular}{l | c c c}
\toprule
\textbf{Method} & \textbf{Essay} & \textbf{WP} & \textbf{Reuters} \\
\midrule
Log-Likelihood  & 63 & 60 & 62  \\
Rank            & 75 & 70 & 72 \\
Log-Rank        & 75 & 70 & 73  \\
Entropy         & 50 & 47 & 52 \\
GLTR            & 88 & 83 & 85 \\
LRR             & 138 & 130 & 134 \\
NPR             & 2307 & 2257 & 2391 \\
DetectGPT       & 2185 & 2142 & 2272 \\
ConDA        & 16 & 16 & 16 \\
OpenAI-D        & 16 & 16 & 16 \\
ChatGPT-D       & 16 & 15 & 15 \\
LM-D            & 59  & 59 & 59 \\
\bottomrule
\end{tabular}
}
\end{table}

\subsection{Ablation Studies}

We now conduct several ablation studies to investigate how different factors would affect detection performance.

\mypara{Effect of Text Length}
We first present the distribution of \#. words for HWTs and MGTs on different datasets (shown in \autoref{figure:ablation_word_cnt}).
We observe that, although we consider to prompt the LLM with \#. words that are expected to be generated (see \autoref{table:dataset_prompts}), different LLMs demonstrate varying degrees of proficiency in matching the specified word count.
For instance, ChatGPT-turbo and Claude exhibit the highest performance, while Dolly and StableLM exhibit the lowest.
We then take a step further to understand how the number of words in texts would affect the detection performance.
Concretely, we train each detection method with the full-length texts and test it with the texts that have been truncated to have maximum $K$ words, where $K \in [10, 20, 50, 100, 200, 500]$.
The results on Essay are shown in \autoref{figure:ablation_fewer_words_Essay}.
First, we observe that, in general, larger $K$ leads to better detection performance, which is in line with previous work~\cite{SBCAHWRW19,VFTK23,CBZAMH23}.
For instance, the detection performance (F1-score) of ChatGLM with Entropy (\autoref{figure:Essay_ChatGLM_fewer_words}) increases from 0.150 to 0.809 when $K$ increases from 10 to 200.
Also, we find that 200 words are generally enough to achieve the (nearly) best performance in most of the cases.
For instance, for Claude (\autoref{figure:Essay_Claude_fewer_words}), Log-Likelihood reaches F1-score of 0.800 and 0.834 with 200-word texts and full-length texts.
This discovery provides valuable insights into the selection of HWTs and MGTs for the development of MGT detection methods.

\mypara{Fine-tune with Fewer Samples}
We then investigate the detection effectiveness with fewer training samples and the results on Essay are shown in \autoref{figure:ablation_fewer_samples_Essay}.
Note that different from the previous evaluation, here we also fine-tune ConDA, OpenAI Detector, and the ChatGPT Detector with those training samples.

We can observe that, in most cases, 10 training samples are sufficient for metric-based methods to achieve good detection performance.
For instance, when the LLM is ChatGLM (\autoref{figure:ablation_fewer_sample_Essay_ChatGLM}), Log-Likelihood reaches 0.967 F1-score with only 10 training samples, which is close to the performance with full training data, i.e., 0.970.
On the other hand, for model-based methods, more training samples can better facilitate the detection performance.
For instance, the F1-score of LM-D increases from 0.121 (with 10 training samples) to 0.984 (with full training data).
This difference is expected, as metric-based methods primarily need to determine a threshold to separate metric values, whereas model-based methods require the optimization of a large number of parameters, typically benefiting from a larger training dataset.

\begin{figure*}[!t]
\centering
\begin{subfigure}{0.39\columnwidth}
\includegraphics[width=\columnwidth]{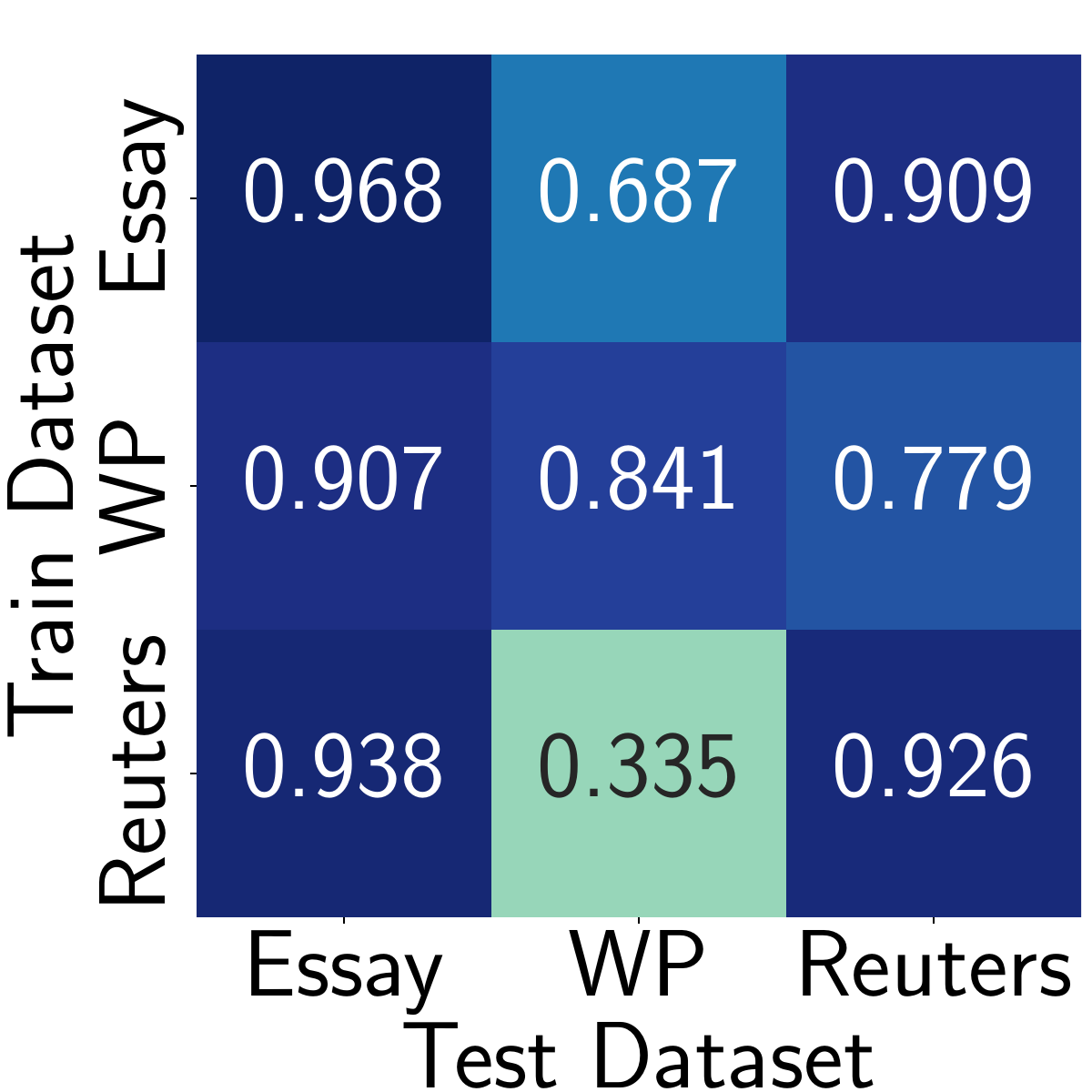}
\caption{Log-Likelihood}
\label{figure:ablation_transfer_Log-Likelihood_ChatGPT-turbo}
\end{subfigure}
\begin{subfigure}{0.39\columnwidth}
\includegraphics[width=\columnwidth]{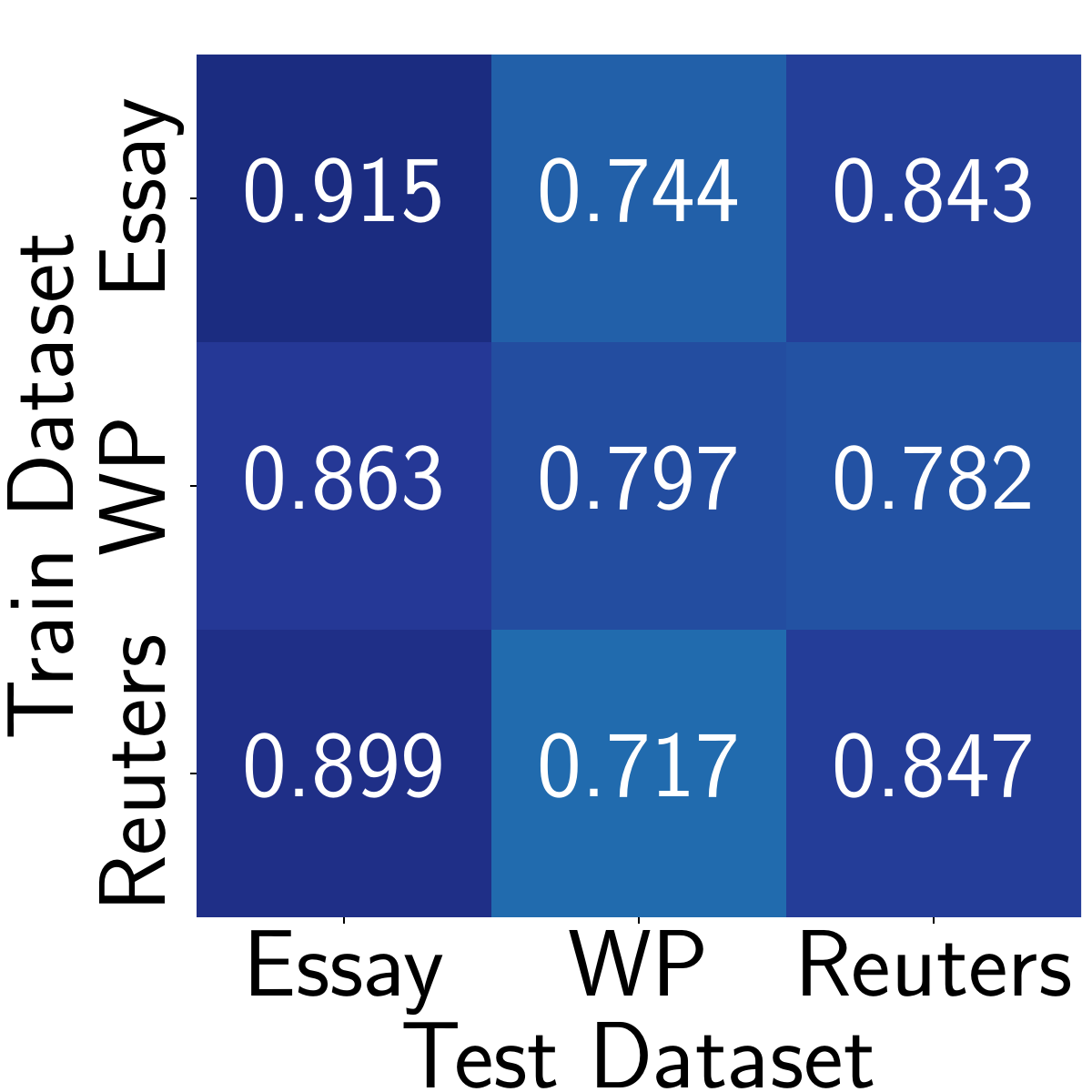}
\caption{Rank}
\label{figure:ablation_transfer_Rank_ChatGPT-turbo}
\end{subfigure}
\begin{subfigure}{0.39\columnwidth}
\includegraphics[width=\columnwidth]{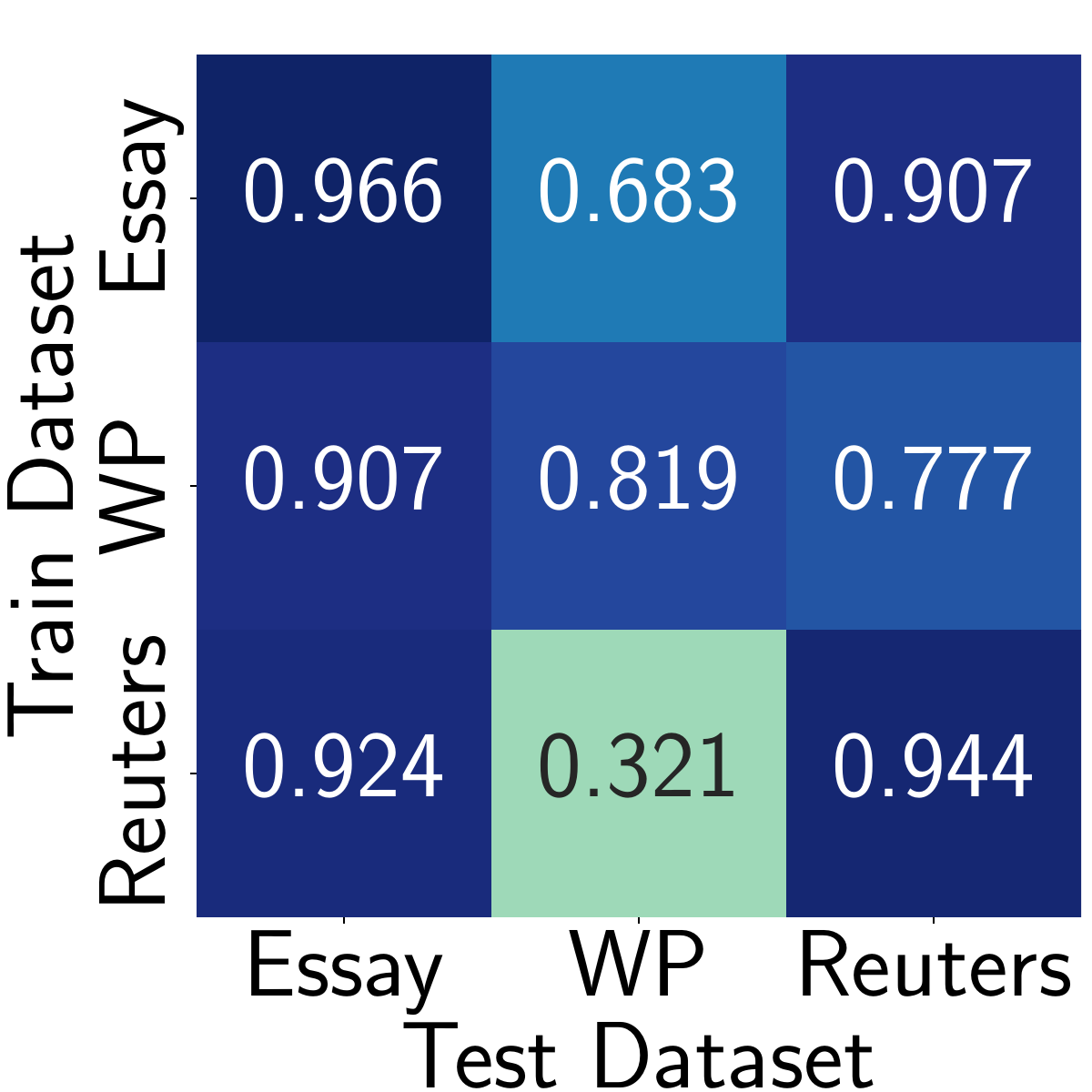}
\caption{Log-Rank}
\label{figure:ablation_transfer_Log-Rank_ChatGPT-turbo}
\end{subfigure}
\begin{subfigure}{0.39\columnwidth}
\includegraphics[width=\columnwidth]{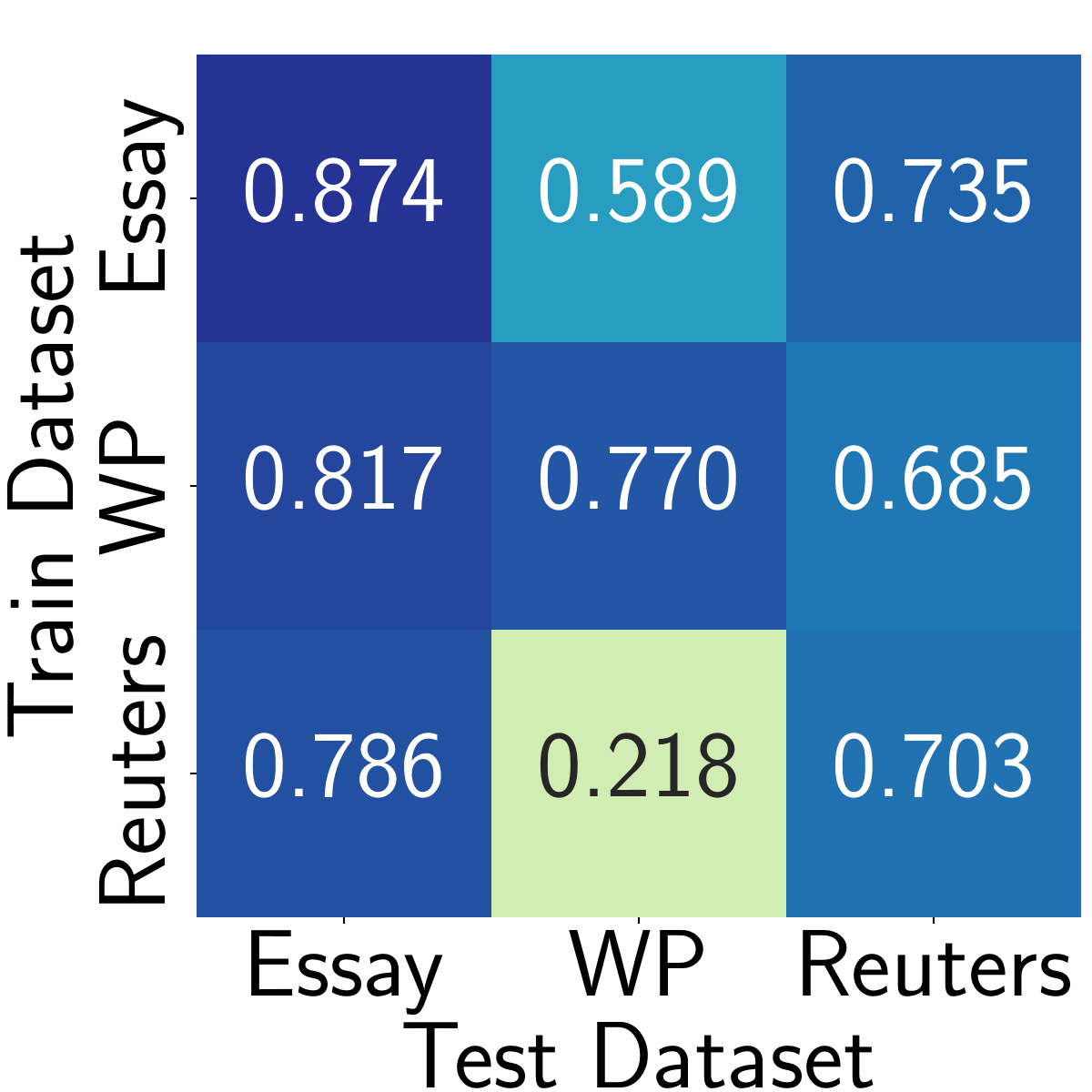}
\caption{Entropy}
\label{figure:ablation_transfer_Entropy_ChatGPT-turbo}
\end{subfigure}
\begin{subfigure}{0.39\columnwidth}
\includegraphics[width=\columnwidth]{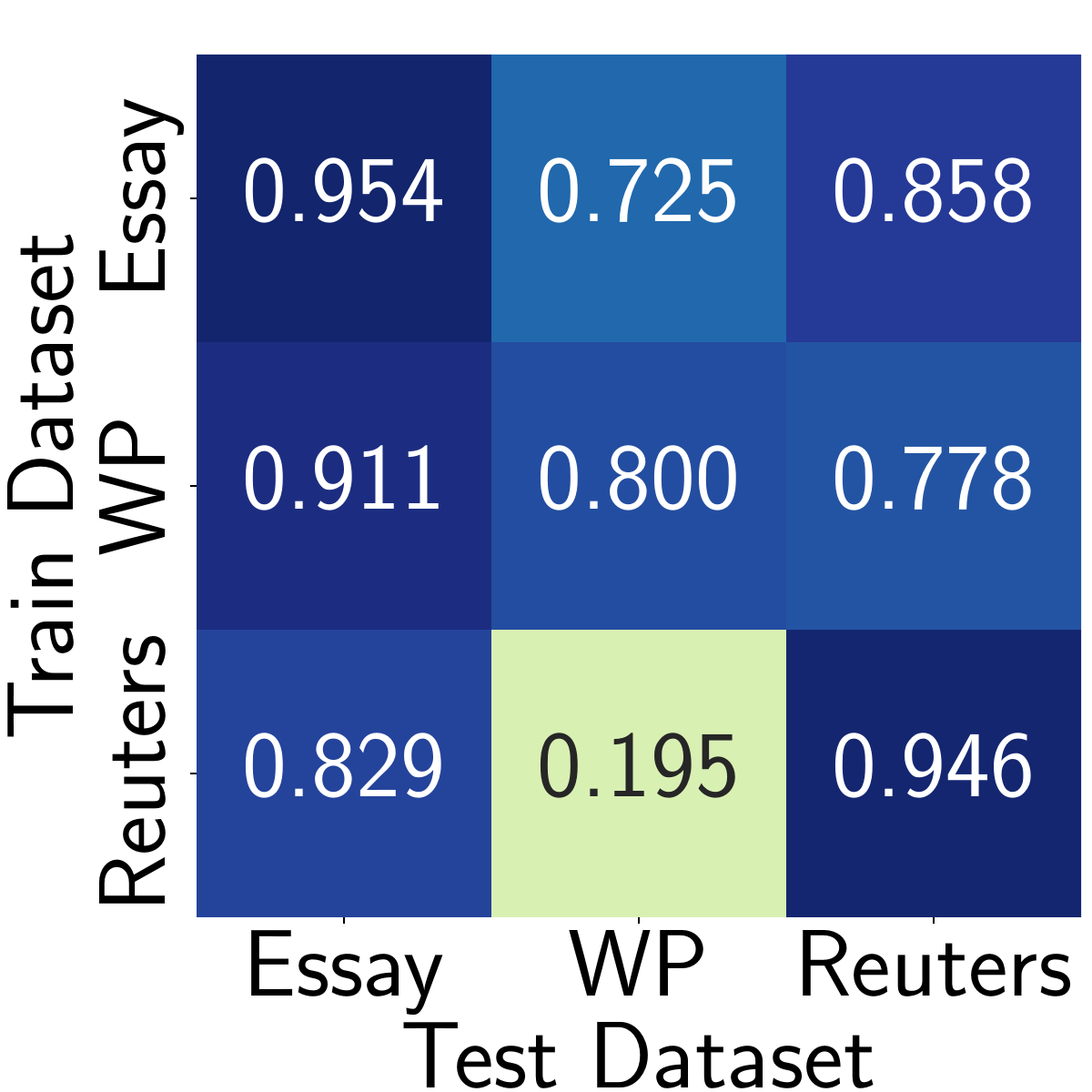}
\caption{GLTR}
\label{figure:ablation_transfer_GLTR_ChatGPT-turbo}
\end{subfigure}
\begin{subfigure}{0.39\columnwidth}
\includegraphics[width=\columnwidth]{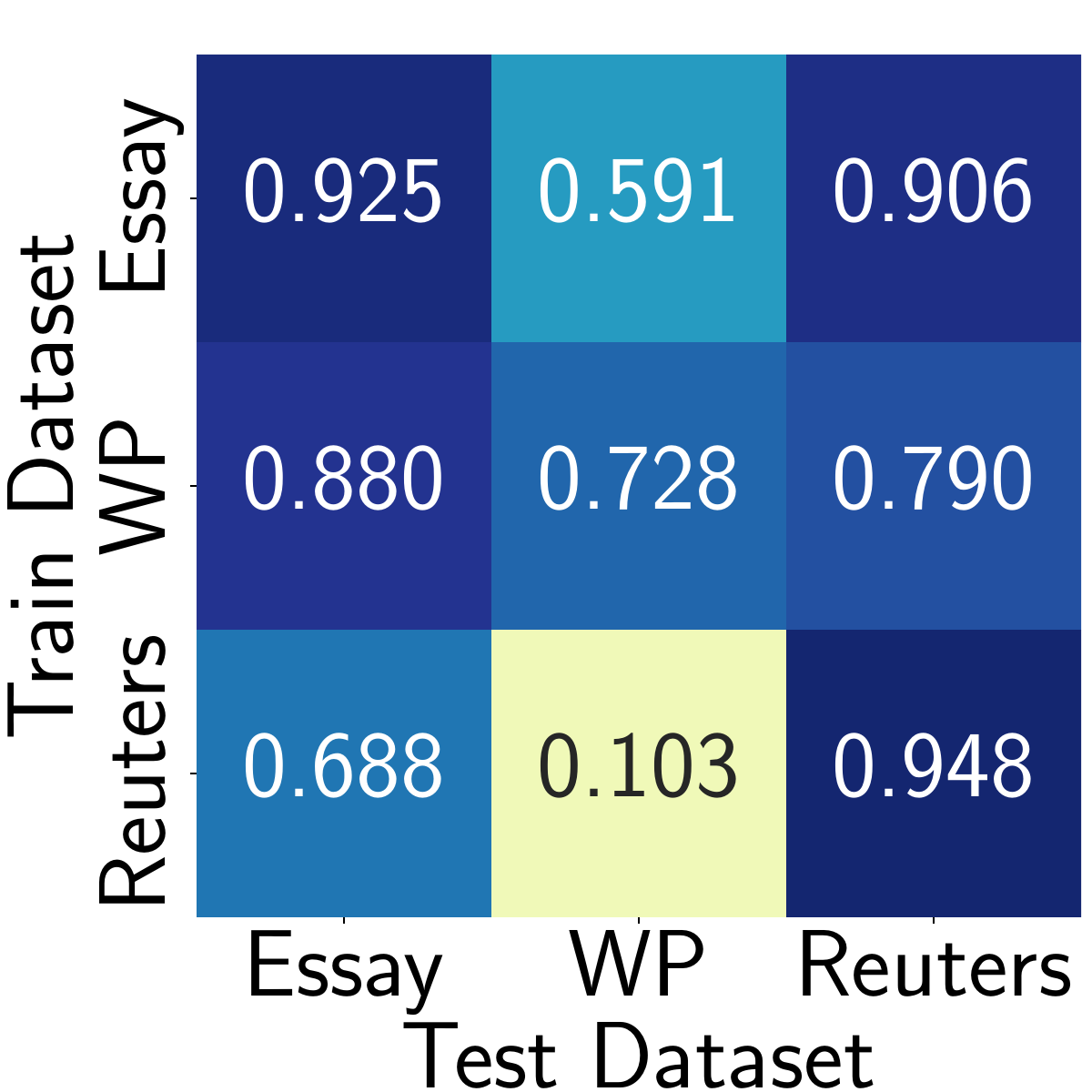}
\caption{LRR}
\label{figure:ablation_transfer_LRR_ChatGPT-turbo}
\end{subfigure}
\begin{subfigure}{0.39\columnwidth}
\includegraphics[width=\columnwidth]{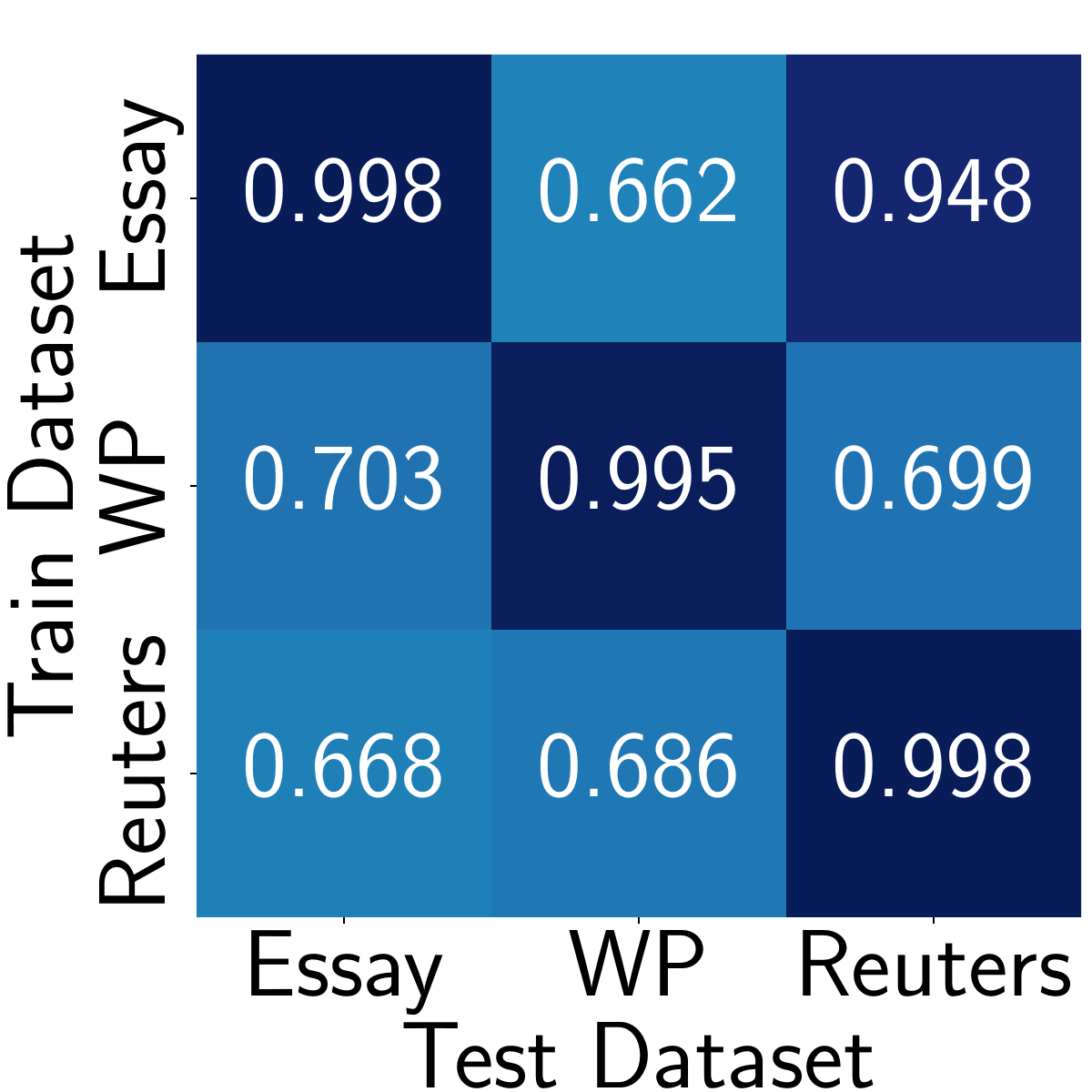}
\caption{ConDA}
\label{figure:ablation_transfer_ConDA_ChatGPT-turbo}
\end{subfigure}
\begin{subfigure}{0.39\columnwidth}
\includegraphics[width=\columnwidth]{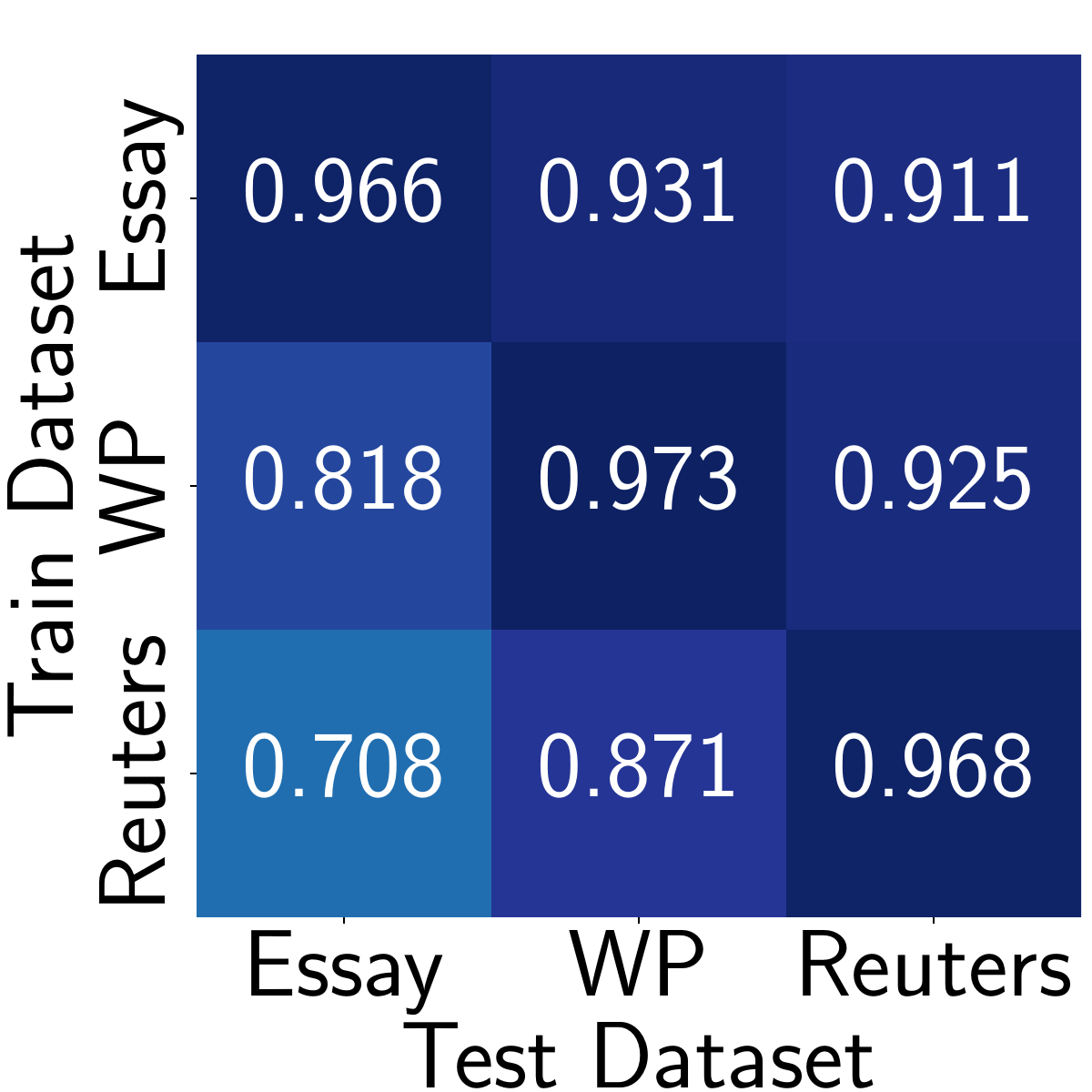}
\caption{OpenAI-D}
\label{figure:ablation_transfer_OpenAI-D_ChatGPT-turbo}
\end{subfigure}
\begin{subfigure}{0.39\columnwidth}
\includegraphics[width=\columnwidth]{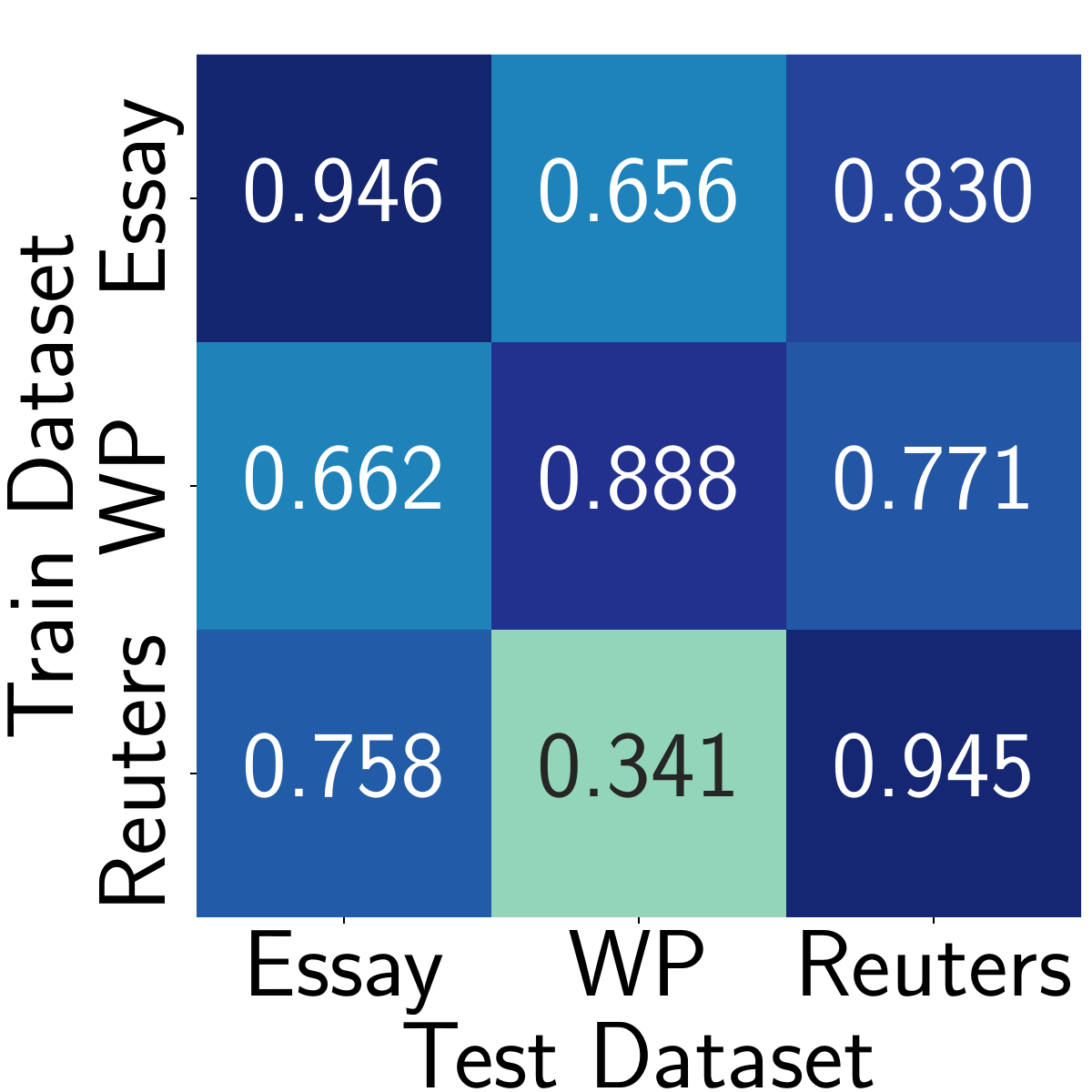}
\caption{ChatGPT-D}
\label{figure:ablation_transfer_ChatGPT-D_ChatGPT-turbo}
\end{subfigure}
\begin{subfigure}{0.39\columnwidth}
\includegraphics[width=\columnwidth]{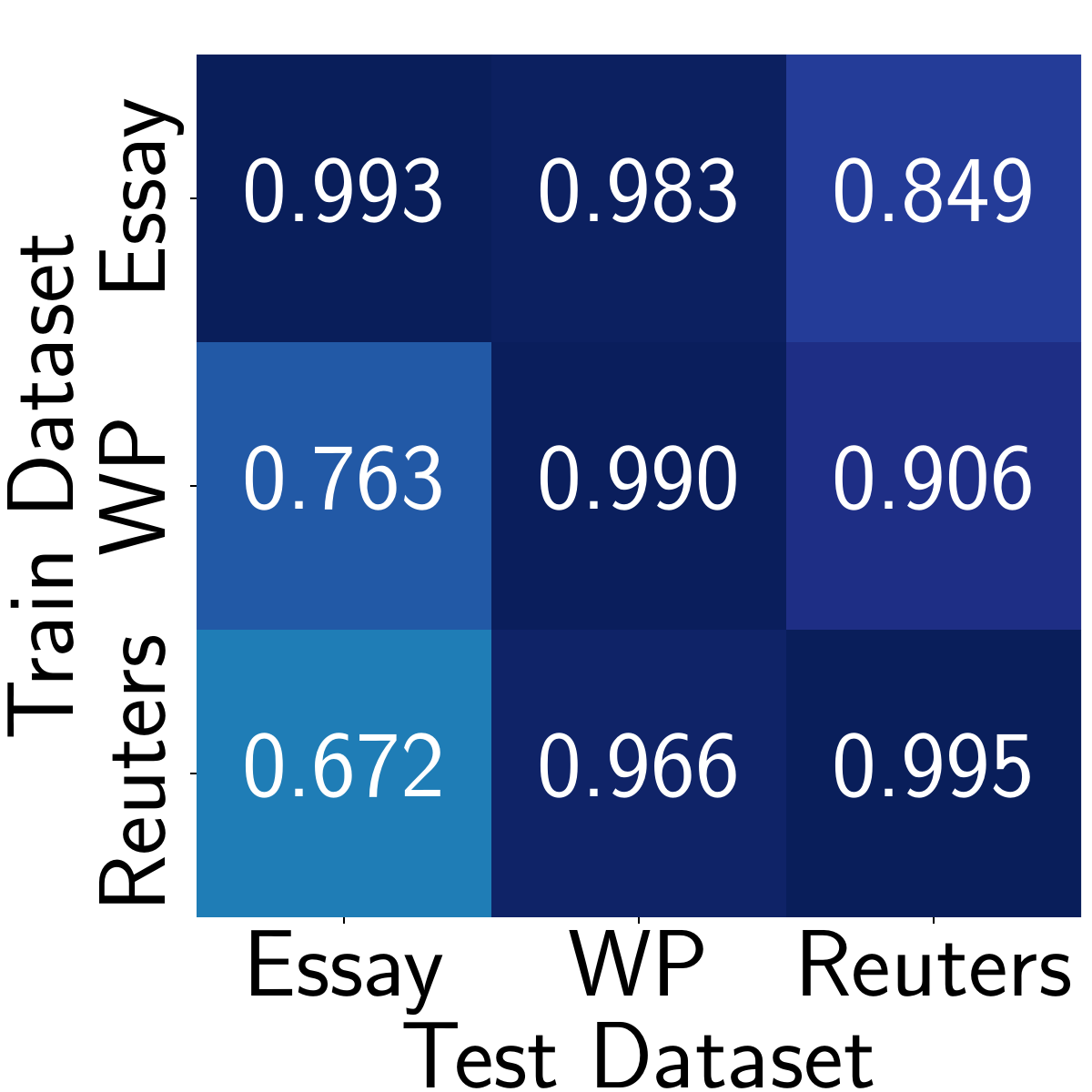}
\caption{LM-D}
\label{figure:ablation_transfer_LM-D_ChatGPT-turbo}
\end{subfigure}
\caption{The F1-score of different detection methods when the training dataset and the testing dataset are different. Here The MGTs are generated by ChatGPT-turbo.}
\label{figure:ablation_transfer_dataset}
\end{figure*}

\begin{figure*}[!t]
\centering
\begin{subfigure}{0.39\columnwidth}
\includegraphics[width=\columnwidth]{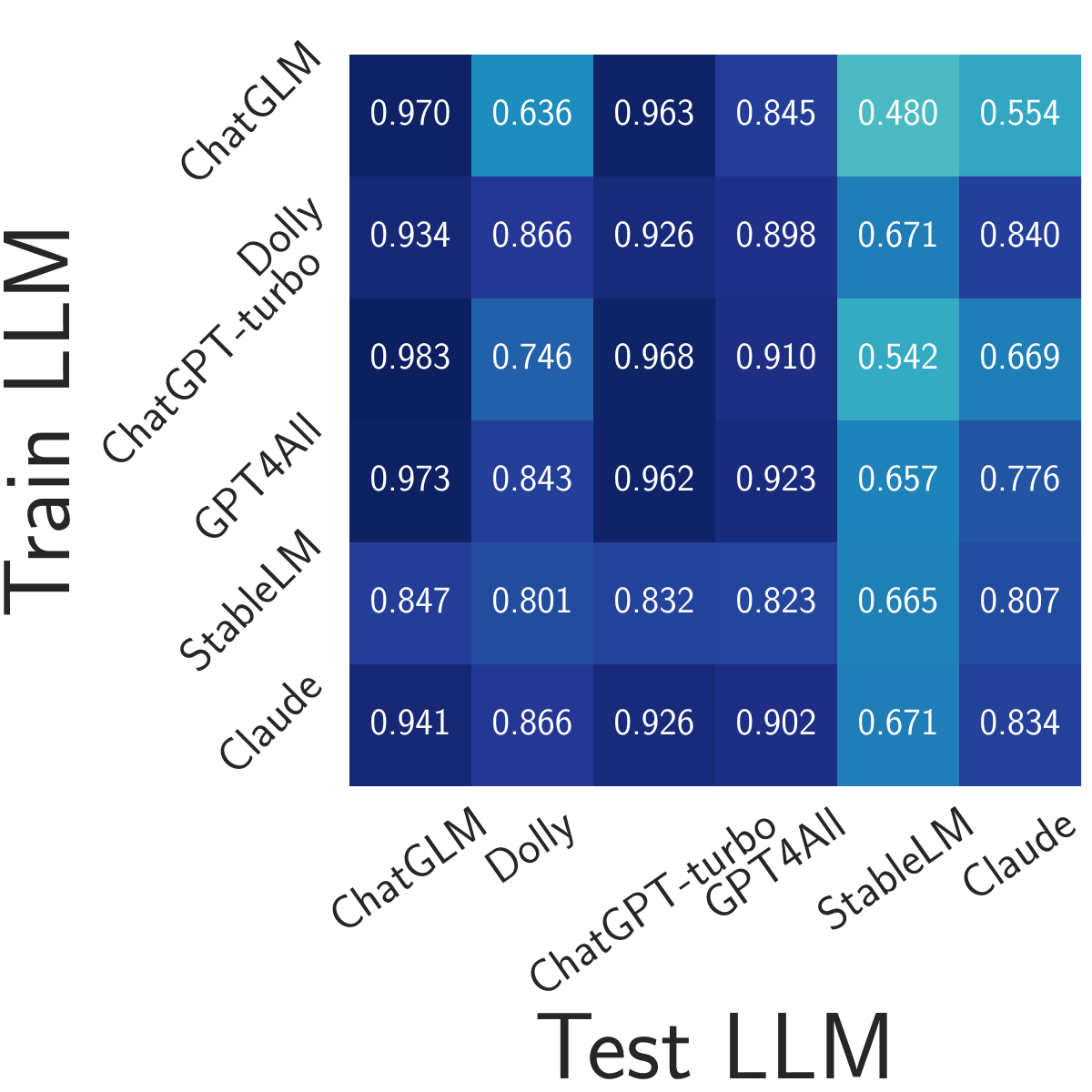}
\caption{Log-Likelihood}
\label{figure:ablation_transfer_llm_Log-Likelihood_Essay}
\end{subfigure}
\begin{subfigure}{0.39\columnwidth}
\includegraphics[width=\columnwidth]{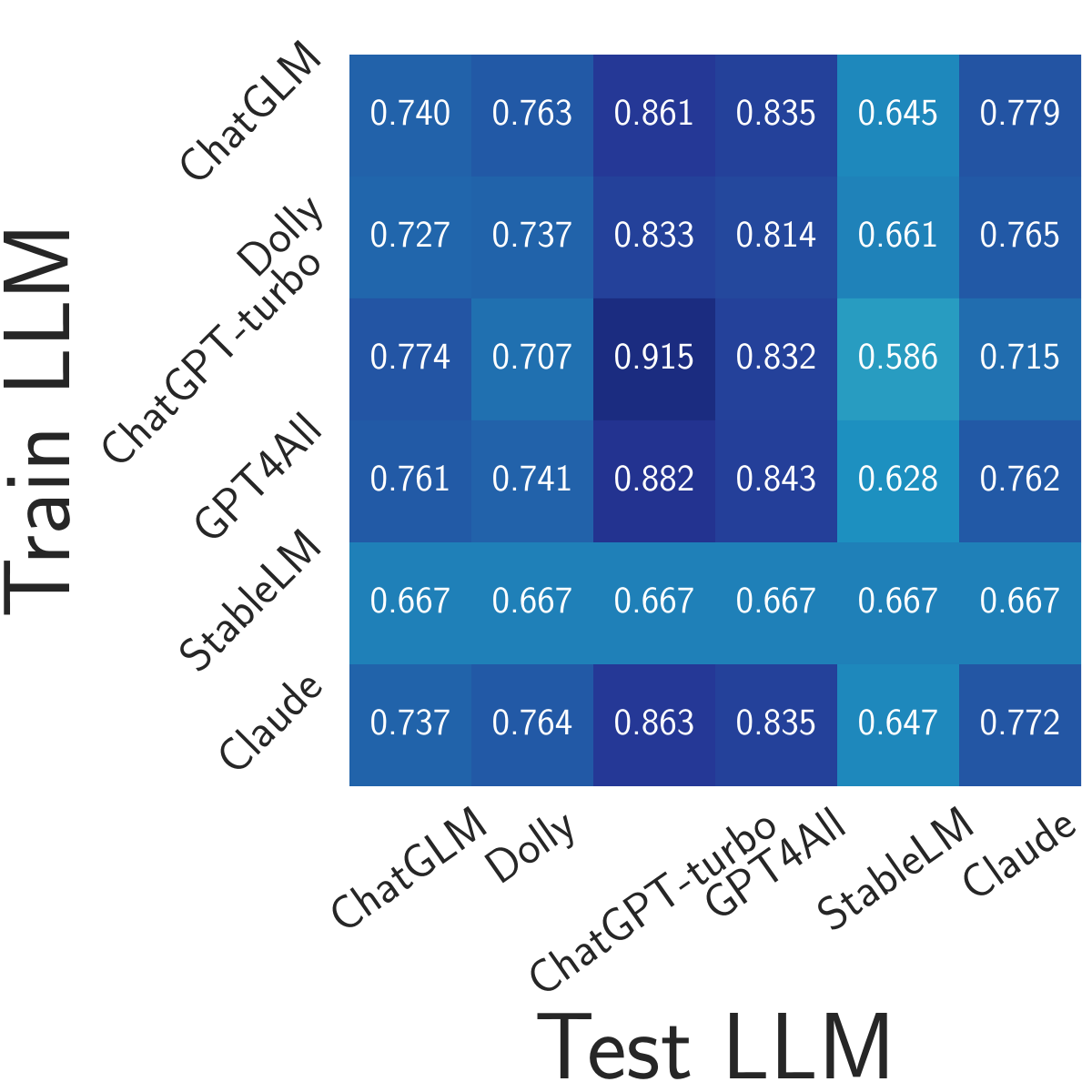}
\caption{Rank}
\label{figure:ablation_transfer_llm_Rank_Essay}
\end{subfigure}
\begin{subfigure}{0.39\columnwidth}
\includegraphics[width=\columnwidth]{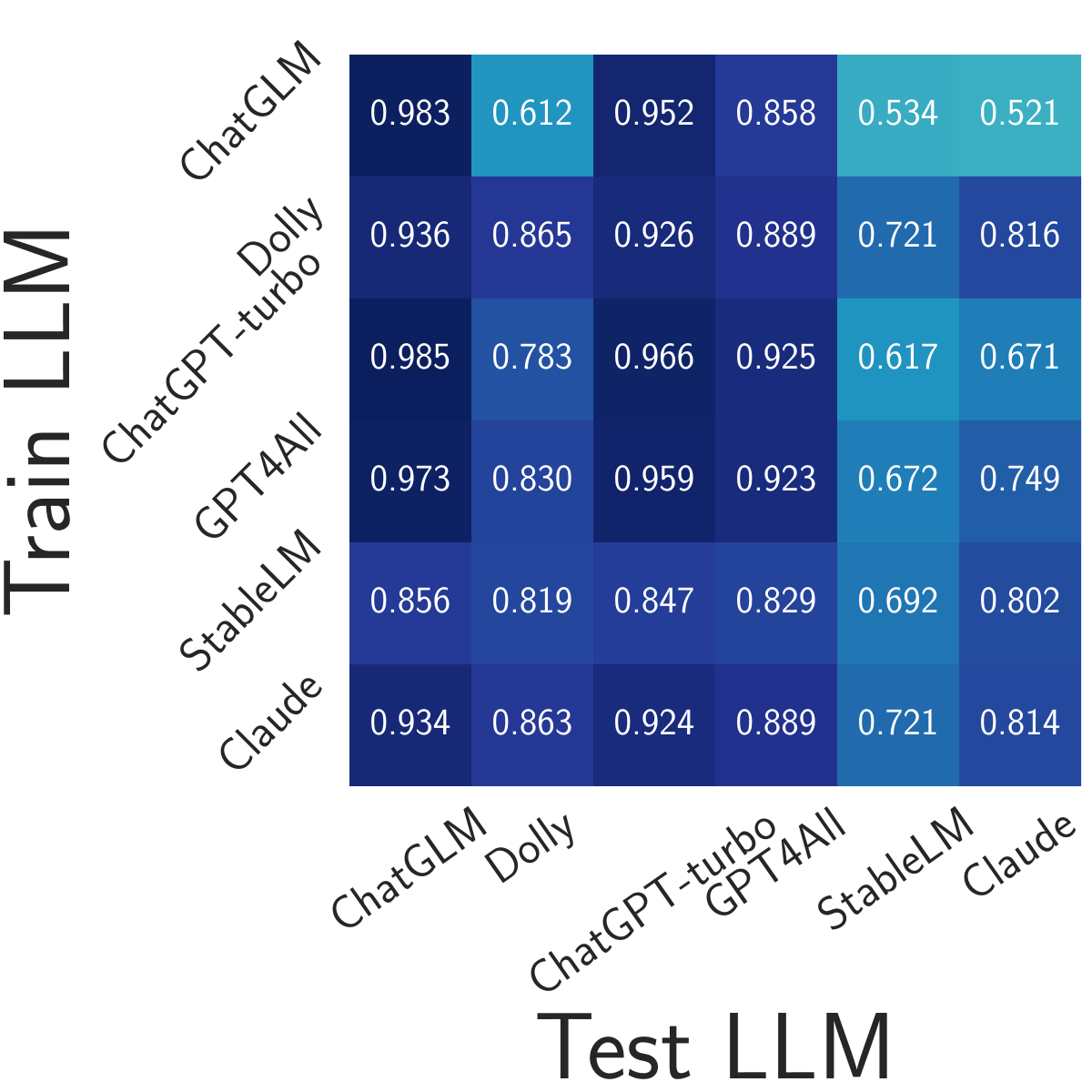}
\caption{Log-Rank}
\label{figure:ablation_transfer_llm_Log-Rank_Essay}
\end{subfigure}
\begin{subfigure}{0.39\columnwidth}
\includegraphics[width=\columnwidth]{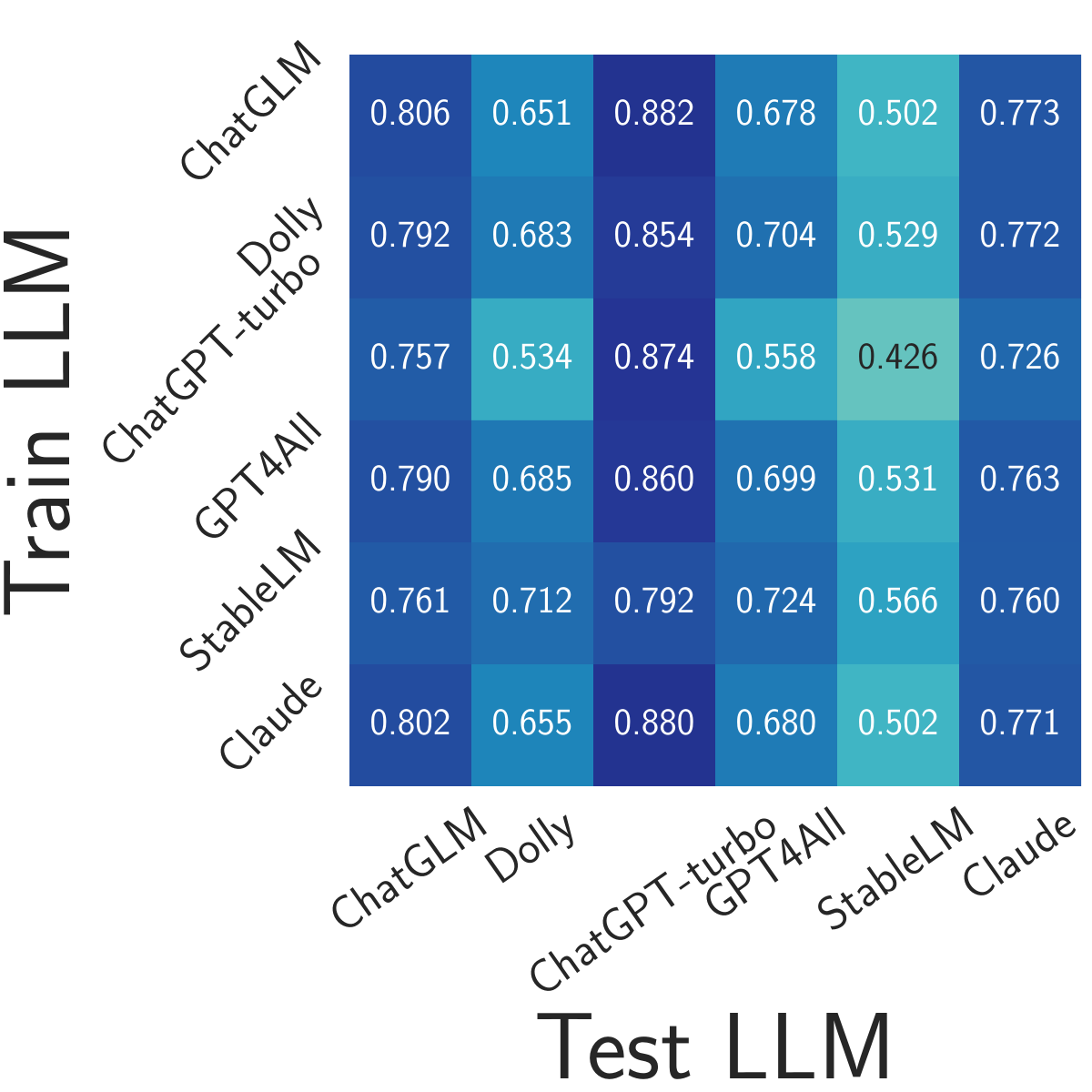}
\caption{Entropy}
\label{figure:ablation_transfer_llm_Entropy_Essay}
\end{subfigure}
\begin{subfigure}{0.39\columnwidth}
\includegraphics[width=\columnwidth]{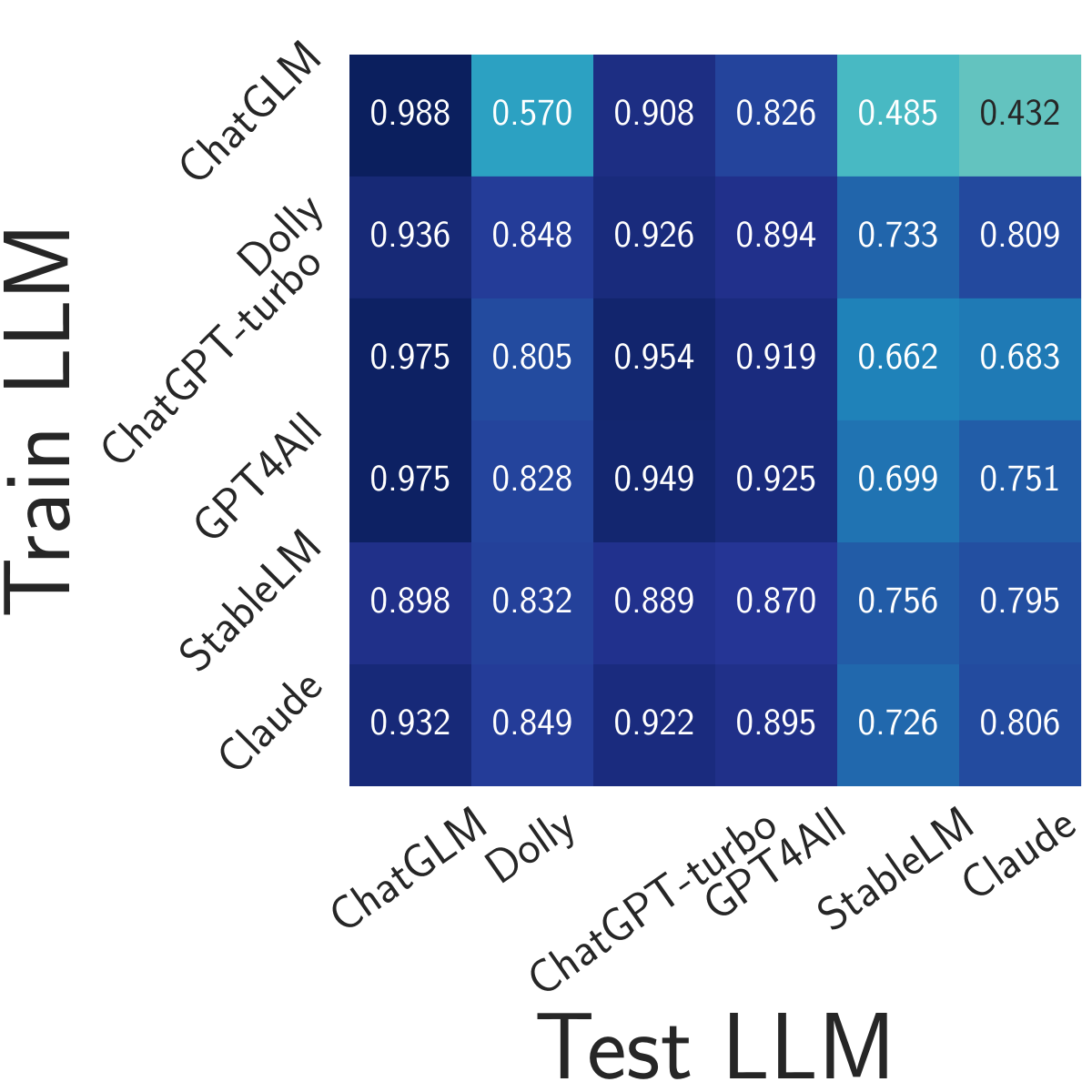}
\caption{GLTR}
\label{figure:ablation_transfer_llm_GLTR_Essay}
\end{subfigure}
\begin{subfigure}{0.39\columnwidth}
\includegraphics[width=\columnwidth]{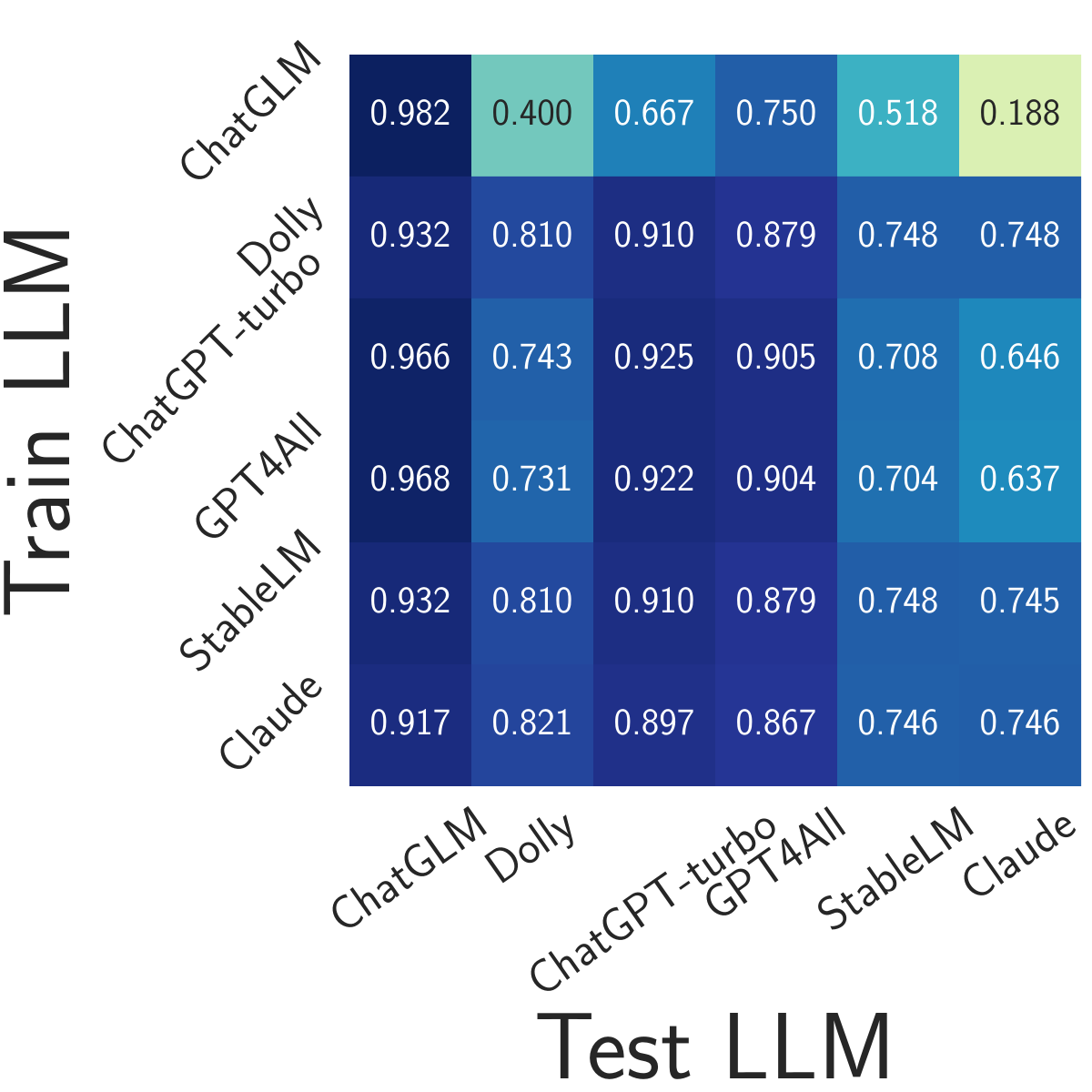}
\caption{LRR}
\label{figure:ablation_transfer_llm_LRR_Essay}
\end{subfigure}
\begin{subfigure}{0.39\columnwidth}
\includegraphics[width=\columnwidth]{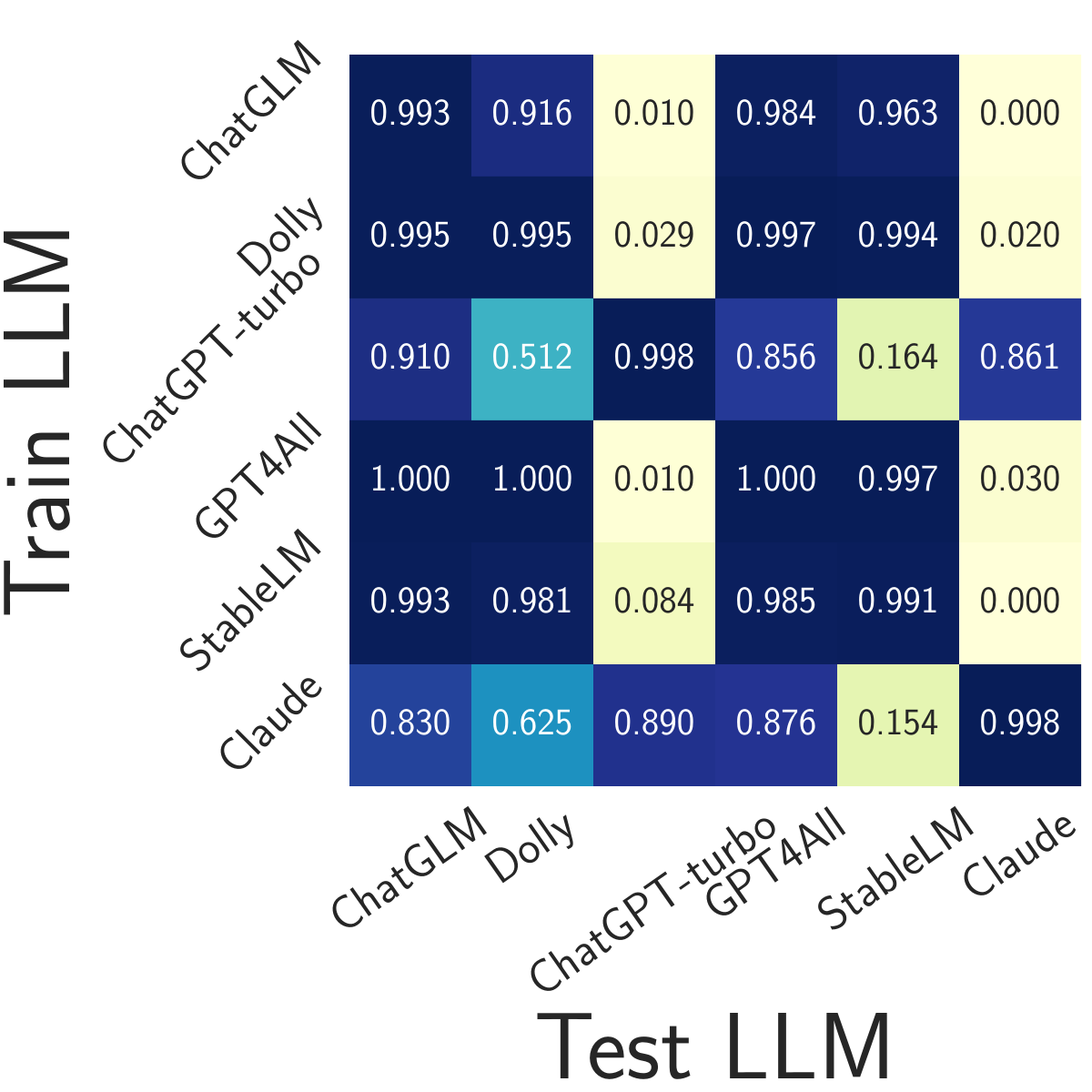}
\caption{ConDA}
\label{figure:ablation_transfer_llm_ConDA_Essay}
\end{subfigure}
\begin{subfigure}{0.39\columnwidth}
\includegraphics[width=\columnwidth]{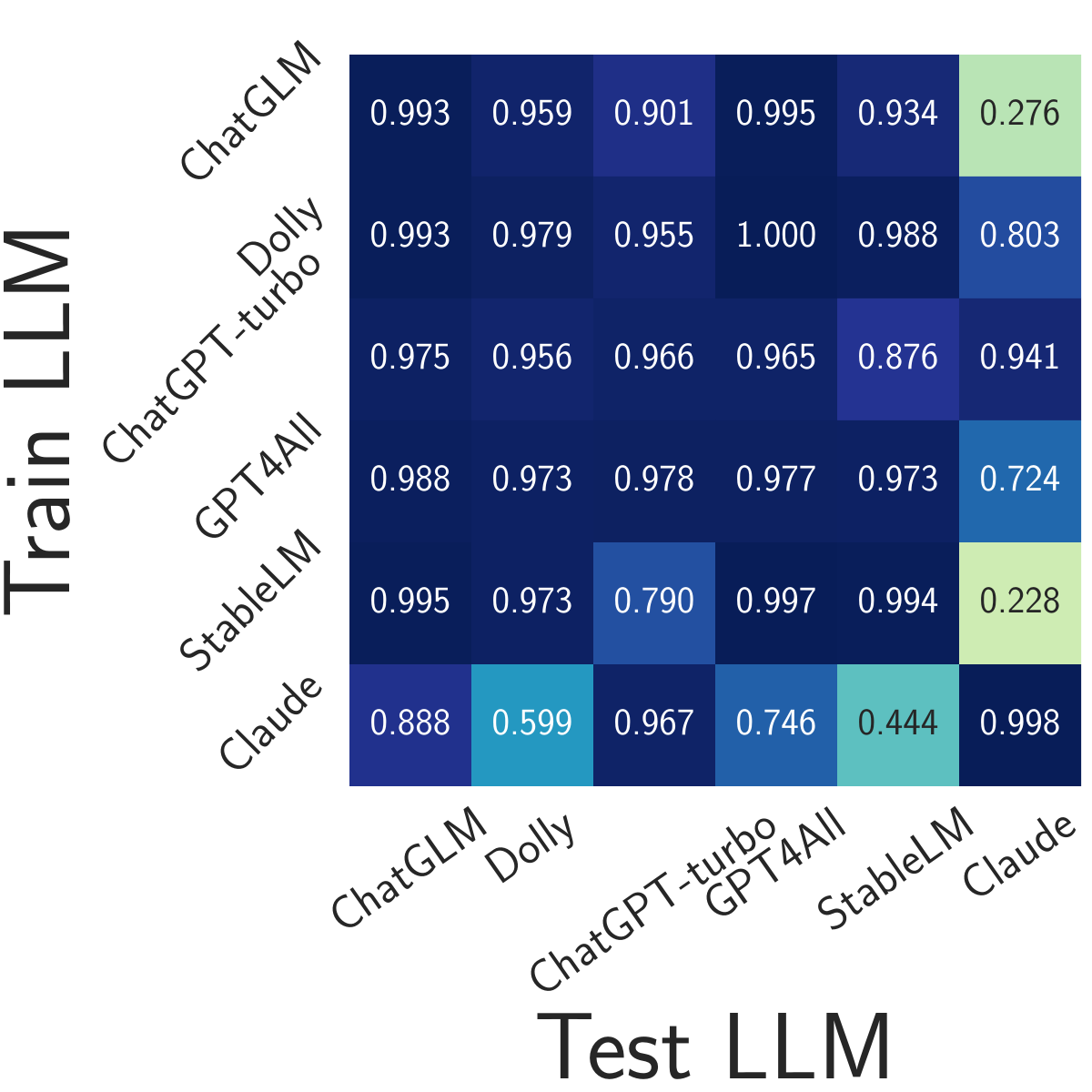}
\caption{OpenAI-D}
\label{figure:ablation_transfer_llm_OpenAI-D_Essay}
\end{subfigure}
\begin{subfigure}{0.39\columnwidth}
\includegraphics[width=\columnwidth]{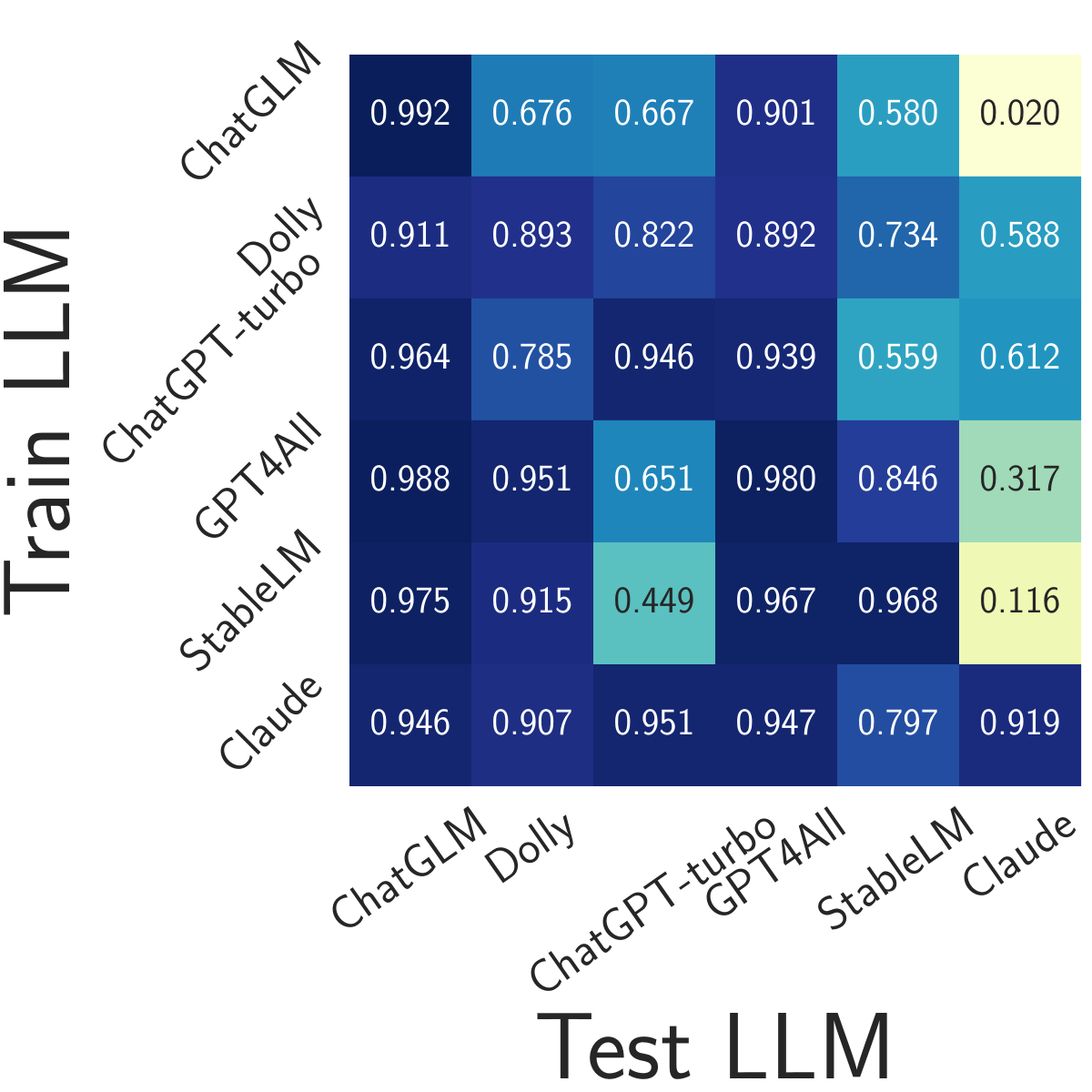}
\caption{ChatGPT-D}
\label{figure:ablation_transfer_llm_ChatGPT-D_Essay}
\end{subfigure}
\begin{subfigure}{0.39\columnwidth}
\includegraphics[width=\columnwidth]{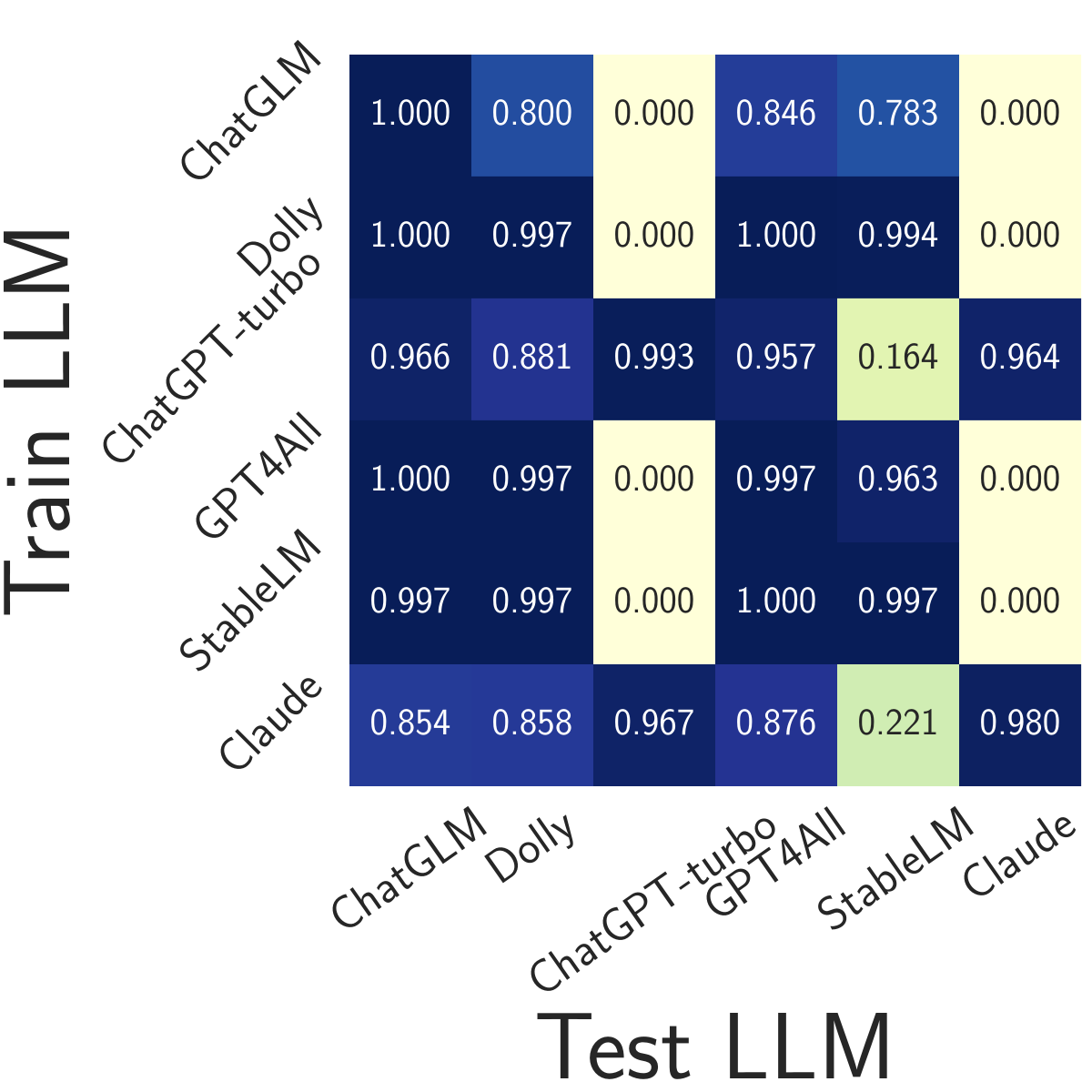}
\caption{LM-D}
\label{figure:ablation_transfer_llm_LM-D_Essay}
\end{subfigure}
\caption{The F1-score of different detection methods on Essay when the train LLM and the test LLM are different.}
\label{figure:ablation_transfer_llm}
\end{figure*}

\mypara{Transfer Setting}
Given the necessity of the training procedure for most detection methods, we aim to investigate the efficacy of transferring these methods to different datasets and LLMs.
Note that here we also fine-tune ConDA, OpenAI Detector, and ChatGPT Detector using the same data with LM Detector for a fair comparison.

We first investigate whether the detection methods trained on one dataset can be transferred to another dataset and the performance is shown in \autoref{figure:ablation_transfer_dataset}.
We can observe that detection methods trained on different datasets may have different transferability to the other datasets.
Concretely, WP has the highest transferability while Reuters has the lowest.
For instance, for Log-Likelihood trained on WP, the test F1-score is 0.907 and 0.779 on Essay and Reuters.
However, for Log-Likelihood trained on Reuters, the test F1-score is 0.938 and 0.335 on Essay and WP.
One possible reason is that Reuters only contains news articles from 50 authors, representing a relatively narrow domain.
In contrast, WP covers a wide range of topics, providing a more diverse dataset.
This diversity can better guide the detection method to acquire general features that facilitate the differentiation between MGTs and HWTs.

\begin{figure}[!htbp]
\centering
\begin{subfigure}{0.9\columnwidth}
\includegraphics[width=\columnwidth]{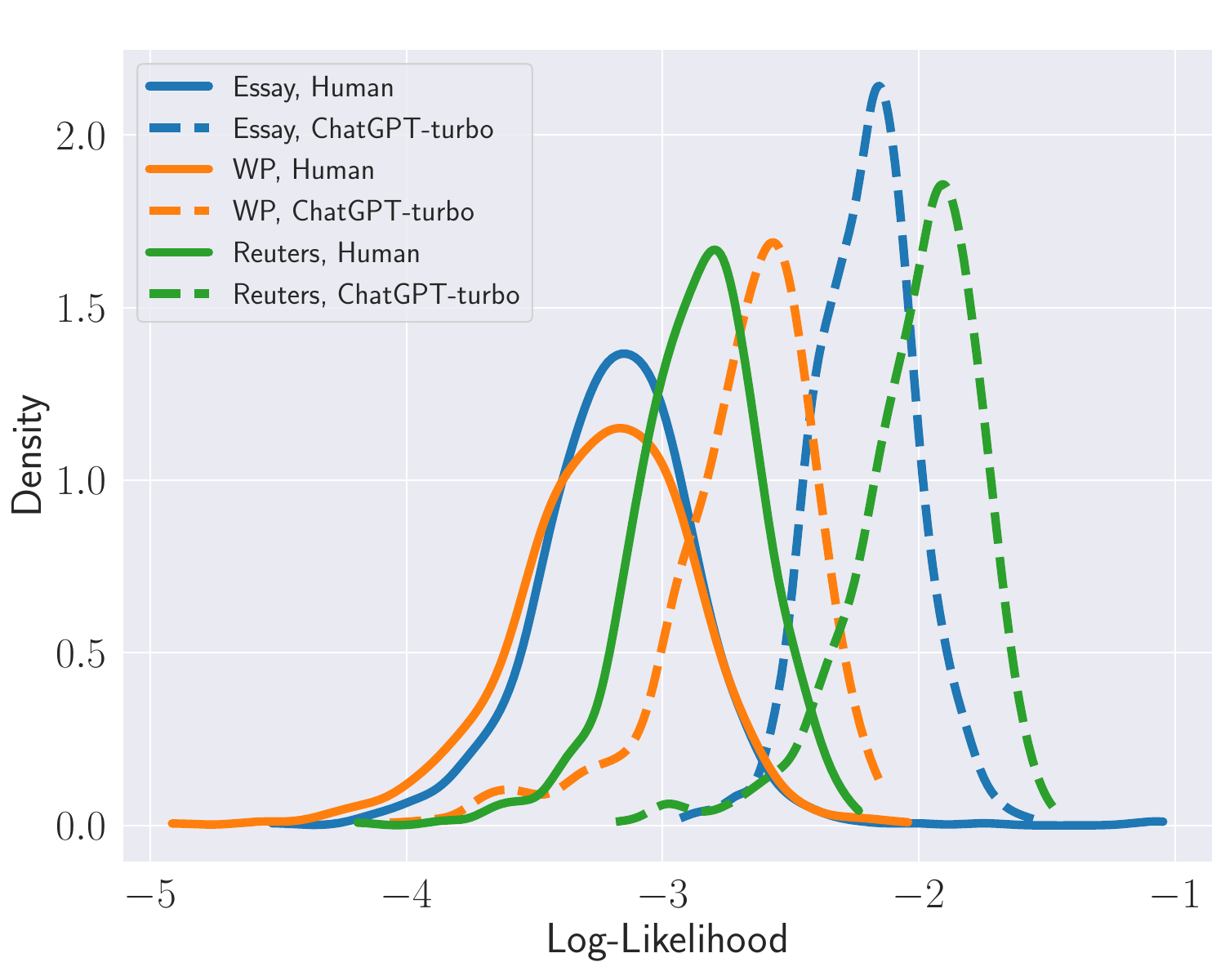}
\caption{Different Datasets}
\label{figure:distribution_Log-Likelihood_dataset}
\end{subfigure}
\begin{subfigure}{0.9\columnwidth}
\includegraphics[width=\columnwidth]{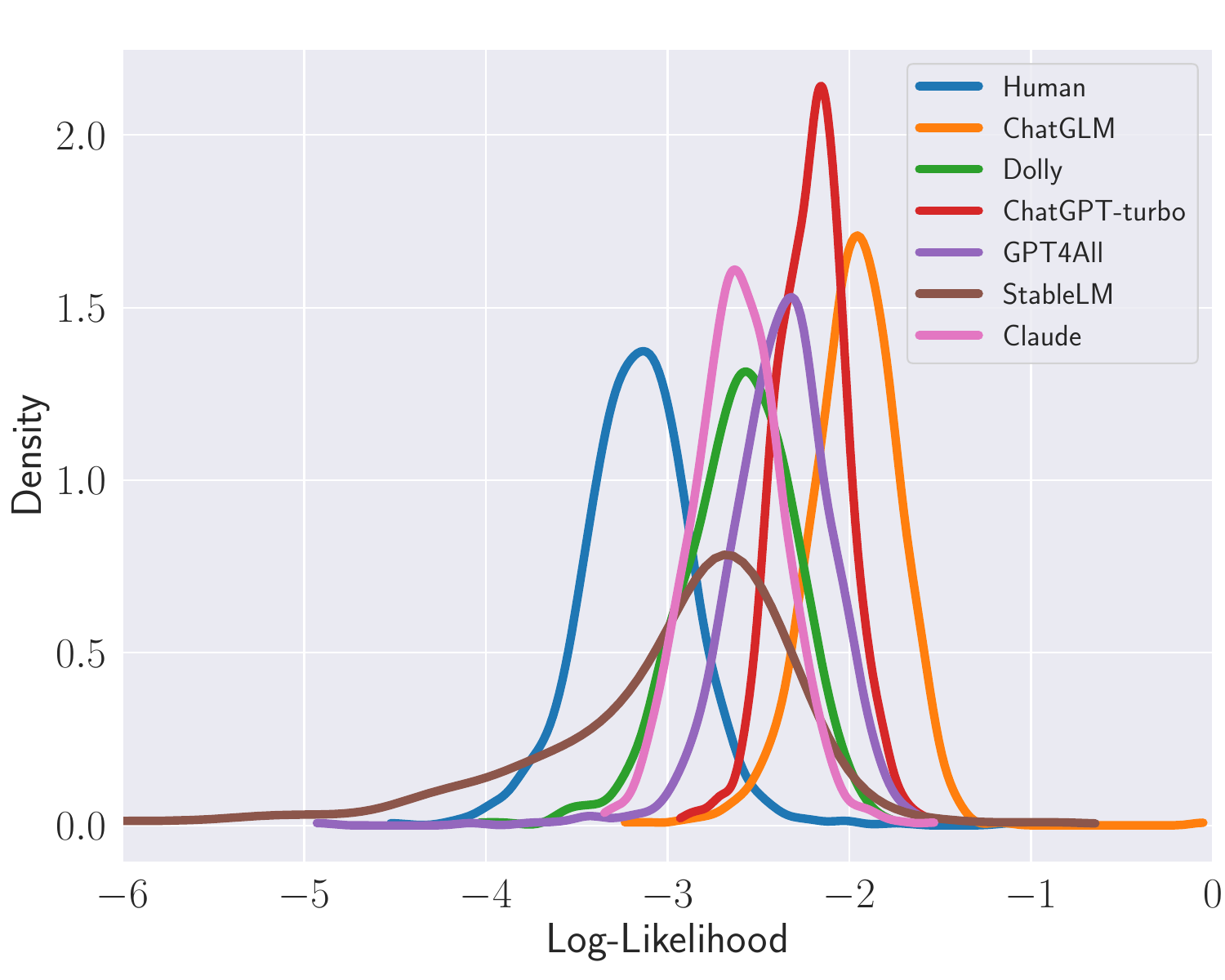}
\caption{Different LLMs}
\label{figure:distribution_Log-Likelihood_llm}
\end{subfigure}
\caption{The log-likelihood distributions of HWTs and MGTs. Note that (b) is derived from the Essay dataset.}
\label{figure:distribution_Log-Likelihood}
\end{figure}

Also, we find that compared to the metric-based methods, model-based methods like OpenAI Detector and LM Detector are relatively robust across different datasets.
For example, given the LM Detector trained on Essay (\autoref{figure:ablation_transfer_LM-D_ChatGPT-turbo}), the test F1-score only drops 0.010 and 0.144 on WP and Reuters.
However, for Entropy trained on Essay (\autoref{figure:ablation_transfer_Entropy_ChatGPT-turbo}), the test F1-score drops 0.285 and 0.139 on WP and Reuters.
This is because metric-based detection methods usually rely on one specific metric to perform the detection, which might not be robust enough for dataset distribution shifts.
We take Log-Likelihood as a case study and visualize its distributions for HWTs and MGTs (generated by ChatGPT-turbo) across different datasets in \autoref{figure:distribution_Log-Likelihood_dataset}.
We can observe that for different datasets, the best threshold to separate HWTs and MGTs is usually different, making it hard to transfer.
On the contrary, model-based methods can extract various features related to both LLM and human patterns.
This feature extraction process could enhance the detection performance and make it more robust to the dataset distribution shift.

\begin{figure}[!htbp]
\centering
\includegraphics[width=0.9\columnwidth]{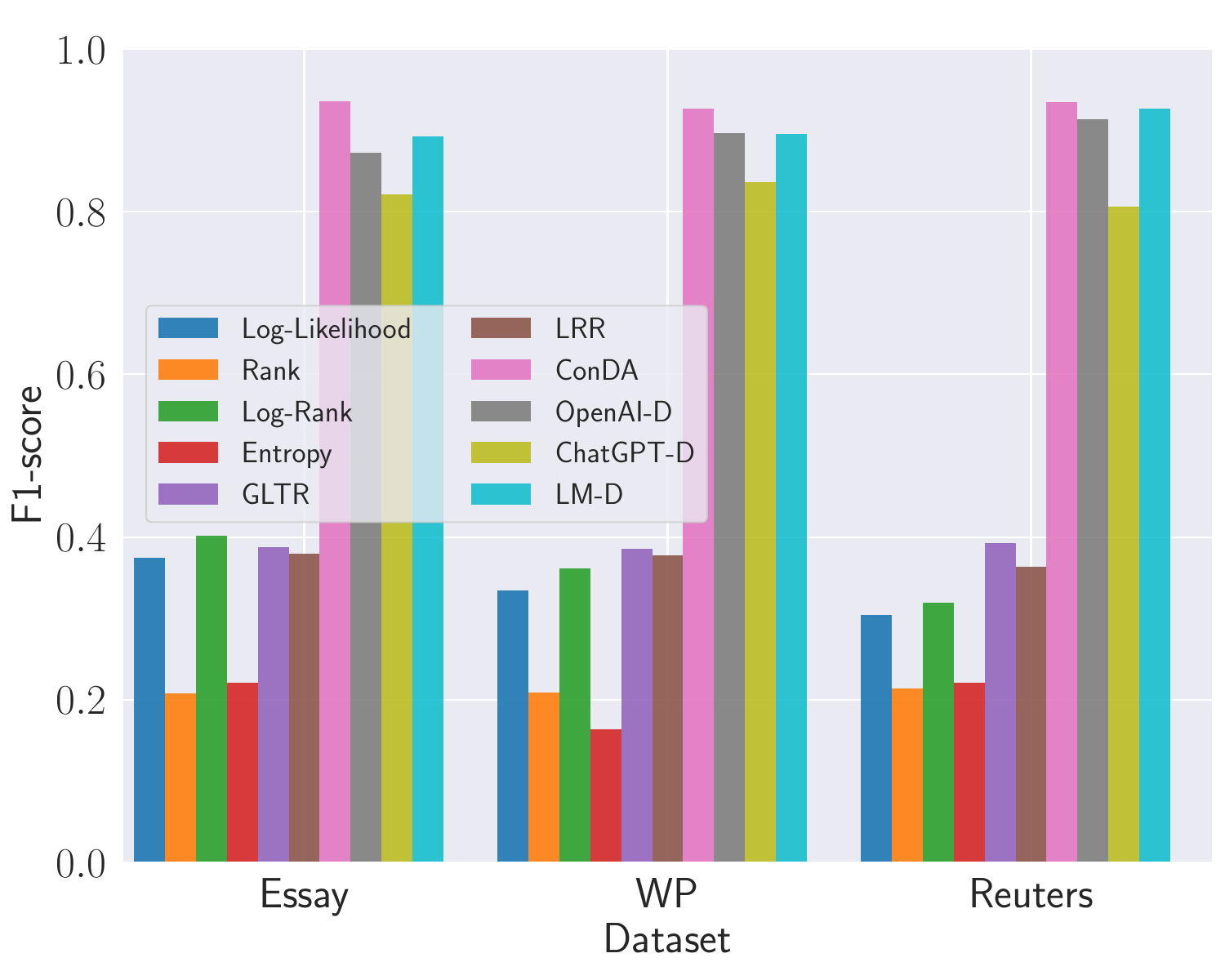}
\caption{The F1-score of different detection methods on the text attribution task.}
\label{figure:attribution_performance}
\end{figure}

\begin{figure*}[!t]
\centering
\begin{subfigure}{0.64\columnwidth}
\includegraphics[width=\columnwidth]{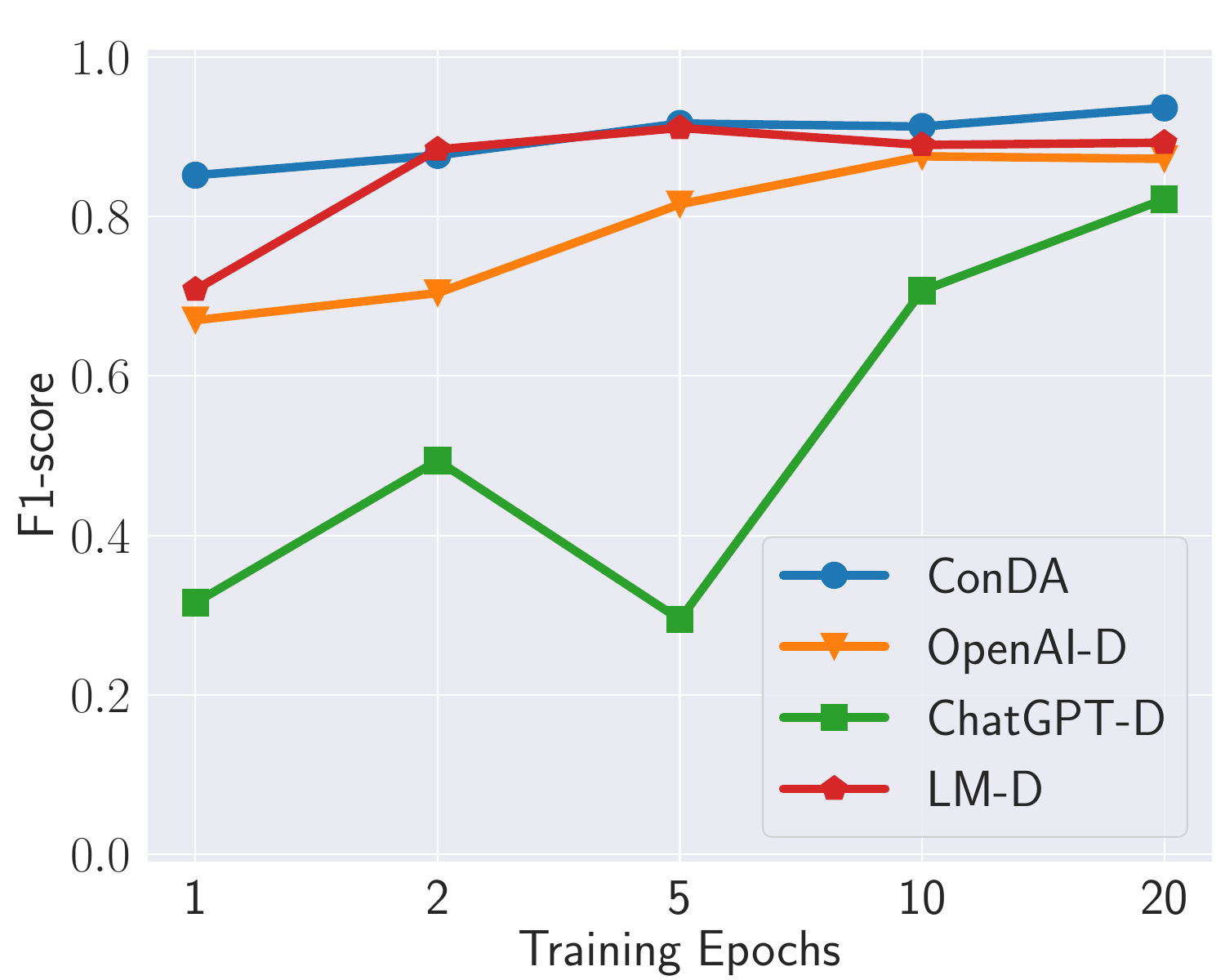}
\caption{Essay}
\label{figure:attribution_epoch_Essay}
\end{subfigure}
\begin{subfigure}{0.64\columnwidth}
\includegraphics[width=\columnwidth]{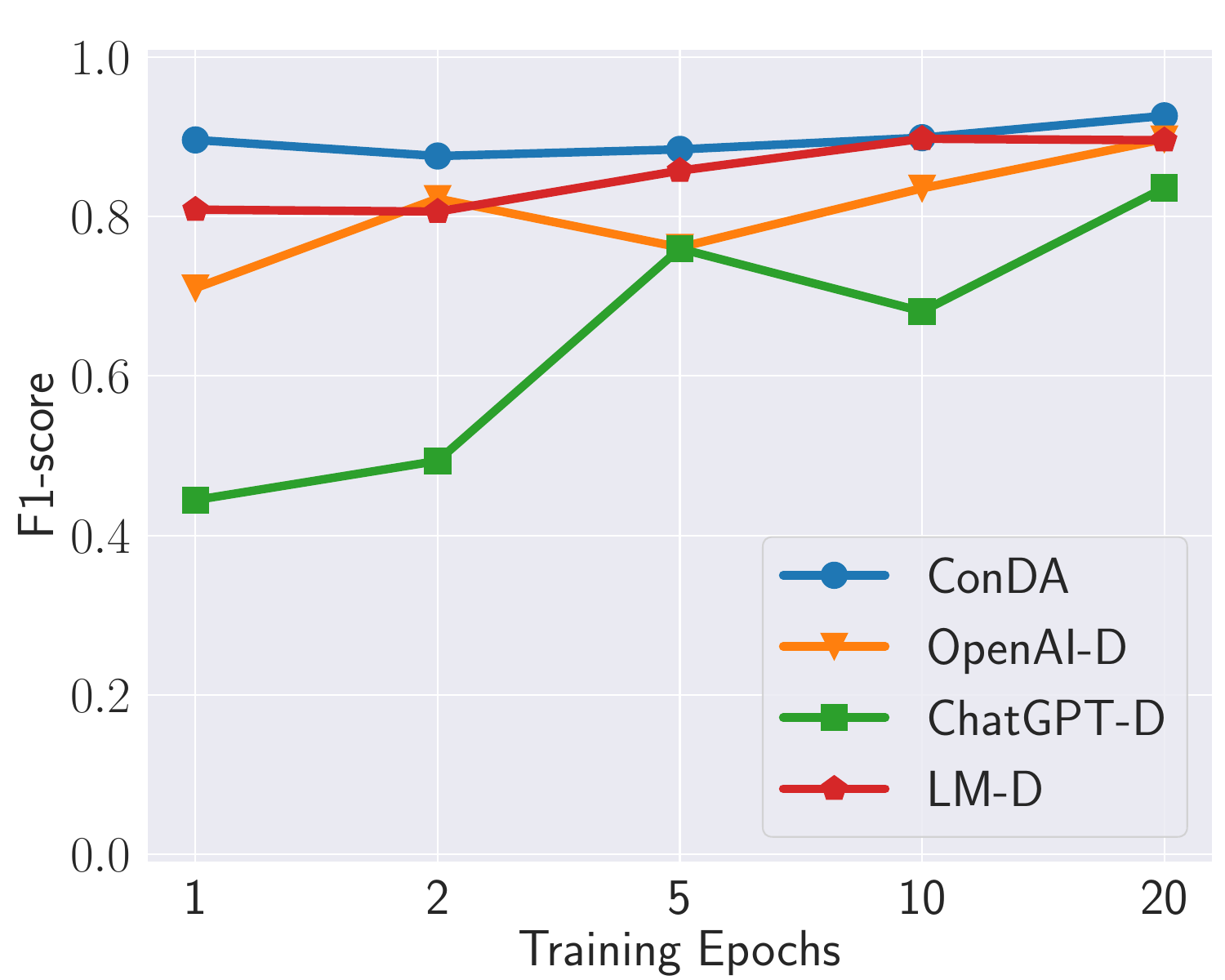}
\caption{WP}
\label{figure:attribution_epoch_WP}
\end{subfigure}
\begin{subfigure}{0.64\columnwidth}
\includegraphics[width=\columnwidth]{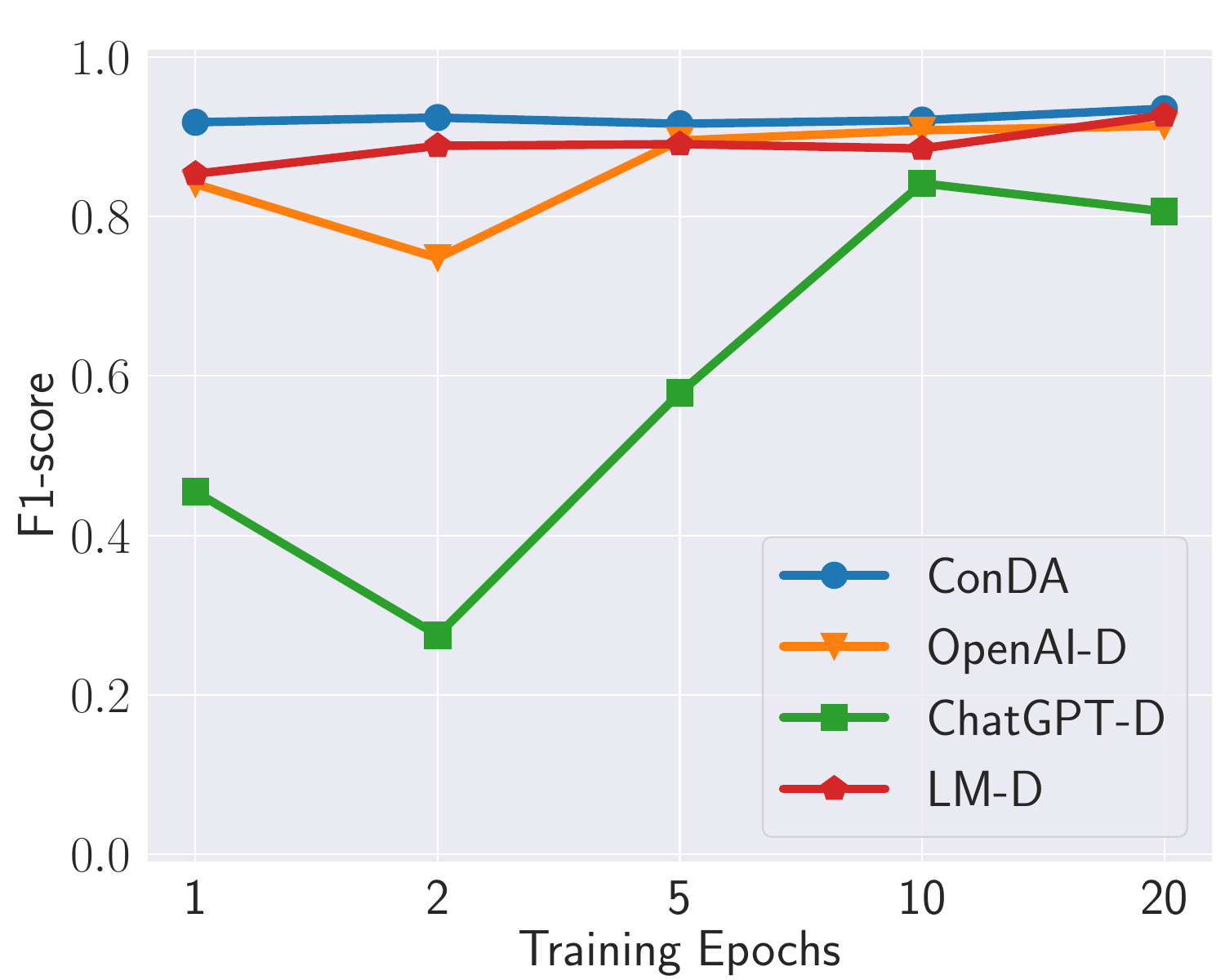}
\caption{Reuters}
\label{figure:attribution_epoch_Reuters}
\end{subfigure}
\caption{The F1-score of text attribution performance with different training epochs.}
\label{figure:attribution_epoch}
\end{figure*}

We then investigate whether the detection methods trained on one LLM can be transferred to the other LLMs as well.
The performance on Essay is summarized in \autoref{figure:ablation_transfer_llm}.
Other datasets show similar trends.
We first observe that the OpenAI Detector in general has better performance when the training and testing LLMs are different (\autoref{figure:ablation_transfer_llm_OpenAI-D_Essay}).
For instance, the OpenAI Detector trained to detect MGTs generated by ChatGPT-turbo can also detect MGTs generated by Claude with 0.941 F1-score, which only drops 0.034.
Another interesting observation is that, compared to model-based methods, metric-based methods are more robust against LLM shifts, i.e., the variance across different LLMs is smaller.

To delve deeper into the underlying reasons, we visualize the log-likelihood distributions for HWTs and MGTs (generated by different LLMs) on Essay (shown in \autoref{figure:distribution_Log-Likelihood_llm}).
We can observe that, although different LLMs have varied log-likelihood distributions, they have significantly higher log-likelihood values than HWTs, and -3.0 log-likelihood is a good split point to separate HWTs and MGTs from different LLMs.
This implies that MGTs generated by diverse LLMs differ from HWTs when evaluated using carefully designed metrics such as log-likelihood.

\subsection{Text Attribution}
\label{section:text_attribution}

\begin{figure*}[!htbp]
\centering
\begin{subfigure}{0.49\columnwidth}
\includegraphics[width=\columnwidth]{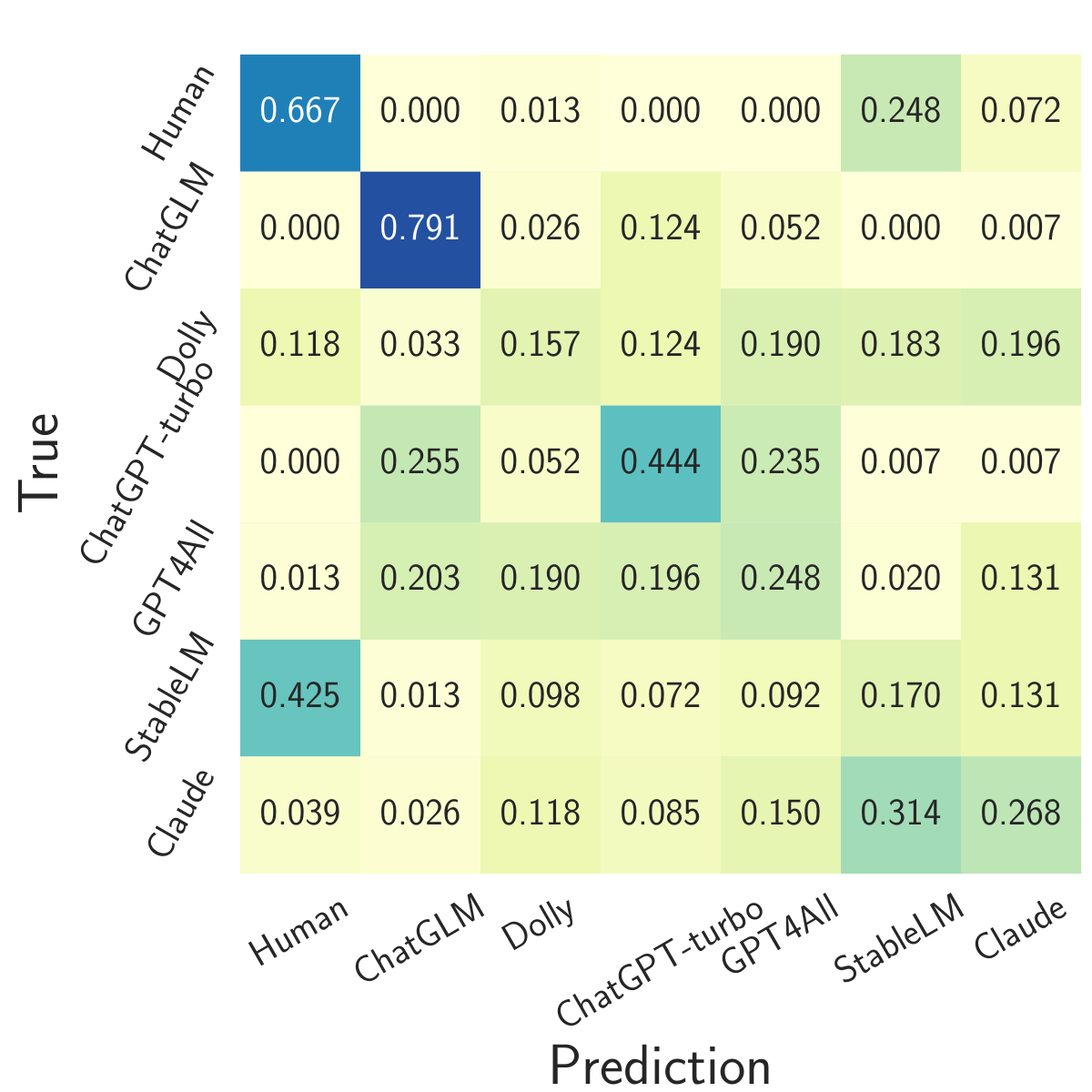}
\caption{Log-Likelihood}
\label{figure:attribution_Essay_Log-Likelihood}
\end{subfigure}
\begin{subfigure}{0.49\columnwidth}
\includegraphics[width=\columnwidth]{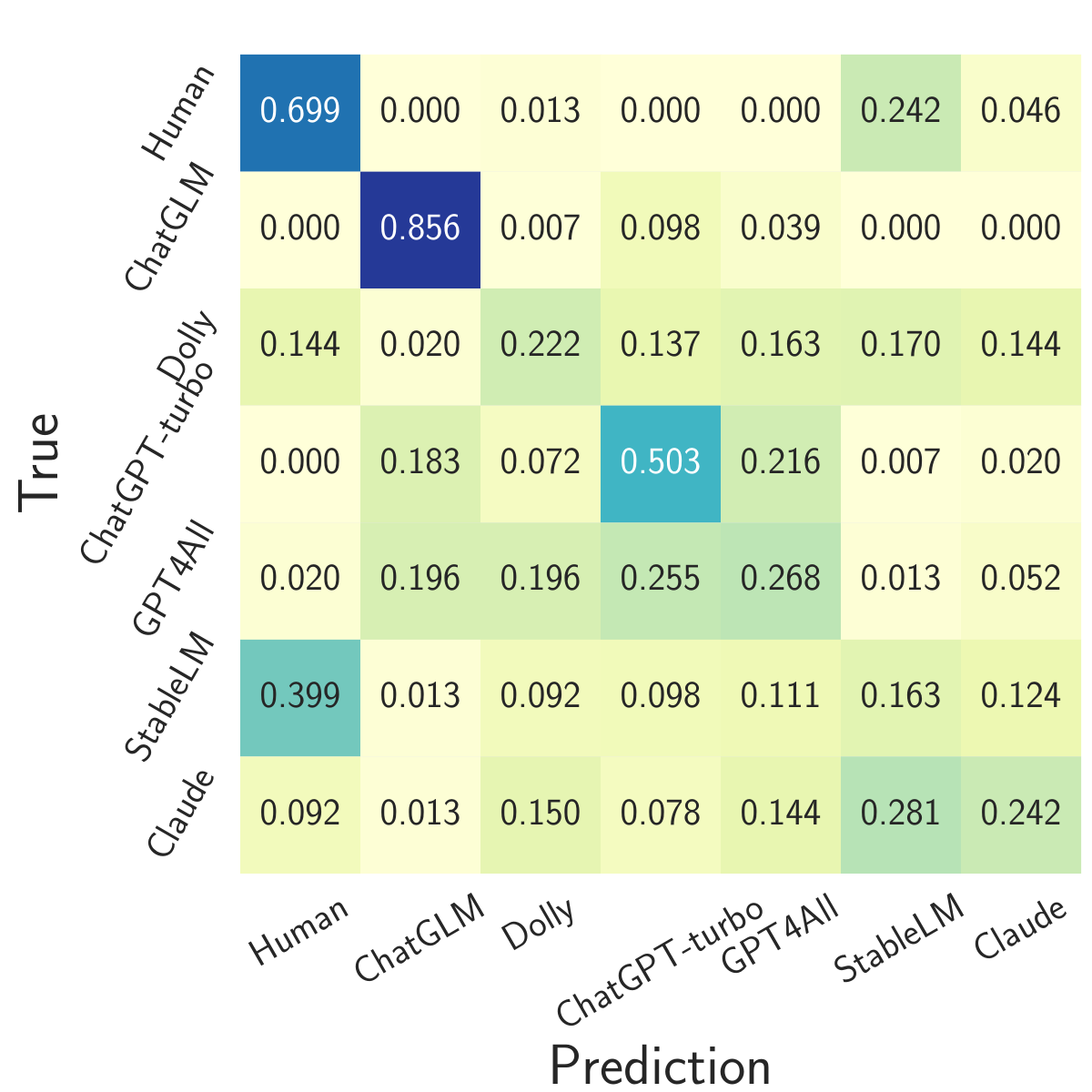}
\caption{Log-Rank}
\label{figure:attribution_Essay_Log-Rank}
\end{subfigure}
\begin{subfigure}{0.49\columnwidth}
\includegraphics[width=\columnwidth]{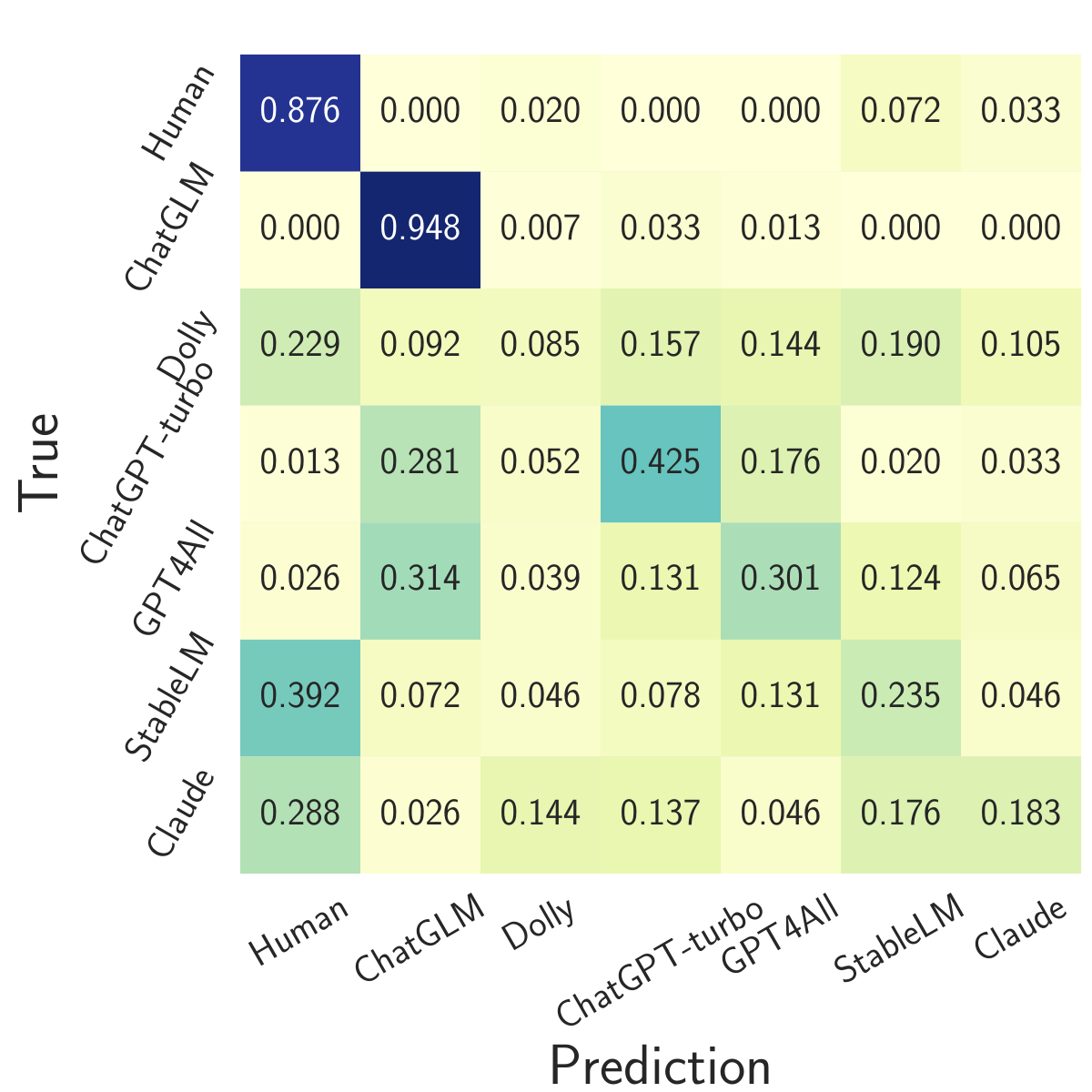}
\caption{GLTR}
\label{figure:attribution_Essay_GLTR}
\end{subfigure}
\begin{subfigure}{0.49\columnwidth}
\includegraphics[width=\columnwidth]{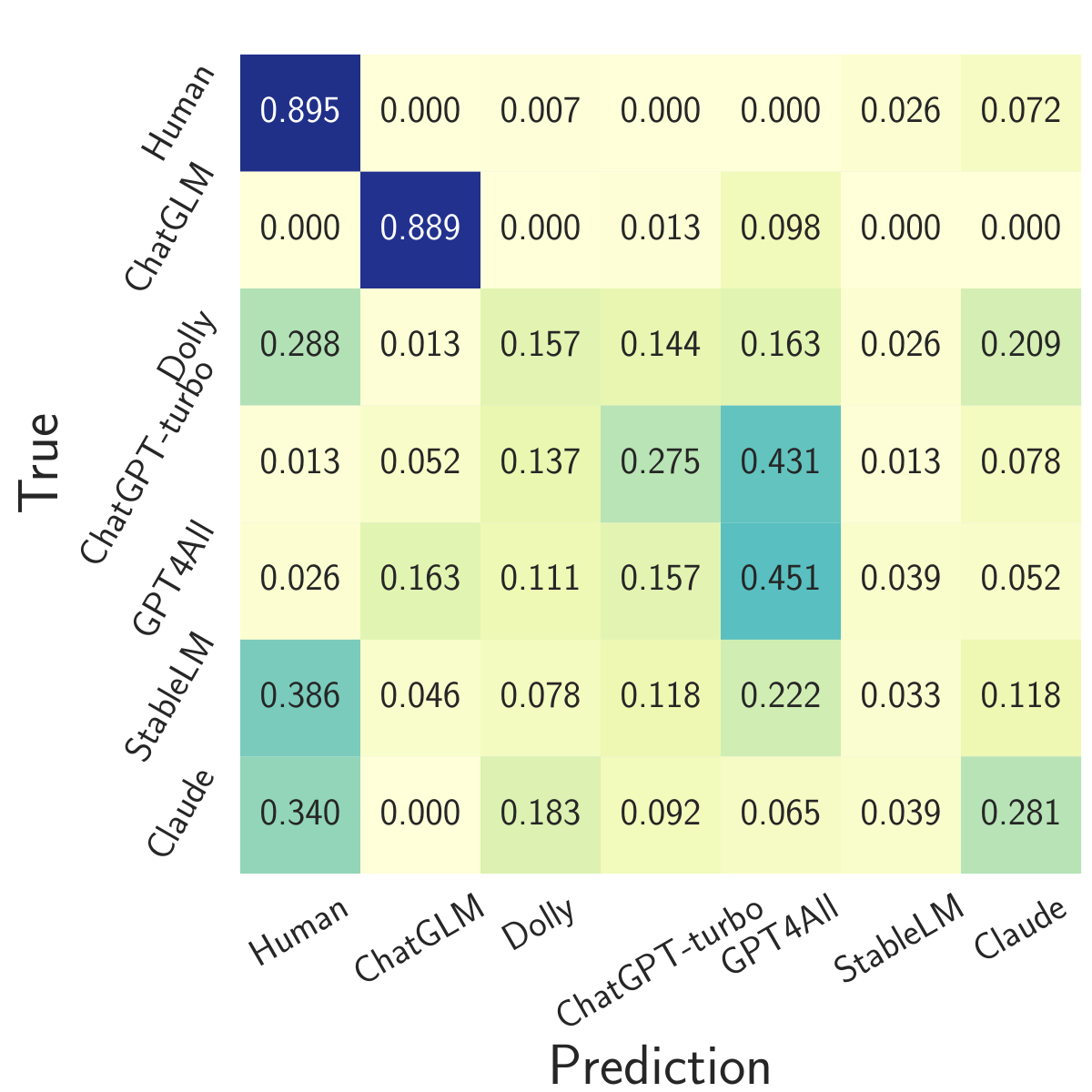}
\caption{LRR}
\label{figure:attribution_Essay_LRR}
\end{subfigure}
\begin{subfigure}{0.49\columnwidth}
\includegraphics[width=\columnwidth]{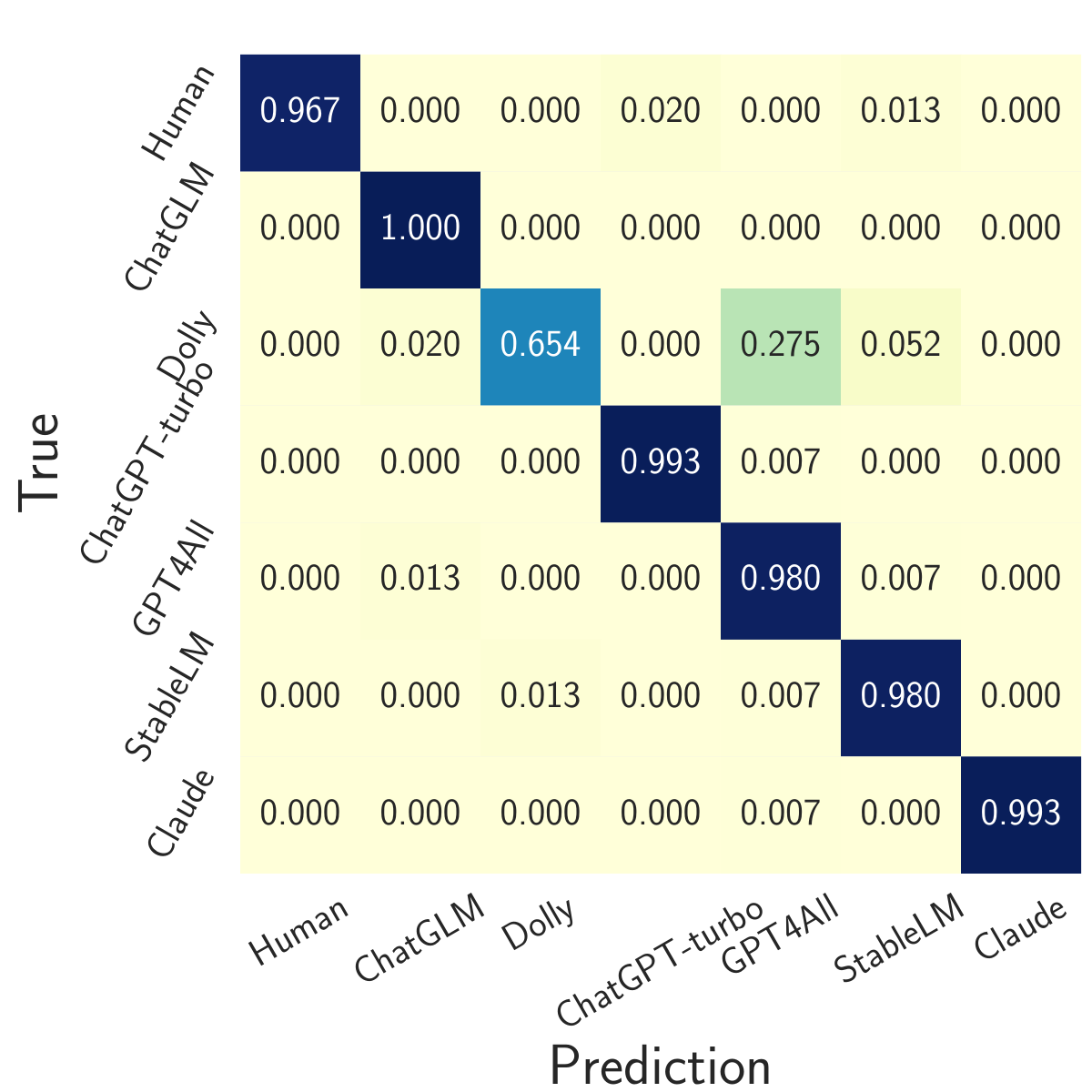}
\caption{ConDA}
\label{figure:attribution_Essay_ConDA}
\end{subfigure}
\begin{subfigure}{0.49\columnwidth}
\includegraphics[width=\columnwidth]{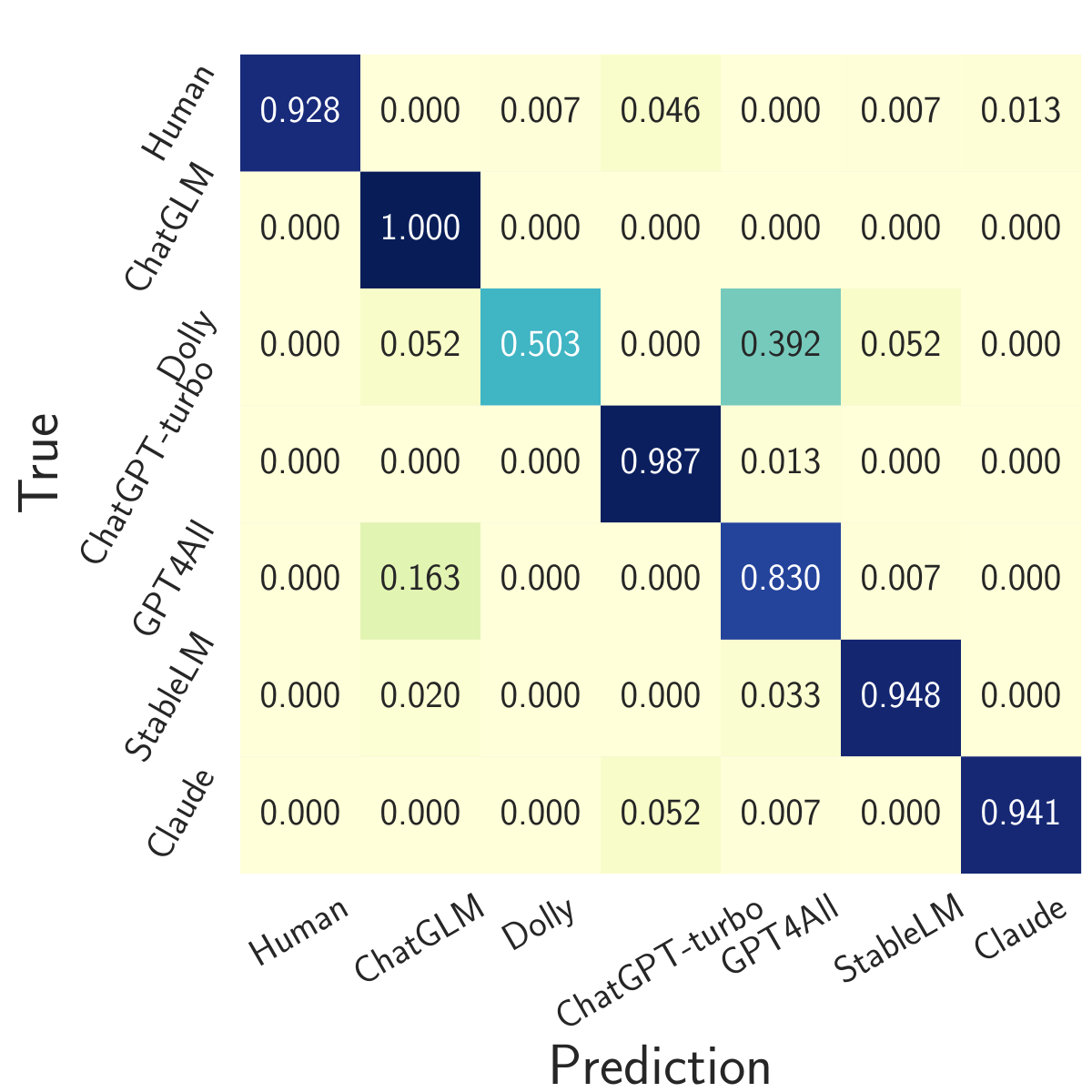}
\caption{OpenAI-D}
\label{figure:attribution_Essay_OpenAI-D}
\end{subfigure}
\begin{subfigure}{0.49\columnwidth}
\includegraphics[width=\columnwidth]{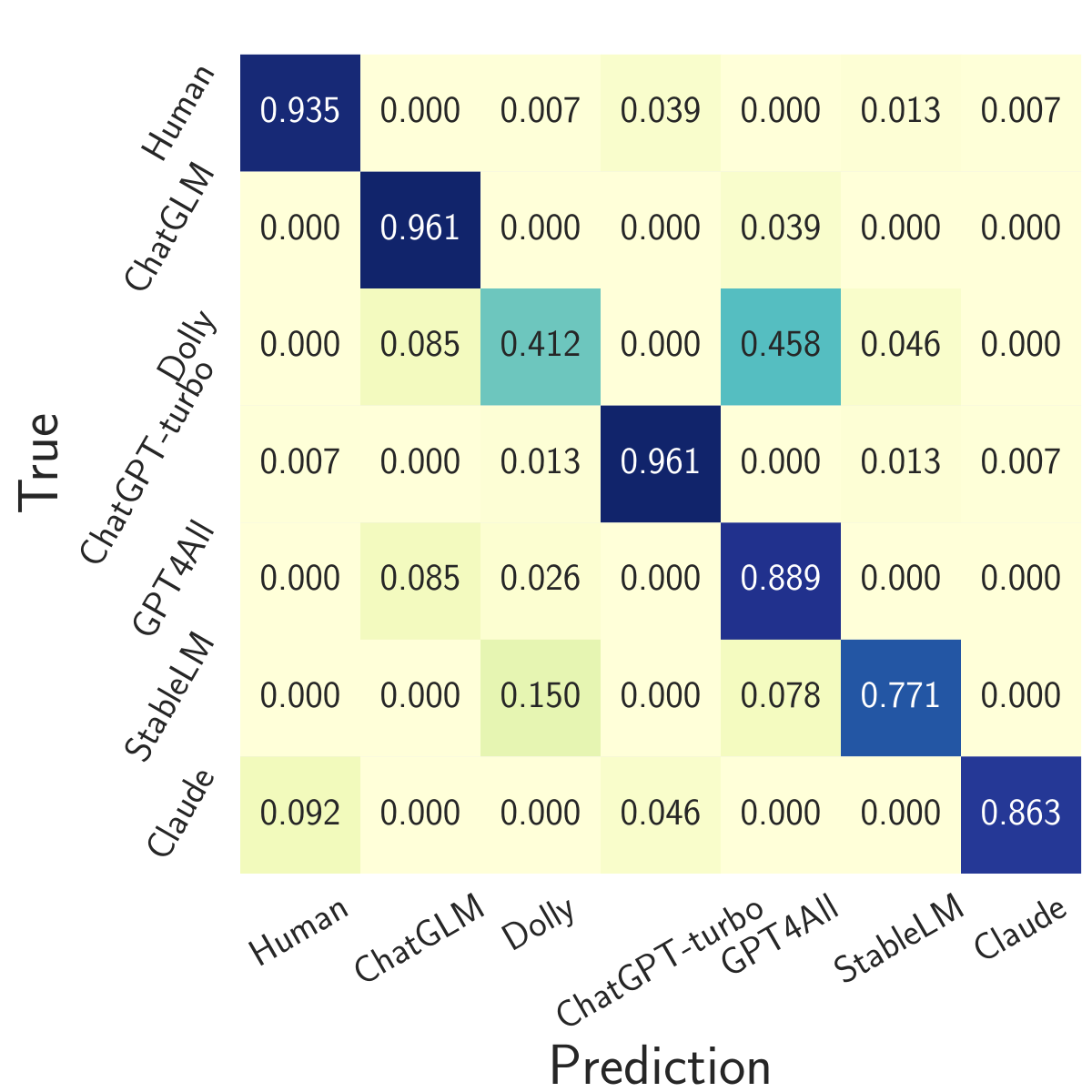}
\caption{ChatGPT-D}
\label{figure:attribution_Essay_ChatGPT-D}
\end{subfigure}
\begin{subfigure}{0.49\columnwidth}
\includegraphics[width=\columnwidth]{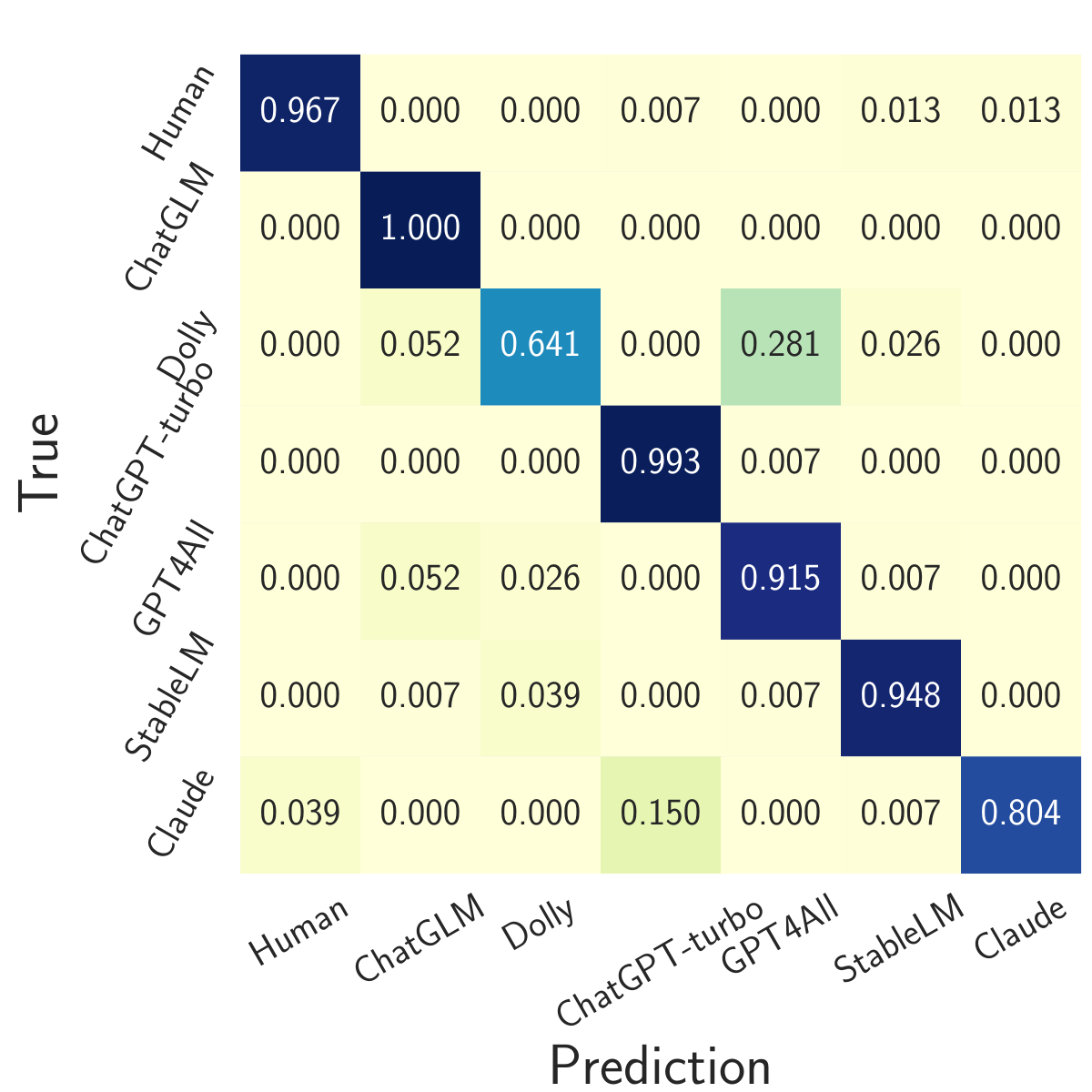}
\caption{LM-D}
\label{figure:attribution_Essay_LM-D}
\end{subfigure}
\caption{The normalized confusion matrix of text attribution with different methods on Essay. Note that the values in the diagonal represent the class-wise accuracy. The first row presents 4 (best-performing) metric-based methods. The second row shows 4 model-based methods.}
\label{figure:attribution_confusion_Essay}
\end{figure*}

Previous evaluations have shown the remarkable performance of MGT detection, especially with the model-based methods.
We now explore a more difficult task, i.e., text attribution.
Here the goal is to examine whether human-generated texts or those generated by different LLMs can be accurately attributed to their respective source models.

We extend previous MGT detection methods to the text attribution task by increasing the number of classes from 2 to 7 in the classification layer and further fine-tuning the model.

The performance of text attribution with different methods is summarized in \autoref{figure:attribution_performance}.
We find that, compared to the MGT detection task, metric-based detection methods have less satisfying performance on the text attribution task.
For instance, on Essay, Log-Likelihood reaches 0.970 F1-score in distinguishing ChatGLM-generated texts from HWTs (\autoref{table:performance}).
However, the F1-score of Log-Likelihood is only 0.374 in the text attribution task (\autoref{figure:attribution_performance}).
This is expected as the text attribution task not only needs to distinguish whether the text is generated by humans or machines but also requires the precise prediction of the source model.
However, the specific characteristics among texts generated by different LLMs cannot be captured precisely by the metric-based methods.

We also observe that model-based methods have significantly better performance than metric-based methods.
For instance, on WP, ConDA achieves a 0.926 F1-score while the F1-score is only 0.361, 0.386, and 0.378 for Log-Rank, GLTR, and LRR, respectively.
We take log-likelihood as an example (\autoref{figure:distribution_Log-Likelihood_llm}), the log-likelihood distributions for MGTs from different LLMs are hard to separate since they are overlapped.
Conversely, model-based methods can better capture the semantic and syntactic relationships between words and phrases, which greatly improves the attribution performance on different LLMs.

\mypara{Training Epochs}
We then investigate how the \#. training epochs affects the text attribution performance.
Note that here we only consider the model-based methods as they have significantly better performance than the metric-based methods.
The results are shown in \autoref{figure:attribution_epoch}.

We find that the performance keeps improving when the training epoch increases from 1 to 5, but plateaus from 5 to 20.
For instance, on WP, the F1-score of ChatGPT Detector increases from 0.444 to 0.760 when the training epoch increases from 1 to 5, while the F1-score is 0.836 with 20 epochs.
Therefore, we stop the training after 20 epochs as the performance is satisfying and the training cost is reasonable.

\mypara{Class-Wise Performance}
To better investigate the detection performance on different classes, we visualize the normalized confusion matrix of different detection methods in \autoref{figure:attribution_confusion_Essay}.
Note that here we only present the results on Essay as other datasets share similar trends.

We can see that, although metric-based methods have acceptable performance in identifying HWTs and MGTs generated by ChatGLM, the performance on attributing MGTs from other LLMs is largely limited.
Take LRR as an example (see \autoref{figure:attribution_Essay_LRR}), the prediction accuracy of Human and ChatGLM is 0.895 and 0.889, while the prediction accuracy of Dolly, ChatGPT-turbo, and Claude is only 0.157, 0.275, and 0.281, respectively.
This is expected due to potential overlap in the distribution of the metric among various LLMs, which introduces extra challenges in attribution.

On the other hand, model-based methods have almost perfect performance in identifying HWTs (over 0.928 accuracy), and better performance in attributing the source model of MGTs.
For instance, ConDA achieves 0.993, 0.980, and 0.0.993 accuracy in attributing texts generated by ChatGPT-turbo, GP4All, and Claude, respectively.
This suggests that model-based methods are more suitable for the text attribution task, as they excel in capturing context, coherence, and long-range dependencies.

Broadly speaking, our findings suggest that the model-based methods excel in the text attribution task.
However, they do exhibit challenges in accurately classifying MGTs within a specific category, e.g., Dolly.
This underscores the necessity for the development of more advanced and effective text attribution techniques.

\begin{table*}[!ht]
\centering
\caption{The performance degradation (F1-score) caused by the three attack strategies. Each cell contains three values. The first, second, and third values represent performance degradation caused by paraphrasing, random spacing, and adversarial perturbation, respectively. The best strategy in each cell is highlighted in \textbf{bold}. Note that we round the value to two decimal places to ease the reading process.}
\label{table:adv_attack_performance}
\scalebox{0.8}{
\begin{tabular}{l l | c c c c c c}
\toprule
\textbf{Dataset} & \textbf{Method} & \textbf{ChatGLM} & \textbf{Dolly} & \textbf{ChatGPT-turbo} & \textbf{GPT4All} & \textbf{StableLM} & \textbf{Claude} \\
\midrule
\multirow{10}{*}{\textbf{Essay}}   
 & Log-Likelihood & 0.37$\vert$0.56$\vert$\textbf{0.77} & 0.24$\vert$0.54$\vert$\textbf{0.65} & 0.22$\vert$\textbf{0.75}$\vert$0.72 & 0.29$\vert$\textbf{0.76}$\vert$0.70 & 0.14$\vert$\textbf{0.30}$\vert$0.22 & 0.24$\vert$\textbf{0.65}$\vert$0.50  \\
 & Rank & -0.10$\vert$0.45$\vert$\textbf{0.59} & 0.11$\vert$0.52$\vert$\textbf{0.65} & 0.09$\vert$0.83$\vert$\textbf{0.87} & 0.07$\vert$\textbf{0.67}$\vert$0.62 & 0.00$\vert$0.00$\vert$0.00 & 0.04$\vert$0.66$\vert$\textbf{0.67}  \\
 & Log-Rank & 0.47$\vert$0.41$\vert$\textbf{0.76} & 0.23$\vert$0.51$\vert$\textbf{0.63} & 0.22$\vert$0.70$\vert$\textbf{0.71} & 0.28$\vert$0.61$\vert$\textbf{0.70} & 0.18$\vert$\textbf{0.27}$\vert$0.22 & 0.21$\vert$\textbf{0.56}$\vert$0.43  \\
 & Entropy & 0.05$\vert$0.37$\vert$\textbf{0.59} & 0.20$\vert$0.37$\vert$\textbf{0.45} & 0.26$\vert$0.61$\vert$\textbf{0.71} & 0.21$\vert$0.48$\vert$\textbf{0.55} & 0.18$\vert$\textbf{0.26}$\vert$0.16 & 0.28$\vert$\textbf{0.50}$\vert$0.47  \\
 & GLTR & 0.61$\vert$0.21$\vert$\textbf{0.68} & 0.21$\vert$0.43$\vert$\textbf{0.50} & 0.19$\vert$0.52$\vert$\textbf{0.57} & 0.27$\vert$0.50$\vert$\textbf{0.67} & 0.21$\vert$\textbf{0.22}$\vert$\textbf{0.22} & 0.23$\vert$\textbf{0.44}$\vert$0.37  \\
 & LRR & \textbf{0.83}$\vert$0.27$\vert$0.77 & 0.22$\vert$0.34$\vert$\textbf{0.54} & 0.25$\vert$0.55$\vert$\textbf{0.70} & 0.32$\vert$0.39$\vert$\textbf{0.68} & \textbf{0.28}$\vert$\textbf{0.28}$\vert$0.27 & 0.20$\vert$\textbf{0.35}$\vert$\textbf{0.35}  \\
 & ConDA & \textbf{0.99}$\vert$0.00$\vert$0.47 & 0.00$\vert$0.00$\vert$0.00 & 0.03$\vert$0.01$\vert$\textbf{0.54} & 0.00$\vert$0.01$\vert$\textbf{0.48} & 0.00$\vert$0.00$\vert$\textbf{0.10} & \textbf{0.32}$\vert$0.02$\vert$0.12  \\
 & OpenAI-D & 0.12$\vert$0.18$\vert$\textbf{0.73} & -0.00$\vert$0.10$\vert$\textbf{0.57} & 0.00$\vert$0.13$\vert$\textbf{0.84} & 0.00$\vert$0.31$\vert$\textbf{0.68} & 0.00$\vert$\textbf{0.20}$\vert$0.16 & 0.21$\vert$0.00$\vert$\textbf{0.38}  \\
 & ChatGPT-D & 0.30$\vert$\textbf{0.94}$\vert$0.89 & 0.00$\vert$0.10$\vert$\textbf{0.23} & -0.01$\vert$\textbf{0.86}$\vert$0.33 & 0.01$\vert$0.18$\vert$\textbf{0.57} & -0.01$\vert$\textbf{0.17}$\vert$0.12 & -0.02$\vert$\textbf{0.23}$\vert$0.17  \\
 & LM-D & \textbf{1.00}$\vert$0.00$\vert$0.04 & 0.00$\vert$0.07$\vert$\textbf{0.88} & 0.00$\vert$0.00$\vert$\textbf{0.97} & 0.00$\vert$0.01$\vert$\textbf{0.69} & 0.00$\vert$0.00$\vert$\textbf{0.16} & 0.00$\vert$0.01$\vert$\textbf{0.94}  \\
\midrule
\multirow{10}{*}{\textbf{WP}}
 & Log-Likelihood & 0.75$\vert$0.43$\vert$\textbf{0.90} & 0.49$\vert$0.45$\vert$\textbf{0.62} & 0.53$\vert$\textbf{0.60}$\vert$0.47 & 0.61$\vert$0.63$\vert$\textbf{0.82} & \textbf{0.50}$\vert$0.45$\vert$0.15 & \textbf{0.65}$\vert$0.61$\vert$0.40  \\
 & Rank & 0.04$\vert$0.33$\vert$\textbf{0.84} & 0.18$\vert$0.36$\vert$\textbf{0.70} & 0.28$\vert$\textbf{0.63}$\vert$0.48 & 0.23$\vert$0.55$\vert$\textbf{0.86} & 0.25$\vert$0.33$\vert$\textbf{0.50} & 0.26$\vert$\textbf{0.48}$\vert$\textbf{0.48}  \\
 & Log-Rank & 0.75$\vert$0.35$\vert$\textbf{0.87} & 0.48$\vert$0.41$\vert$\textbf{0.63} & 0.51$\vert$\textbf{0.56}$\vert$0.47 & 0.58$\vert$0.44$\vert$\textbf{0.77} & \textbf{0.53}$\vert$0.43$\vert$0.14 & \textbf{0.60}$\vert$0.52$\vert$0.36  \\
 & Entropy & 0.34$\vert$0.25$\vert$\textbf{0.56} & 0.29$\vert$0.25$\vert$\textbf{0.43} & \textbf{0.44}$\vert$0.40$\vert$0.42 & 0.41$\vert$0.40$\vert$\textbf{0.58} & \textbf{0.38}$\vert$0.26$\vert$0.10 & \textbf{0.50}$\vert$0.41$\vert$0.32  \\
 & GLTR & \textbf{0.73}$\vert$0.16$\vert$0.60 & 0.42$\vert$0.31$\vert$\textbf{0.50} & 0.45$\vert$\textbf{0.46}$\vert$\textbf{0.46} & 0.53$\vert$0.38$\vert$\textbf{0.63} & \textbf{0.56}$\vert$0.34$\vert$0.09 & \textbf{0.56}$\vert$0.44$\vert$0.27  \\
 & LRR & \textbf{0.79}$\vert$0.20$\vert$0.78 & 0.44$\vert$0.22$\vert$\textbf{0.59} & 0.42$\vert$0.41$\vert$\textbf{0.47} & 0.60$\vert$0.27$\vert$\textbf{0.64} & \textbf{0.64}$\vert$0.33$\vert$0.14 & \textbf{0.48}$\vert$0.31$\vert$0.21  \\
 & ConDA & 0.00$\vert$0.00$\vert$\textbf{0.01} & 0.00$\vert$0.00$\vert$\textbf{0.07} & 0.00$\vert$0.00$\vert$\textbf{0.02} & 0.01$\vert$0.01$\vert$\textbf{0.09} & \textbf{0.01}$\vert$0.00$\vert$0.00 & \textbf{0.86}$\vert$0.12$\vert$0.18  \\
 & OpenAI-D & 0.01$\vert$0.03$\vert$\textbf{0.69} & 0.01$\vert$0.10$\vert$\textbf{0.59} & 0.00$\vert$0.00$\vert$\textbf{0.36} & 0.00$\vert$0.08$\vert$\textbf{0.58} & 0.01$\vert$\textbf{0.11}$\vert$0.09 & \textbf{0.18}$\vert$0.00$\vert$0.05  \\
 & ChatGPT-D & -0.01$\vert$0.90$\vert$\textbf{0.91} & -0.01$\vert$0.16$\vert$\textbf{0.21} & -0.00$\vert$\textbf{0.74}$\vert$0.30 & -0.00$\vert$\textbf{0.66}$\vert$0.48 & -0.01$\vert$\textbf{0.50}$\vert$0.05 & 0.04$\vert$\textbf{0.37}$\vert$0.07  \\
 & LM-D & 0.00$\vert$0.02$\vert$\textbf{0.99} & -0.00$\vert$0.04$\vert$\textbf{0.89} & 0.00$\vert$0.00$\vert$0.55 & 0.00$\vert$0.11$\vert$\textbf{0.96} & 0.14$\vert$0.42$\vert$\textbf{0.94} & -0.00$\vert$0.06$\vert$\textbf{0.97}  \\
\midrule
\multirow{10}{*}{\textbf{Reuters}}
 & Log-Likelihood & 0.57$\vert$\textbf{0.78}$\vert$0.69 & -0.23$\vert$-0.38$\vert$\textbf{-0.21} & 0.31$\vert$0.62$\vert$\textbf{0.76} & 0.31$\vert$\textbf{0.64}$\vert$0.51 & 0.35$\vert$\textbf{0.53}$\vert$0.34 & 0.44$\vert$\textbf{0.77}$\vert$0.62  \\
 & Rank & 0.02$\vert$0.46$\vert$\textbf{0.49} & \textbf{-0.09}$\vert$-0.40$\vert$-0.22 & 0.11$\vert$0.80$\vert$\textbf{0.82} & 0.13$\vert$\textbf{0.56}$\vert$0.46 & 0.12$\vert$\textbf{0.47}$\vert$0.32 & 0.04$\vert$\textbf{0.62}$\vert$0.52  \\
 & Log-Rank & 0.54$\vert$0.59$\vert$\textbf{0.70} & \textbf{-0.21}$\vert$-0.37$\vert$-0.22 & 0.28$\vert$0.49$\vert$\textbf{0.76} & 0.33$\vert$\textbf{0.63}$\vert$0.52 & 0.37$\vert$\textbf{0.56}$\vert$0.36 & 0.41$\vert$\textbf{0.74}$\vert$0.62  \\
 & Entropy & 0.19$\vert$\textbf{0.37}$\vert$\textbf{0.37} & -0.16$\vert$-0.23$\vert$\textbf{-0.14} & 0.32$\vert$\textbf{0.63}$\vert$0.61 & \textbf{-0.09}$\vert$-0.16$\vert$-0.17 & \textbf{-0.11}$\vert$-0.16$\vert$-0.15 & 0.42$\vert$\textbf{0.63}$\vert$0.52  \\
 & GLTR & 0.52$\vert$0.32$\vert$\textbf{0.68} & \textbf{0.13}$\vert$-0.02$\vert$-0.10 & 0.33$\vert$0.33$\vert$\textbf{0.67} & 0.30$\vert$\textbf{0.56}$\vert$0.50 & 0.33$\vert$\textbf{0.50}$\vert$0.36 & 0.39$\vert$\textbf{0.65}$\vert$0.61  \\
 & LRR & 0.54$\vert$0.32$\vert$\textbf{0.70} & 0.22$\vert$\textbf{0.33}$\vert$0.19 & 0.35$\vert$0.31$\vert$\textbf{0.74} & 0.30$\vert$0.48$\vert$\textbf{0.54} & 0.39$\vert$\textbf{0.47}$\vert$0.46 & 0.29$\vert$0.54$\vert$\textbf{0.57}  \\
 & ConDA & 0.00$\vert$0.00$\vert$0.00 & 0.00$\vert$0.00$\vert$\textbf{0.01} & 0.00$\vert$0.00$\vert$0.00 & 0.00$\vert$0.00$\vert$0.00 & -0.01$\vert$-0.01$\vert$\textbf{0.06} & 0.00$\vert$0.00$\vert$0.00  \\
 & OpenAI-D & 0.01$\vert$0.10$\vert$\textbf{0.51} & 0.00$\vert$\textbf{0.23}$\vert$0.13 & -0.00$\vert$0.18$\vert$\textbf{0.80} & 0.00$\vert$0.22$\vert$\textbf{0.35} & 0.00$\vert$\textbf{0.18}$\vert$0.14 & 0.05$\vert$0.00$\vert$\textbf{0.58}  \\
 & ChatGPT-D & 0.01$\vert$\textbf{0.90}$\vert$0.79 & -0.02$\vert$0.07$\vert$\textbf{0.15} & 0.00$\vert$\textbf{0.74}$\vert$0.53 & 0.00$\vert$\textbf{0.55}$\vert$0.50 & -0.00$\vert$\textbf{0.41}$\vert$0.20 & -0.08$\vert$0.29$\vert$\textbf{0.47}  \\
 & LM-D & 0.00$\vert$0.00$\vert$\textbf{0.71} & 0.00$\vert$0.00$\vert$\textbf{0.20} & -0.01$\vert$0.01$\vert$\textbf{0.99} & 0.00$\vert$0.00$\vert$\textbf{0.59} & -0.01$\vert$0.01$\vert$\textbf{0.36} & -0.00$\vert$0.01$\vert$\textbf{0.74}  \\
\bottomrule
\end{tabular}
}
\end{table*}

\subsection{Adversarial Attacks}

Our previous evaluation demonstrates the effectiveness of MGT detection methods.
We then take a step further to evaluate whether those methods are robust against adversarial attacks.
Concretely, we consider three different attack strategies:
\begin{itemize}
    \item \mypara{Paraphrasing} This attack replaces each sentence with its paraphrase using a paraphrasing model.
    \item \mypara{Random Spacing} This attack randomly inserts spaces into the given texts.    
    \item \mypara{Adversarial Perturbation} This attack optimizes perturbations to deceive a target detector.
\end{itemize}
we evaluate the attack performance on the MGT detection task and we consider the fine-tuned version of all model-based methods.
For paraphrasing, we consider ChatGPT Paraphraser\footnote{\url{https://huggingface.co/datasets/humarin/chatgpt-paraphrases}.} as the paraphrasing model.
As to random spacing, we consider inserting space in each position of a text with 1\% probability.
Regarding adversarial perturbation, we consider the attack proposed by Ren et al.~\cite{RDHC19} and leverage the TextAttack\footnote{\url{https://github.com/QData/TextAttack/}.} library to implement the attack.
We consider the LM Detector as the target detector for adversarial perturbation.
Note that the attack only perturbs MGTs by applying one attack strategy.
The HWTs remain unchanged.

We evaluate the effectiveness of the attack by measuring the F1-score degradation after applying the attack.
The results are summarized in \autoref{table:adv_attack_performance}.
We observe that, in general, adversarial perturbation is the most effective strategy.
For instance, on Essay with ChatGPT-turbo, the adversarial perturbation degrades 0.97 F1-score for LM Detector.
Also, this perturbation can transfer to other detection methods as well, e.g., the F1-score degradation is 0.84, 0.54, and 0.70 for OpenAI Detector, ConDA, and LRR on Essay with ChatGPT-turbo.
Another observation is that paraphrasing and random spacing are more effective against metric-based methods than model-based methods.
For example, on Reuters with Claude, paraphrasing and random spacing degrade the F1-score of Log-Likelihood by 0.44 and 0.77, while only 0.05 and 0.00 for OpenAI-D.
This could be attributed to the fact that paraphrasing or random spacing alters word distribution, potentially influencing the metrics calculated by metric-based methods.
However, since these methods do not significantly change the semantic meaning, their effect on model-based methods is comparatively lower.

Overall, our findings indicate that MGT detection methods are vulnerable to potential adversarial attacks.
This prompts the need for developing more robust detection methods.
Possible directions could be combining input filtering, data augmentation, and adversarial training into existing detection methods.

\section{Limitations and Discussions}

\mypara{Choice of LLM/Methods/Datasets}
In our current study, we have concentrated on 6 representative LLMs and 13 detection methods on 3 benchmarking datasets.
We agree that experiments with new LLMs with billions of parameters would be fantastic, but they would exhaust our technical capabilities and provide little further insights.
Regarding methods, our implemented methods are mainly based on pre-defined metrics (metric-based) or pre-trained LMs (model-based).
Nevertheless, as LLMs continue to advance, we recognize the potential of incorporating novel approaches such as prompt tuning and in-context learning into MGTBench as well.
These methodologies have demonstrated remarkable efficacy and efficiency in adapting to new tasks, suggesting a valuable direction for future exploration and integration within our framework.
Our modular design of MGTBench is a significant strength, enabling seamless scalability and adaptability.
This flexibility allows for the incorporation of an array of new detection methods and LLMs as they emerge, ensuring that MGTBench remains at the forefront of developments in this rapidly evolving field.
Regarding the datasets, our current datasets cover aspects ranging from news articles (Reuters) to creative prompts (WP).
We acknowledge that there are also other domains of datasets with varied topics, word counts, and even languages.
Expanding our datasets to encompass this diversity would undoubtedly enrich the development of MGT detection methods.

\mypara{Security Implication}
Designing and benchmarking MGT detection methods has important security implications in the following aspects.
First, detecting MGT effectively can stifle the spread of automated misinformation campaigns, which can be used to manipulate public sentiment, sway elections, or incite unrest.
Second, Phishing emails or scam messages can be tailored using generative AI.
A robust MGT detection system can help in identifying and blocking such attempts, safeguarding individuals and institutions from potential financial losses.
Also, for sectors like journalism, academia, and legal processes, the authenticity of information is paramount
MGT detection ensures the integrity of content in these fields, thereby preserving trust.
Moreover, as AI systems become more integral in various applications, adversarial attacks, where AI-generated content is used to fool other AI systems, will become more prevalent.
An MGT detection framework can act as a first line of defense against such attacks.
Another implication is that as users become more aware of AI's capabilities, there's a growing skepticism towards digital content.
Effective MGT detection can alleviate these concerns, ensuring users can trust the platforms they engage with.

\mypara{Broader Implication in Generative AI}
MGTBench provides a uniform platform where different detection methodologies can be evaluated and compared.
This standardization can drive the development of more effective, efficient, and universally accepted MGT detection techniques.
Also, knowing the strengths and weaknesses of generative models through a detection benchmark can guide researchers in refining these models, potentially leading to more advanced and indistinguishable MGTs.
Moreover, our work signifies the importance of the MGT detection issue, raising awareness among the public.
This can lead to better-informed consumers of digital content and a more nuanced understanding of generative AI's capabilities and limitations.

\section{Conclusion}

In this paper, we perform the first systematic quantification of existing MGT detection methods under the representative powerful LLMs.
Specifically, we consider 8 metric-based detection methods and 5 model-based detection methods.
Our analysis reveals that the LM Detector consistently excels across various datasets.
Additionally, our ablation studies indicate a shortcoming of current methods in accurately distinguishing MGTs with a limited number of words.
Interestingly, most detection methods demonstrate the ability to achieve satisfactory performance with considerably fewer training samples.
Moreover, we observe that metric-based methods exhibit better adaptability to different LLMs, whereas model-based methods show superior flexibility in transitioning across diverse datasets.

We also explore the feasibility of applying existing MGT detection methods to text attribution, a notably more complex task.
Our findings indicate that model-based methods significantly outperform their metric-based counterparts.
This effectiveness is attributed to the model-based approaches' proficiency in capturing semantic and syntactic nuances within texts, whereas metric-based methods, dependent mainly on specific metrics, struggle to differentiate between source LLMs in the text attribution task.

Further, we assess the robustness of MGT detection methods by introducing three different adversarial attacks, namely paraphrasing, random spacing, and adversarial perturbation.
Our evaluation uncovers a pronounced vulnerability across various detection methods to these attacks.
This assessment highlights the need to develop more robust MGT detection methods.

To facilitate research in this domain, we integrate the detection methods as well as datasets into a modular-designed framework named MGTBench.
We envision that MGTBench will serve as a benchmark tool to expedite future research endeavors in enhancing MGT detection methodologies and refining the training processes of LLMs.

\newpage
\bibliographystyle{plain}
\bibliography{normal_generated_py3,others}

\end{document}